\documentclass[prd,aps,eqsecnum,amsmath,floatfix,nofootinbib,preprint,tightenlines]{revtex4}

\usepackage{latexsym}
\usepackage{graphicx}
\usepackage{multirow}
\usepackage[dvipsnames]{xcolor}
\usepackage{hyperref}

\def\ttl#1{{\it #1}}

\def\floatcaption#1#2{ \caption{#2 \label{#1}} }

\def\bibi{\bibitem}

\def\a{\alpha}
\def\b{\beta}
\def\c{\chi}
\def\d{\delta}
\def\e{\epsilon}                
\def\f{\phi}                    
\def\g{\gamma}
\def\h{\eta}

\def\j{\psi}
\def\k{\kappa}
\def\l{\lambda}
\def\m{\mu}
\def\n{\nu}

\def\p{\pi}                     
\def\r{\rho}                    
\def\s{\sigma}                  
\def\t{\tau}

\def\z{\zeta}

\def\F{\Phi}
\def\G{\Gamma}
\def\J{\Psi}
\def\L{\Lambda}
\def\O{\Omega}
\def\P{\Pi}

\def\S{\Sigma}
\def\U{\Upsilon}

\def\ca{{\cal A}}

\def\cc{{\cal C}}
\def\cd{{\cal D}}

\def\cf{{\cal F}}

\def\cl{{\cal L}}

\def\cn{{\cal N}}

\def\cq{{\cal Q}}

\def\cs{{\cal S}}
\def\ct{{\cal T}}

\def\cbo{{\,\raise-.15ex\Sc [\,}}                       

\def\Sl#1{\rlap{\hbox{$\mskip 3 mu /$}}#1}      
\def\vev#1{\Big\langle #1 \Big\rangle}           

\def\svev#1{\left\langle #1\right\rangle}       

\def\ddt#1{{\buildrel {\hbox{\LARGE .\kern-2pt.}} \over {#1}}}

\def\ie{\mbox{\it i.e.}}
\def\eg{\mbox{\it e.g.}}

\def\leqx{\,\raisebox{-1.0ex}{$\stackrel{\textstyle <}{\sim}$}\,}

\def\tr{{\rm tr}\,}

\def\hc{{\rm h.c.\,}}

\def\half{{1\over 2}}
\def\Re{{\rm Re\,}}
\def\Im{{\rm Im\,}}

\def\seef{{\it cf.\  }}

\def\tr{\,{\rm tr}}
\def\vev#1{\langle #1\rangle}

\def\Vefftop{V_{\rm eff}^{\rm top}}
\def\Veff{V_{\rm eff}}
\def\Ctop{C^{\rm top}}
\def\one{\mbox{\bf 1}}
\def\three{\mbox{\bf 3}}
\def\threebar{\mbox{${\bf \overline{3}}$}}
\def\five{\mbox{\bf 5}}
\def\fivebar{\mbox{${\bf \overline{5}}$}}
\def\sixbar{\mbox{${\bf \overline{6}}$}}
\def\sone{{\bf 1}}
\def\sthree{{\bf 3}}
\def\sthreebar{{\bf \overline{3}}}
\def\sfive{{\bf 5}}
\def\sfivebar{{\bf \overline{5}}}
\def\irrep{\textit{irrep}}
\def\irreps{\textit{irreps}}

\def\tA{\tilde{A}}
\def\tB{\tilde{B}}

\def\tileta{\tilde{\eta}}
\def\tPhi{\tilde{\Phi}}
\def\tl{\tilde{\l}}

\def\tx{\tilde{x}}

\def\tileta{\tilde\eta}

\def\ttr{\widehat{\tr}}
\def\CPSM{\widehat{CP}}

\def\hatb{\hat{b}}
\def\hatc{\hat{c}}

\def\hs{\hat{s}}
\def\hatt{\hat{t}}

\def\hP{\hat{\Pi}}
\def\hF{\hat{\Phi}}
\def\hX{\hat{X}}
\def\hbX{\hat{\overline{X}}}

\def\ba{\bar{a}}
\def\bb{\overline{b}}
\def\bq{\overline{q}}
\def\bj{\overline{\j}}
\def\bJ{\overline{\J}}
\def\bh{\overline{\eta}}
\def\bc{\overline{\chi}}

\def\bU{\overline{\U}}
\def\bTheta{\overline{\Theta}}
\def\bB{\overline{B}}
\def\bt{\overline{t}}
\def\bA{\overline{A}}
\def\bS{\overline{S}}
\def\bD{\overline{D}}
\def\bN{\overline{N}}
\def\bX{\overline{X}}

\def\bca{\overline{\ca}}
\def\bcd{\overline{\cd}}
\def\bcs{\overline{\cs}}

\def\bT{\overline{T}}

\begin{document}

\begin{center}
{\large\bf Effective potential in 
  ultraviolet completions for composite Higgs models}\\[8mm]
Maarten Golterman$^a$ and Yigal Shamir$^b$\\[10 mm]
{\small
$^a$Department of Physics and Astronomy, San Francisco State University,\\
San Francisco, CA 94132, USA\\
$^b$Raymond and Beverly Sackler School of Physics and Astronomy,\\
Tel~Aviv University, 69978, Tel~Aviv, Israel}\\[10mm]
\end{center}

\begin{quotation}
We consider a class of composite Higgs models based on
asymptotically free $SO(d)$ gauge theories with $d$ odd,
with fermions in two irreducible representations, and in which
the Higgs field arises as a pseudo Nambu-Goldstone boson
and the top quark is partially composite.  The Nambu-Goldstone coset
containing the Higgs field, or Higgs coset,
is either $SU(4)/Sp(4)$ or $SU(5)/SO(5)$, whereas the
top partners live in two-index representations of the relevant flavor group
($SU(4)$ or $SU(5)$).  In both cases,  there is a large number of terms
in the most general four-fermion lagrangian describing the interaction
of third-generation quarks with the top partners.  We derive the
top-induced effective potential for the
Higgs coset together with the singlet pseudo Nambu-Goldstone boson
associated with the non-anomalous axial symmetry,
to leading order in the couplings between the third-generation quarks and
the composite sector.  We obtain expressions for the low-energy constants
in terms of top-partner two-point functions.
We revisit the effective potential of another composite Higgs model
that we have studied previously, which is based on an $SU(4)$ gauge theory
and provides a different realization of the $SU(5)/SO(5)$ coset.
The top partners of this model live in the fundamental representation
of $SU(5)$, and, as a result, the effective potential
of this model is qualitatively different from the $SO(d)$ gauge theories.
We also discuss the role of the isospin-triplet fields
contained in the $SU(5)/SO(5)$ coset, and show that, without
further constraints on the four-fermion couplings, an expectation value
for the the Higgs field will trigger the subsequent condensation
of an isospin-triplet field.
\end{quotation}

\newpage
\section{\label{introduction} Introduction}
Among the mechanisms that have been proposed to keep the Higgs particle
naturally light, the so-called composite Higgs paradigm \cite{GK,DGK}
postulates the existence of a new strong sector, perhaps in the few TeV range,
based on an asymptotically free gauge theory
that we will call hypercolor.  Spontaneous chiral symmetry breaking
in the hypercolor theory produces a set of Nambu-Goldstone bosons (NGBs).
When we couple the Standard Model and the hypercolor theory,
this breaks explicitly the flavor symmetry group of the hypercolor theory to a
smaller group, thereby generating an effective potential for the now pseudo
Nambu-Goldstone bosons (pNGBs).  The Higgs doublet is composed of four
of these pNGBs, and it is assumed that minimizing the effective potential
triggers electroweak symmetry breaking.

As the only fermion in the Standard Model
with a mass comparable to the electroweak scale,
the top quark is usually given a special role in composite-Higgs models.
We will assume that the top quark couples linearly to baryons
of the hypercolor sector, which we will refer to as hyperbaryons.  The observed
top-quark mass eigenstate is a linear superposition of
the elementary top quark and the composite hyperbaryons,
and this partial compositeness is responsible for its large mass \cite{DBK}.

The literature on composite Higgs largely leaves unspecified the details
of the new dynamics, and focuses on its low-energy sector
containing the pNGBs, which can be studied as a non-linear sigma model
(for reviews, see Refs.~\cite{RC,BCS,PW}).
Nevertheless, a number of
ultraviolet completions of composite Higgs models have been proposed
\cite{FK,ferretti16}.
All these models are asymptotically free gauge theories with fermionic matter,
sharing some additional basic features.
The models are free of gauge anomalies, both the hypercolor theory by itself,
and the coupled system of hypercolor together with
the Standard Model, including all their gauge interactions.
While ultimately only a lattice calculation can settle it,\footnote{%
  For reviews of recent lattice work, see Refs.~\cite{TDreview,NP,Pica,bqs}.
}
analytic considerations suggest that all candidate models exist in
a chirally broken phase, and are not infrared conformal.

As for the spectrum of the hypercolor theory,
the Nambu-Goldstone coset must contain an $SU(2)_L$ doublet that can
be identified as the Higgs field.  In addition, the hyperbaryon spectrum
must contain top partners, states with the same Standard-Model
quantum numbers as left-handed or right-handed quarks, that can couple
linearly to third-generation quarks.  In this paper we will consider
mass generation for the top quark only,\footnote{%
  For a discussion of mass generation for the bottom quark,
  see Ref.~\cite{ferretti}.  For general remarks on
  mass generation for other Standard Model fermions, see Sec.~\ref{discconc}.
}
and therefore we need hyperbaryons that can couple to $q_L=(t_L,b_L)$
and to $t_R$, but not to $b_R$.
From a ``low-energy''
point of view, the differences between the various models are mainly
in the Nambu-Goldstone coset, and in which \irreps\ of the flavor symmetry group
of the hypercolor theory the top partners live \cite{FK,ferretti16}.
As we will see, different hyperbaryon spectra can give rise to
very different effective potentials even when
the Nambu-Goldstone coset is the same.

A further assumption with important dynamical implications is that
the Standard Model gauge symmetries are embedded into the unbroken
flavor symmetry group of the hypercolor theory.  This gives rise to the
vacuum alignment phenomenon \cite{MP1980,EW1983,Wvac}.
In particular, the effective potential
induced by the coupling of electro-weak gauge bosons to the hypercolor theory
has its minimum at the origin for the Higgs field.
As a result, the top-sector effective potential is instrumental
in generating the non-trivial minimum for the Higgs field
that will trigger electro-weak symmetry breaking.

In this paper we discuss mainly (but not only)
composite Higgs models based on an $SO(d)$ gauge group with $d$ odd.
Each model will contain fermions in the vector
and in the spinor \irreps.  Since we will choose $d$ to be odd,
the spinor \irrep\ is irreducible.
The number of fermions of each \irrep\ is just enough
to accommodate the Standard Model's symmetries into the unbroken flavor
symmetry group, while having pNGBs with the correct quantum numbers
to be identified as the Higgs field.  When the spinor \irrep\
is pseudoreal, the symmetry breaking pattern is assumed to be
\cite{BGR,CCDFLP,Knechtetal}
\begin{equation}
\label{GH4}
  \frac{G}{H} = \frac{SU(4)}{Sp(4)}\times\frac{SU(6)}{SO(6)} \times U(1)_A \ ,
\end{equation}
which corresponds to 6 Weyl (equivalently, Majorana) fermions
in the (real) vector \irrep,
plus 4 Weyl fermions (or 2 Diracs) in the spinor \irrep.
$U(1)_A$ is the non-anomalous axial symmetry whose generator is
a linear combination of the axial charges of the two \irreps.
Demanding that the $SO(d)$ theory will be asymptotically free
allows us to choose $d=5$ or $d=11$ \cite{FK,ferretti16}.\footnote{%
  See Sec.~\ref{su4sp4} for $d=13$.
}
In the case that the spinor \irrep\ is real, the symmetry breaking pattern is
\begin{equation}
\label{GH5}
  \frac{G}{H} = \frac{SU(5)}{SO(5)}\times\frac{SU(6)}{SO(6)} \times U(1)_A \ ,
\end{equation}
which requires 5 Majorana fermions in the spinor \irrep.
The asymptotically free cases are $d=7$ and $d=9$.

The Standard-Model symmetries are embedded into the unbroken subgroup $H$
as follows.  The QCD gauge symmetry $SU(3)_c$
together with (ordinary) baryon number $B$
are embedded into the unbroken $SO(6)$, while $SU(2)_L$ and
$SU(2)_R$ are embedded into the $Sp(4)$ subgroup
of $SU(4)$, or into the $SO(5)$ subgroup of $SU(5)$.
For all the fields of the hypercolor theory, as well as for the quark fields
that will couple to it (namely, $t_L$, $b_L$ and $t_R$),
the usual Standard-Model hypercharge is given by
$Y=T_R^3+2B$, where $T_R^3$ is the third $SU(2)_R$ generator,
and baryon number has the usual normalization with $B=1/3$
for a single quark.  With these conventions,
the electric charge is $Q=T_L^3+Y=T_L^3+T_R^3+2B$.
The simplest hyperbaryons which can play the role of top partners
are hypercolor singlet states made out of two $SO(d)$ spinors
and one $SO(d)$ vector,
which belong to two-index \irreps\ of the $SU(4)$ or $SU(5)$ flavor group.

We comment that $Sp(4)$ is the covering group of $SO(5)$,
and $SU(4)$ of $SO(6)$.\footnote{
  There exist isomorphisms $SO(5)\simeq Sp(4)/Z_2$
  and $SO(6)\simeq SU(4)/Z_2$.
}
For the purpose of this paper it does not matter if the gauge group is $Sp(4)$
or $SO(5)$, and we opt for $SO(5)$ just so that most of the gauge groups we deal
with (except in Sec.~\ref{fix}) will be $SO(d)$ groups with $d$ odd.\footnote{%
  For lattice work on the $Sp(4)$ gauge group, see Refs.~\cite{HPW,LATSp4}.
}
The symmetry-breaking cosets $SU(4)/Sp(4)$ and $SO(6)/SO(5)$ are isomorphic,
and following Ref.~\cite{ferretti16} we opt for the former.

This paper is organized as follows.  In Sec.~\ref{sod}
we introduce our notation for the $SO(d)$ gauge theories, and construct all
the dimension-9/2 hyperbaryons that can serve as top partners.
In Sec.~\ref{su4sp4} we proceed to study
the case of a pseudoreal spinor \irrep.  Of the 5 pNGBs in the $SU(4)/Sp(4)$
coset, four make up the Higgs doublet, while the last one, $\h$, is inert
under all the Standard Model symmetries.  We begin by listing
all the possible embeddings of the quark fields $q_L=(t_L,b_L)$ and $t_R$
into spurions belonging to two-index \irreps\ of $SU(4)$.
We write down the most general four-fermion lagrangian
describing the interaction of these spurions with the hyperbaryons,
finding that it contains 15 independent terms.  We then
work out the resulting effective potential
for the pNGBs.  Thanks to the simplicity of the $SU(4)/Sp(4)$ coset,
this potential can be obtained in closed form.  We also work out
all the low-energy constants, which can be expressed in terms of
hyperbaryons two-point functions.
A summary of our results for this coset is given in Sec.~\ref{su4disc}.

In Sec.~\ref{su5so5} we deal with the case that the spinor \irrep\ is real.
The coset $SU(5)/SO(5)$ contains 14 NGBs, five of which are the same as
before: a (2,2) of $SU(2)_L\times SU(2)_R$ that constitutes the Higgs doublet,
and the singlet $\h$.  The remaining nine NGBs fill up the (3,3)
representation.  Again there are 15 independent couplings in the
four-fermion lagrangian.  The presence of the isospin-triplet pNGBs
makes the analysis technically more involved, and we calculate
the full effective potential only to third order in the pNGB fields.
As before, we also discuss the low-energy constants.

We then turn to the following important issue (Sec.~\ref{su5Veff}).
The $SU(5)/SO(5)$ effective potential will in general
contain cubic terms of the form $\sim h^2\varphi$, where $h$
is the physical Higgs, and $\varphi$ is one of the nine new pNGBs.\footnote{%
  The precise definitions are given in App.~\ref{Veffphi}.
}
The effective potential for $\varphi$ takes the form
\begin{equation}
  V(\varphi) = c_1 f h^2 \varphi + c_2 f^2 \varphi^2 + O(\varphi^3) \ ,
\label{Vtriplet}
\end{equation}
where $f$ represents the scale of the hypercolor theory,
and the coefficients $c_{1,2}$ are dimensionless.
When the Higgs field $h$ condenses, the cubic term (the first term on the
right-hand side of Eq.~(\ref{Vtriplet})) induces a linear term for $\varphi$.
This, in turn, forces the subsequent condensation
of the $\varphi$ field \cite{ferretti16}.
Assuming\footnote{%
  If $c_2<0$, this is likely to lead to a larger expectation value
  for $\varphi$.
}
$c_2>0$ (and neglecting the $O(\varphi^3)$ terms),
the minimum of this potential is $\varphi=-(c_1/(2c_2))h^2/f$.
If the coefficients $c_{1,2}$ have a similar magnitude,
the $\varphi$ expectation value will be suppressed by only one power
of $h/f$ relative to $h$ itself.
This is problematic, because $\varphi$ transforms non-trivially
under $SU(2)_L\times SU(2)_R$, and
an expectation value for $\varphi$ does not preserve the
custodial symmetry.  This diagonal subgroup of $SU(2)_L\times SU(2)_R$
is needed in order to protect the $\r$-parameter \cite{GM},
for which there are tight
experimental constraints.  To shed more light on this issue
we also calculate the full potential in the case that all the $SU(5)/SO(5)$
pNGBs are turned off except for $h$ and $\varphi$, and we discuss
whether, and if so, how, those problematic cubic terms might be avoided.

In Sec.~\ref{fix} we revisit the $SU(4)$ composite Higgs model that was
previously studied by Ferretti in Ref.~\cite{ferretti}, and by us in
Ref.~\cite{topsect}.  In the latter work, we made rather restrictive assumptions
that lead to a four-fermion lagrangian containing just two terms,
and to an effective potential that is quartic in the four-fermion couplings.
Here we take essentially the opposite approach, and make only the most
minimal assumptions, which lead to a four-fermion lagrangian containing
six terms.  We find that, in general, an effective potential
is then generated already at second order in the four-fermion couplings.
However, as we explain in the concluding section (Sec.~\ref{discconc}),
this potential may suffer from a serious drawback.
In addition, for the four-fermion lagrangian we studied in Ref.~\cite{topsect}
we find that the effective potential contains two more terms that we overlooked
in Ref.~\cite{topsect}.

Because of the length of this paper, we have collected the main
phenomenological lessons that can be drawn from all our analyses
in Sec.~\ref{discconc}.  The appendices cover various technical points.

\begin{boldmath}
\section{\label{sod} $SO(d)$ gauge theories}
\end{boldmath}
The $SO(d)$ gauge theories we study in this paper have fermions
in the vector and spinor \irreps.  Since $d$ is always odd,
the spinor \irrep\ is irreducible.  The Higgs field is
identified with pNGBs that arise from chiral symmetry breaking
of the spinor-\irrep\ fermions.  We denote the Weyl fermions in the spinor
\irrep\ as $\U_i$, where $i$ is the flavor index.  There will be 4 of them
when the spinor \irrep\ is pseudoreal, and 5 when it is real.
The flavor symmetry group is, correspondingly, $G_\U=SU(4)$ or $SU(5)$.
We find it convenient to construct the hypercolor baryons in terms of
4-component fields
\begin{equation}
\label{majlike}
\c_i = \left( \begin{array}{c}
    \U_i \\
    \cc\,\e\bU^T_i
  \end{array} \right) \ ,
\end{equation}
and
\begin{equation}
\label{maj}
  \bc_i = \c_i^T C \cc \ .
\end{equation}
Here $C$ is the four-dimensional charge-conjugation matrix, and
$\cc$ is the charge-conjugation matrix in $d=2n+1$ dimensions.
For our notation, Dirac algebra conventions,
and for the properties of the charge conjugation matrix
in various dimensions, see App.~\ref{dconj}.
(When the spinor \irrep\ is real, as in Sec.~\ref{su5so5} below,
the $\c_i$ are Majorana fermions.)  For $g\in G_\U$,
a flavor transformation acts as
$\U\to g\U$, $\bU\to\bU g^\dagger$, or, in terms of the 4-component fields,
\begin{equation}
\label{Ftrans}
  \c \to (g P_R + g^* P_L) \c \ , \qquad
  \bc \to \bc (g^T P_R + g^\dagger P_L) \ .
\end{equation}
The infinitesimal form is
\begin{equation}
\label{flavormaj}
  \d\c = i (P_R T_a-P_L T_a^T)\c \ , \qquad
  \d\bc =  i\bc(P_R T_a^T -P_L T_a) \ ,
\end{equation}
with $T_a$ the hermitian generators.
As we will discuss in the following sections, the $SU(2)_L$ and
$SU(2)_R$ symmetries of the Standard Model are embedded into $H_\U$,
the unbroken flavor symmetry group of the spinor-\irrep\ fermions.

In addition, all models will contain 6 Majorana fermions in the real,
vector \irrep, with an assumed associated chiral symmetry breaking pattern
$SU(6)\to SO(6)$.
As already mentioned, the Standard Model symmetries
$SU(3)_c$ and $U(1)_B$, where $B$ is ordinary baryon number,
are both subgroups of the unbroken $SO(6)$.  We find it convenient to
regroup the 6 Majorana fermions into 3 Dirac fermions,
$\j_{Ia},\bj_{Ia}$, where $I=1,2,\ldots,d$ is the $SO(d)$ vector index,
while $a=1,2,3$ indexes ordinary color.  Like quarks,
the baryon number of these Dirac fermions is $1/3$.
The baryon number of the $\c$ fermions is zero.\footnote{%
  According to our naming conventions the roles of $\c$ and $\j$
  are flipped relative to Refs.~\cite{FK,ferretti16}.
}

The embedding of the Standard Model symmetries is such that
the pNGBs in the $SU(6)/SO(6)$ coset carry ordinary color,
but no $SU(2)_L\times SU(2)_R$ quantum numbers.  Since in this paper we are
mainly interested in the Higgs potential, we will mostly ignore the
$SU(6)/SO(6)$ pNGBs.

\begin{table}[t]
\vspace*{2ex}
\begin{center}
\begin{tabular}{c|c|c|c} \hline\hline
 \irrep\ & $d$ & $B_{ij,X}^r$ & $\bB_{ji,X}^r$
\\ \hline
 \multirow{2}{*}{$A,A^c$} & 5,7
 & $(\bc_i\,P_{R,L}\,\G_I\,\c_j)\,P_X\,\j_{Ia}$
 & $(\bc_i\,P_{L,R}\,\G_I\,\c_j)\,\bj_{Ia} (1-P_X)$
\\
 & 9,11
 & $(\bc_i\,P_{R,L}\,\G_I\s_{\m\n}\,\c_j)\,P_X\,\s_{\m\n}\j_{Ia}$
 & $(\bc_i\,P_{L,R}\,\G_I\s_{\m\n}\,\c_j)\,\bj_{Ia} \s_{\m\n}(1-P_X)$
\\ \hline
 \multirow{2}{*}{$S,S^c$} & 5,7
 & $(\bc_i\,P_{R,L}\,\G_I\s_{\m\n}\,\c_j)\,P_X\,\s_{\m\n}\j_{Ia}$
 & $(\bc_i\,P_{L,R}\,\G_I\s_{\m\n}\,\c_j)\,\bj_{Ia} \s_{\m\n}(1-P_X)$
\\
 & 9,11
 & $(\bc_i\,P_{R,L}\,\G_I\,\c_j)\,P_X\,\j_{Ia}$
 & $(\bc_i\,P_{L,R}\,\G_I\,\c_j)\,\bj_{Ia} (1-P_X)$
\\ \hline
 \multirow{2}{*}{$D$} & 5,11
 & $\{\bc_i\,P_R\,\G_I\g_\m\,\c_j\}\,P_X\,\g_\m\j_{Ia}$
 & $-\{\bc_i\,P_L\,\G_I\g_\m\,\c_j\}\,\bj_{Ia}\,\g_\m (1-P_X)$
\\
 & 7,9
 & $\{\bc_i\,P_R\,\G_I\g_\m\,\c_j\}\,P_X\,\g_\m\j_{Ia}$
 & $\{\bc_i\,P_L\,\G_I\g_\m\,\c_j\}\,\bj_{Ia}\,\g_\m (1-P_X)$
\\ \hline
 \multirow{2}{*}{$N$} & 5,11
 & $\d_{ij}(\bc_k\,P_R\,\G_I\g_\m\,\c_k)\,P_X\,\g_\m\j_{Ia}$
 & $-\d_{ij}(\bc_k\,P_L\,\G_I\g_\m\,\c_k)\,\bj_{Ia}\,\g_\m (1-P_X)$
\\
 & 7,9
 & $\d_{ij}(\bc_k\,P_R\,\G_I\g_\m\,\c_k)\,P_X\,\g_\m\j_{Ia}$
 & $\d_{ij}(\bc_k\,P_L\,\G_I\g_\m\,\c_k)\,\bj_{Ia}\,\g_\m (1-P_X)$
\\ \hline\hline
\end{tabular}
\end{center}
\begin{quotation}
\floatcaption{tabantihB}{Top partners.
The first column defines the \irrep\ of the flavor group
to which the hyperbaryon belongs, which
can be a two-index \irrep, or a singlet.
The second column defines the $SO(d)$ gauge theory.
The next two columns give the hyperbaryon and anti-hyperbaryon operators.
The chiral projector $P_X,$ $X=L,R,$ is always acting on the open Dirac index.
The notation $\{\bc_i \cdots \c_j\}$ for the adjoint \irrep\
refers to the traceless part of the bilinear.  Our choice of signs
for the anti-hyperbaryons is explained in Sec.~\ref{secCP}.
The minus sign in $\bB_{ji,X}^r$ for the $D$ and $N$ \irreps\ for $d=5,11$
arises because of the difference
between the $CP$ transformation rules~(\ref{CPc}) and~(\ref{CPd}).
}
\end{quotation}
\vspace*{-4ex}
\end{table}

\vspace{2ex}

\subsection{\label{hb} Top-partner hypercolor baryons}
We will restrict the discussion to the simplest top partners,
which are created by local 3-fermion operators
constructed as follows.
We first assemble two $SO(d)$-spinor fermions into a bilinear transforming
as an $SO(d)$ vector, and then contract this bilinear
with an $SO(d)$-vector fermion to form an $SO(d)$-singlet state.
The resulting hyperbaryon and anti-hyperbaryon fields are tabulated
in Table~\ref{tabantihB}.  Unless it forms a singlet,
the $SO(d)$-spinor bilinear belongs to one of the two-index \irreps\ of the
flavor group $G_\U$, which, we recall, can be $SU(4)$ or $SU(5)$.
When a single four-dimensional Dirac matrix (aside from the chiral projector)
is sandwiched between the two $\c$ fermions,
we encounter the adjoint \irrep\ ($D$), or a singlet ($N$).
When the number of four-dimensional Dirac matrices is zero or two,
the same chiral projector is applied to both of the $\c$ fermions,
and the bilinear then has definite symmetry properties on its spin index.
Taking into account also the symmetry properties on the $SO(d)$ index
(see the last column of Table~\ref{tabC})
fixes the symmetry on the flavor index.  In view of Eq.~(\ref{Ftrans}),
when the chiral projector is $P_R$ we encounter
the two-index symmetric ($S$) or two-index antisymmetric ($A$) representations,
whereas for $P_L$ we obtain the complex conjugate representations
$S^c$ and $A^c$.\footnote{%
  For the $SO(5)\sim Sp(4)$ gauge theory, the $N$, $D$ and $A$ hyperbaryons
  were previously considered in Refs.~\cite{BGR,CCDFLP}.
}

We use the following notation.  A generic hyperbaryon is denoted
$B_{ij,X}^r$, where $i$ and $j$ are flavor indices,
and the optional subscript $X=L,R$ denotes the projector applied
to the open Dirac index, which in turn is always carried by the $\j$ fermion.
$r$ labels the \irrep, which can be one of $D$, $N$, $S$, $S^c$, $A$ or $A^c$.
Our notation is such that the anti-hyperbaryon of $B_{ij}^r$
is denoted $\bB_{ji}^r$, with the flavor indices flipped.
This will prove convenient when using matrix notation in flavor space.

We comment in passing that the Ferretti--Karateev list of requirements is
fairly restrictive \cite{FK,ferretti16}.  Models that satisfy all
the requirements and have a prescribed coset structure of the effective theory
are so few, that in effect, knowing the coset structure essentially fixes
the model, and thus, ultimately, also the top-partner content.
However, by itself, the coset structure does not
tell us what will be the \irreps\ to which
the top-partners belong.  For example, the models of Sec.~\ref{su5so5}
and Sec.~\ref{fix} both share an $SU(5)/SO(5)$ coset.
But in Sec.~\ref{su5so5} the hyperbaryons belong to 2-index \irreps\ of $SU(5)$,
whereas in Sec.~\ref{fix} they belong to the (anti)fundamental \irrep.
Thus, the straightforward way to find the top partners of a given model
is to explicitly construct the relevant gauge invariant operators.
Of course, the explicit form of the hyperbaryon operators will also be needed
for the derivation of the low-energy constants.

\subsection{\label{secCP} $CP$ symmetry}
As a stand-alone theory, all the hypercolor theories we study in this paper
are invariant under $C$ and $P$.  Because we couple the hypercolor theory
to $q_L=(t_L,b_L)$ and to $t_R$, but not to $b_R$, the four-fermion lagrangian
can be invariant only under the combined $CP$ transformation.\footnote{%
  Provided that all the four-fermion coupling constants are real.
}
The $CP$ transformation acts on a gauge field as
\begin{subequations}
\label{CP}
\begin{equation}
\label{CPa}
  A_\m(x) \ \to\ \tilde{A}_\m(\tx) \ ,
\end{equation}
where $\tx_\m=x_\m$ if $\m=4$, and $\tx_\m=-x_\m$ if $\m=1,2,3$,
with a similar definition for $\tilde{A}_\m$.  The $SO(d)$ gauge field
is invariant under charge conjugation,
so that its transformation rule stems from parity only.
The $SO(d)$-vector Dirac fermions transform as
\begin{equation}
\label{CPb}
  \j(x) \ \to\ i\g_4 C\, \bj(\tx)^T  \ , \qquad
  \bj(x) \ \to\ -i\j(\tx)^T C \g_4 \ .
\end{equation}
Except for the choice of phases, which is explained in App.~\ref{appCP},
this is the usual $CP$ transformation rule of a Dirac fermion.
The $\c$ fields transform according to
\begin{equation}
\label{CPc}
  \c_i(x) \ \to \ i\g_4\c_i(\tx) \ , \qquad
  \bc_i(x) \ \to \ -i\bc_i(\tx)\g_4 \ ,
\end{equation}
in the case that the spinor \irrep\ is real ($\cc=\cc^T$),
whereas for the pseudoreal case ($\cc=-\cc^T$) their transformation rule is
\begin{equation}
\label{CPd}
  \c_i(x) \ \to \ i\g_4\g_5\c_i(\tx) \ , \qquad
  \bc_i(x) \ \to \ -i\bc_i(\tx)\g_5\g_4 \ .
\end{equation}
\end{subequations}
The induced transformation of the hyperbaryon fields is
\begin{equation}
\label{CPhB}
  B_{ij}^r(x) \ \to i\g_4 C \bB_{ji}^r(\tx)^T \ , \qquad
  \bB_{ij}^r(x) \ \to\ -i B_{ji}^r(\tx)^T C \g_4 \ .
\end{equation}
The sign choices we have made in Table~\ref{tabantihB} ensure that
all hyperbaryons transform under $CP$ like the $\j$ fermions.
For more details, see App.~\ref{appCP}.

\begin{boldmath}
\section{\label{su4sp4} The $SU(4)/Sp(4)$ coset}
\end{boldmath}
There are two models where the spinor \irrep\ is pseudoreal,
one based on an $SO(5)$ gauge group and the other on $SO(11)$.
The $SO(13)$ theory is asymptotically free as well, but according to
analytic considerations it is probably inside the conformal window,
and not chirally broken \cite{ferretti16,diboson}.  In any event,
since all the relevant properties of the $SO(d)$ theories are periodic
in $d$ modulo 8, the discussion of the $SO(5)$ theory would carry over as is
to the $SO(13)$ case, if the latter were to be chirally broken.
For previous work on the $SU(4)/Sp(4)$ models,
see Refs.~\cite{ferretti16,BGR,CCDFLP,Knechtetal,diboson}.

The order parameter for the spontaneous breaking
of the flavor symmetry $G_\U=SU(4)$ is the expectation value of $\bc_i\c_j$.
This order parameter is antisymmetric on its flavor indices.
We will assume that $\svev{\bc_i\c_j} \propto \e_{0,ij}$,
where the $4\times 4$ matrix $\e_0$ is defined in Eq.~(\ref{eps0}).
With this convention, we may take the order parameter to be $\svev{\bc\e_0\c}$.
Applying an infinitesimal flavor transformation~(\ref{flavormaj})
to the order parameter we get
\begin{equation}
\label{flavorcondpseudo}
  \d_a(\bc\e_0\c)
  = i\bc\left(P_R(\e_0 T_a + T_a^T\e_0) - P_L(T_a\e_0 + \e_0 T_a^T)\right)\c \ .
\end{equation}
Of the 15 generators of $SU(4)$, there are 10 which leave the order parameter
invariant (see Eq.~(\ref{sp4gendef})).
They generate the unbroken group, $H_\U=Sp(4)$.

The remaining 5 generators belong to the coset $G_\U/H_\U=SU(4)/Sp(4)$.
Taking $T_a$ to be a coset generator, the variation of the order parameter
gives rise to an interpolating field for one of the NGBs,
\begin{equation}
\label{condnontrivial}
  \d_a(\bc\e_0\c) = 2i\bc\left(P_R\e_0 T_a-P_L T_a\e_0\right)\c\ .
\end{equation}
Equivalently, the full NGB field is
\begin{equation}
 \hP = 2i\ttr(P_R\c\bc\e_0-P_L\e_0\c\bc) \ ,
\label{micpion}
\end{equation}
where the notation $\ttr$ indicates that the trace is over
the Dirac and color indices, but not over the flavor indices.
It readily follows that $\e_0\hP$ (or $\hP\e_0$) is antisymmetric on its
flavor indices.  The flavor trace of $\hP$ with a coset generator
reproduces Eq.~(\ref{condnontrivial}), while its trace with an $Sp(4)$ generator
vanishes identically, showing that $\hP$ has the correct number of
degrees of freedom.  Using Eqs.~(\ref{Ftrans}) and~(\ref{sp4def}),
we see that $\hP$ transforms in the expected way under the unbroken group
\begin{equation}
  \hP \to g \hP g^\dagger \ , \qquad g\in Sp(4) \ .
\label{PiSp4}
\end{equation}
Under the $CP$ transformation of the hypercolor theory, Eq.~(\ref{CPd}), we have
\begin{equation}
\label{CPSp4}
  \hP(x) \to -\hP^T(\tx) \ ,
\end{equation}
where we have used Eq.~(\ref{maj}).
Notice that (apart from the usual coordinates transformation)
the $CP$ transformation does not merely flip the sign of $\hP$.
Related, when the coset generator $T_a$ commutes with $\e_0$,
the NGB field~(\ref{condnontrivial}) is a pseudoscalar, as in the familiar
QCD case.  But when $T_a$ anticommutes with $\e_0$, the NGB field is a scalar.
We will discuss the phenomenological significance of this result shortly.

In the effective chiral theory, the NGBs of $SU(4)\to Sp(4)$ symmetry breaking
are represented by an antisymmetric unitary field $\S\in SU(4)$,
$\S^T=-\S$.  In addition, the effective theory depends on
an $SU(6)/SO(6)$ non-linear field, which we will not discuss in this paper,
and a field $\F\in U(1)$ associated with the spontaneous breaking
of the non-anomalous $U(1)_A$ symmetry
\cite{FK,BGR,CCDFLP,Knechtetal,topsect,diboson,CLLR,tworeps}.
The axial transformations are
\begin{subequations}
\label{Axtrans}
\begin{eqnarray}
  \d_A\c &=& \frac{i}{2}\g_5 \c \ , \qquad
  \d_A\bc \ = \ \frac{i}{2}\bc\g_5 \ ,
\label{Axtransa}\\
  \d_A\j &=& iq\g_5\j \ ,\qquad
  \d_A\bj \ = \ iq\bj\g_5 \ ,
\label{Axtransb}\\
  \d_A\F &=& i\F \ .
\label{Axtransc}
\end{eqnarray}
\end{subequations}
Eq.~(\ref{Axtransa}) gives the transformation rule of the spinor \irrep,
which sets the normalization of the non-anomalous axial transformation
in the microscopic theory.  Eq.~(\ref{Axtransb}) is the transformation rule
of the vector \irrep,  where $q=-(1/3)T_\c/T_\j$,\footnote{%
  In the case of the $SU(5)/SO(5)$ models of Sec.~\ref{su5so5},
  $q=-(5/12)T_\c/T_\j$.  For more details see, \eg,
  Refs.~\cite{diboson,tworeps,anomdim}.
}
and the group traces are $T_\c=2^{(d-5)/2}$ and $T_\j=2$.
Finally Eq.~(\ref{Axtransc}) sets our normalization for the transformation rule
of the corresponding effective field.  The formal correspondence
between the elementary and the effective fields is then
\begin{equation}
\label{formal}
  \F\S \leftrightarrow \ttr(P_R\c\bc) \ , \qquad
  \F^*\S^* \leftrightarrow \ttr(P_L\c\bc) \ .
\end{equation}
As already mentioned,
we will assume that the vacuum is given by $\svev{\S}=\e_0$ and $\svev{\F}=1$,
and parametrize the non-linear field as
\begin{equation}
\label{vacsu4}
  \S = \exp(i\P/f)\,\e_0\exp(i\P^T/f) = \exp(2i\P/f)\e_0\ ,
\end{equation}
where $f$ is the decay constant.  The effective NGB field $\P$ is hermitian,
traceless, and satisfies $\e_0\P=\P^T\e_0$, just as $\hP$.
Flavor transformations act on the non-linear field as
\begin{equation}
\label{flavorSig}
  \S\to g\S g^T \ , \qquad g\in SU(4) \ .
\end{equation}
For $g\in Sp(4)$, it follows that the effective NBG field $\P$
transforms in the same way as the NGB field of the microscopic theory,
Eq.~(\ref{PiSp4}).  The transformation rule of $\P$ under $CP$ is defined
to be the same as in Eq.~(\ref{CPSp4}).  The leading-order chiral lagrangian
is invariant under these transformations.

The embedding of $SU(2)_L$ and $SU(2)_R$ in $Sp(4)$ is given in Eq.~(\ref{su2s}),
and the parametrization of the effective field $\P$ is given in Eq.~(\ref{pions}).
Four of the NGBs are identified with the Higgs doublet, $H=(H_+,H_0)$,
whereas the fifth, $\h$, is a singlet under $SU(2)_L\times SU(2)_R$.
Using the parametrization~(\ref{pions}), a $CP$ transformation acts as
\begin{equation}
\label{transH}
  H_0\to H_0^*\ , \qquad H_+\to H_+^* \ , \qquad \h\to -\h \ .
\end{equation}
This correctly reproduces the $CP$ transformation of
the Higgs field in the Standard Model.

The rest of this section is organized as follows.
In Sec.~\ref{su4spurions} we obtain all the spurion embeddings
of the quark fields.  In Sec.~\ref{su4LEHC} we write down
the four-fermion lagrangian $\cl_{EHC}$, and in Sec.~\ref{su4yeff}
we list all the effective top Yukawa couplings allowed by it.
In Sec.~\ref{su4Veff} we begin the discussion of the effective potential
of the pNGBs, $\Veff$.  We group the various contributions
into twelve ``template'' forms, and then work out all the contributions
to $\Veff$ in closed form.  In Sec.~\ref{su4LECs} we derive the
low-energy constants.  We summarize our findings in Sec.~\ref{su4disc},
which also contains a simple example of a phenomenologically viable potential.
Finally, we discuss spontaneous $CP$ breaking in Sec.~\ref{SCPB}.

\subsection{\label{su4spurions} Spurions}
Much like in technicolor theories, the coupling of the Higgs field to the
gauge bosons of the Standard Model arises naturally when the relevant
global symmetries of the hypercolor theory are gauged;
but a more elaborate setup is needed to generate masses for fermions.
Here we postulate the existence of yet another gauge symmetry,
dubbed ``extended hypercolor'' (EHC).  We assume that the EHC gauge symmetry
breaks spontaneously at some scale $\L_{EHC}$ which is large
relative to the scale of the hypercolor theory, $\L_{HC}$.
The remnant of the EHC interactions at the hypercolor scale is a set of
four-fermion interactions, and
we assume that these four-fermion interactions couple the third generation
quark fields $q_L=(t_L,b_L)$ and $t_R$ to the hyperbaryon fields constructed
in Sec.~\ref{hb}.  The EHC theory will thus generate a mass for the top quark
through the mechanism of partial compositeness.  We comment that this setup
does not necessarily generate a mass for any other Standard Model's fermion.
Their masses may have to involve some other dynamics
(see Sec.~\ref{discconc}).

Unlike the hyperbaryon fields, quark fields
fit into \irreps\ of the smaller, Standard-Model symmetry.
They do not fill up any \irreps\ of the global symmetry group of
the hypercolor theory.  The coupling of quark and hyperbaryon fields
therefore explicitly breaks the flavor symmetry of the hypercolor theory.
This will induce a potential $\Veff$ for the NGBs.

While $\Veff$ is invariant only under Standard-Model symmetries,
it depends on low-energy constants that can be expressed
in terms of correlation functions of the stand-alone hypercolor theory.
When we derive expressions for these low-energy constants, we may benefit from the
full global symmetry of the hypercolor theory, including in particular $G_\U$.
The way to do this is to promote the quark fields to spurion fields
transforming in \irreps\ of $G_\U$.

In the rest of this subsection we construct the spurions explicitly.
Each embedding of $q_L$ is defined by
\begin{eqnarray}
  X_L(x) &=& t_L(x) \hX_{t_L} + b_L(x) \hX_{b_L} \ ,
\label{embedL}\\
  \bX_L(x) &=& \bt_L(x) \hbX_{t_L} + \bb_L(x) \hbX_{b_L} \ ,
\nonumber
\end{eqnarray}
and similarly for $t_R$,
\begin{equation}
\label{embedR}
  X_R(x) = t_R(x) \hX_{t_R}  \ ,\qquad
  \bX_R(x) = \bt_R(x) \hbX_{t_R} \ ,
\end{equation}
where the hatted objects are constant $4\times 4$ matrices.
Because the EHC theory is not known, we will allow for the
most general four-fermion lagrangian which is compatible
with the (spurionized) symmetries of the hypercolor theory, and with $CP$.

In order to build the four-fermion lagrangian
we have to allow for all embeddings of the quark fields into spurions
belonging to two-index \irreps\ of $G_\U=SU(4)$ (or to a singlet),
which are consistent with the embedding of $SU(2)_L$ and $SU(2)_R$ into $SU(4)$.
We begin with the spurion embeddings of $q_L$.
For the adjoint \irrep\ of $SU(4)$ there are two options,
\begin{eqnarray}
\label{topleftchichibar}
D_L^{1}&=&\left(
\begin{array}{cccc}
0&0&t_L&0\\
0&0&b_L&0\\
0&0&0&0\\
0&0&0&0
\end{array}
\right)\ ,\\
D_L^{2}&=&\left(
\begin{array}{cccc}
0&0&0&0\\
0&0&0&0\\
0&0&0&0\\
b_L&t_L&0&0
\end{array}
\right)\ .
\end{eqnarray}
Remembering that $q_L=(t_L,b_L)$ is an $SU(2)_L$ doublet with $T_R^3=-1/2$,
one can check that these spurions are consistent with the Standard-Model
transformation properties of $q_L$.
To this end we use that the adjoint spurions transform
as $D_L^i\to g D_L^i g^\dagger$ under $g\in SU(4)$,
and the embedding~(\ref{su2s}) of $SU(2)_L$ and $SU(2)_R$ into $SU(4)$.
For the two-index antisymmetric \irrep\ we have one embedding,
\begin{equation}
\label{topleftchichiA}
A_L = \left(
\begin{array}{cccc}
0&0&0&t_L\\
0&0&0&b_L\\
0&0&0&0\\
-t_L&-b_L&0&0
\end{array}
\right)\ ,
\end{equation}
and likewise for the two-index symmetric \irrep,
\begin{equation}
\label{topleftchichiS}
S_L = \left(
\begin{array}{cccc}
0&0&0&t_L\\
0&0&0&b_L\\
0&0&0&0\\
t_L&b_L&0&0
\end{array}
\right)\ .
\end{equation}
The $A_L$ and $S_L$ spurions transforms as $X_L\to g X_L g^T$,
$X\in\{A,\ S\}$, under $g\in SU(4)$, and again
one can verify consistency with Standard-Model quantum numbers.
The embeddings for the complex conjugate \irreps\ $A^c$ and $S^c$
may be obtained using the rule
\begin{equation}
\label{Xcvev}
X^c = -\e_0 X \e_0 \ ,
\end{equation}
where again $X\in\{A,\ S\}$.
Let us explain this rule.
We first observe that $X^c$ spurions transform under $g\in SU(4)$
as $X^c\to g^* X^c g^\dagger$.  Restricting to $g\in Sp(4)$,
and using Eqs.~(\ref{Xcvev}) and~(\ref{sp4def}), we have
\begin{equation}
\label{relation}
 g^* X^c g^\dagger = -g^* \e_0 X \e_0 g^\dagger
 = -\e_0 g X g^T \e_0 \ .
\end{equation}
The rightmost expression involves the transformation rule
of a field in the $A$ or $S$ \irreps, and we have already verified
that this correctly reproduces the Standard-Model transformation rules
for the $A_L$ and $S_L$ spurions.
Since $SU(2)_L\times SU(2)_R$ is a subgroup of $Sp(4)$, it follows that
the spurion $X^c$ defined by Eq.~(\ref{Xcvev}) will again reproduce
the correct Standard-Model transformation rules.
Applying Eq.~(\ref{Xcvev}) we find the explicit forms
\begin{eqnarray}
\label{topleftchichiAc}
A_L^c&=&
\left(
\begin{array}{cccc}
0&0&-b_L&0\\
0&0&t_L&0\\
b_L&-t_L&0&0\\
0&0&0&0
\end{array}
\right)\ ,
\\
\label{topleftchichiSc}
S_L^c&=&
\left(
\begin{array}{cccc}
0&0&-b_L&0\\
0&0&t_L&0\\
-b_L&t_L&0&0\\
0&0&0&0
\end{array}
\right)\ .
\end{eqnarray}

Let us move on to $t_R$, which is a singlet of $SU(2)_L$ with $T_R^3=0$
(note that $t_R$ is not required to be invariant under the full $SU(2)_R$,
but only under rotations generated by $T_R^3$).  In this case we have
more options, starting with the singlet
\begin{equation}
\label{tRsinglet}
N_R=t_R\left(
\begin{array}{cccc}
1&0&0&0\\
0&1&0&0\\
0&0&1&0\\
0&0&0&1
\end{array}
\right)\ .
\end{equation}
There are two linearly independent options for the adjoint \irrep,
\begin{eqnarray}
\label{tRadj1}
D_R^{1}&=&t_R\,\left(
\begin{array}{cccc}
1&0&0&0\\
0&1&0&0\\
0&0&-1&0\\
0&0&0&-1
\end{array}\right)\ ,
\\
\label{tRadj2}
D_R^{2}&=&t_R\,\left(
\begin{array}{cccc}
0&0&0&0\\
0&0&0&0\\
0&0&1&0\\
0&0&0&-1
\end{array}\right)\ ,
\end{eqnarray}
another two for the anti-symmetric \irrep,
\begin{eqnarray}
\label{tRantisymm1}
A_R^{1}&=&t_R\,\left(
\begin{array}{cccc}
0&1&0&0\\
-1&0&0&0\\
0&0&0&0\\
0&0&0&0
\end{array}\right)\ ,
\\
\label{tRantisymm2}
A_R^{2}&=&t_R\,\left(
\begin{array}{cccc}
0&0&0&0\\
0&0&0&0\\
0&0&0&1\\
0&0&-1&0
\end{array} \right)\ ,
\end{eqnarray}
and one for the symmetric \irrep,
\begin{equation}
\label{tRsymm}
S_R=t_R\,\left(
\begin{array}{cccc}
0&0&0&0\\
0&0&0&0\\
0&0&0&1\\
0&0&1&0
\end{array}\right) \ .
\end{equation}
The spurion embeddings for the $A^c$ and $S^c$ \irreps\ again follow
using Eq.~(\ref{Xcvev}).  Explicitly,
\begin{eqnarray}
\label{cspurionSMvevs}
A_R^{c1}&=&A_R^1\ ,\\
A_R^{c2}&=&A_R^2\ ,\nonumber\\
S_R^c&=&-S_R\ .
\nonumber
\end{eqnarray}

It remains to construct the anti-spurion embeddings.
Referring to the decompositions~(\ref{embedL}) and~(\ref{embedR}),
we define the $c$-number coefficients of the anti-spurion fields via
\begin{equation}
  \hbX \equiv \hX^\dagger = \hX^T \ .
\label{antiX}
\end{equation}
The last equality follows because we have chosen
all the $c$-number spurions $\hX$ to be real.

\begin{boldmath}
\subsection{\label{su4LEHC} $\cl_{EHC}$}
\end{boldmath}
\hspace{-2ex}
With the top-partner hyperbaryons and the spurions at hand,
the most general four-fermion lagrangian that couples them is given by
\begin{subequations}
\label{LEHCsu4}
\begin{eqnarray}
  \cl_{\rm EHC} &=& \cl_{\rm EHC,1} + \cl_{\rm EHC,2} \ ,
\label{LEHCsu4a}\\
  \cl_{\rm EHC,1} &=&
  \tr\Big(\l_1\bA_L B_R^{A} + \l_2\bA^c_L B_R^{A^c}
  + \l_3\bS_L B_R^{S} + \l_4\bS^c_L B_R^{S^c}
\label{LEHCsu4b}\\
  &&  + (\l_5 \bD_{R}^1 + \l_6 \bD_{R}^2) B_L^D +
\l_7\bN_R B_L^N  +\hc \Big) \ ,
\nonumber\\
  \cl_{\rm EHC,2} &=& \tr \Big(
  (\tl_1\bA_R^1 + \tl_2\bA_R^2) B_L^{A}
  + (\tl_3\bA^{c1}_R + \tl_4\bA^{c2}_R) B_L^{A^c}
\label{LEHCsu4c}\\
  && + \tl_5\bS_R B_L^S + \tl_6\bS^c_R B_L^{S^c}
     + (\tl_7 \bD_{L}^1 + \tl_8 \bD_{L}^2) B_R^D +\hc \Big) \ ,
\nonumber
\end{eqnarray}
\end{subequations}
where the trace is over $SU(4)$ indices.
$\l_1,\ldots,\l_7$ and $\tl_1,\ldots,\tl_8$
are (dimensionful) coupling constants.
We have grouped in $\cl_{\rm EHC,1}$ those terms where $B_L$ belongs
to $D$ or $N$, while $B_R$ belongs to $A,A^c,S$ or $S^c$, and
the other way around for $\cl_{\rm EHC,2}$.
The four-fermion lagrangian is invariant under the spurionized
$SU(4)$ symmetry.  In addition, it is truly invariant under the
Standard-Model gauge symmetries $SU(3)_c$, $SU(2)_L$, and $U(1)_Y$,
and it conserves baryon number, or, which is equivalent, the $T_R^3$ charge.

Assuming that all the coupling constants are real,
the four-fermion lagrangian is also invariant under the combined
$CP$ transformation of the hypercolor theory and the Standard Model,
in which the $c$-number spurions are inert.  How $CP$ works
is best illustrated through an example.  The $CP$ rules of Sec.~\ref{secCP}
imply in particular that $\bt_R B_{L,ij}\leftrightarrow \bB_{L,ji}t_R$.
Remembering that $c$-number spurions don't transform, we have
\begin{equation}
\label{CPcnum}
  \tr(\bX_R B_L) = \bt_R \tr(\hbX_R B_L)
  \leftrightarrow \tr(\hbX_R \bB_L^T) t_R
  = \tr(\hX_R^T \bB_L^T) t_R = \tr(\bB_L X_R) \ ,
\end{equation}
where again the trace and transpose operations are applied to
the flavor indices.
In order to establish the $CP$-invariance of $\cl_{\rm EHC}$ we have
used Eq.~(\ref{antiX}), which in turn relies on the fact that all the
$c$-number spurions are real.  That such a choice can be made, is a special
feature of the $SU(4)/Sp(4)$ coset.  (As we will see in Sec.~\ref{su5so5}, things are
slightly more involved for the $SU(5)/SO(5)$ case.)  Of course,
we could have chosen to multiply some $c$-number spurions by
arbitrary phases.
This would invalidate Eq.~(\ref{antiX}) for those $c$-number spurions,
and, as a result, there would be fewer terms in $\cl_{\rm EHC}$
if we wish to maintain $CP$ invariance.  However, opting to do this
is arbitrary.  Once again, the point is that apart from
some very general assumptions, we do not know the EHC theory.
Therefore, we must consider the most general four-fermion lagrangian
consistent with those general assumptions.
When all the four-fermion couplings are taken to be real,
this requires choosing all the $c$-number spurions to be real as well.

As already noted, in this paper we do not study the $SU(6)/SO(6)$ pNGBs
associated with the vector-\irrep\ fermions, and therefore we only gave
the $SU(3)_c$ quantum numbers of the hyperbaryons.
Requiring full $SU(6)$ invariance will give rise
to the same four-fermion lagrangian once the spurions assume
their Standard Model values.  Indeed, each term in Eq.~(\ref{LEHCsu4})
can be trivially ``lifted'' to an $SU(6)$-invariant form,
as we illustrate through the following examples.
For definiteness, we will refer
to the hyperbaryons of the $SO(5)$ gauge theory.

We begin with the first term
on the right-hand side of Eq.~(\ref{LEHCsu4b}), $\bA_L B_R^A$.
Since the $\c$ fermions play little role,
for brevity we express the hyperbaryon operator as
$B^A_{R,a} = f(\c)^A_I \j_{R,Ia} = f(\c)^A_I \J_{Ia}$,
where in the last equality we have used that the 3 Dirac fermions
introduced earlier are composed of 6 right-handed vector-\irrep\ Weyl fermions
$\J_1,\ldots,\J_6$ according to
\begin{equation}
  \j_a = \left( \begin{array}{c}
    \J_a \\
    \e\bJ^T_{3+a}
  \end{array} \right) \ , \qquad
  \bj_a = \left( \begin{array}{cc}
    -\J^T_{3+a} \e & \bJ_a
  \end{array} \right)\ , \qquad a=1,2,3 \ .
\label{SU6embed}
\end{equation}
A complete $SU(6)$ \irrep\ is now obtained by simply replacing the index
$a=1,2,3,$ with a new index $\ba=1,\ldots,6$, explicitly,
$\tB^A_{R,\ba} = f(\c)^A_I \J_{I\ba}$, where we are using a tilde to refer to
$SU(6)$ \irreps.  The $SU(6)$-invariant interaction
is thus $\tilde{\bA}_{L,\ba} \tB^A_{R,\ba}$.  In order to ensure
equality between the $SU(3)_c$ and $SU(6)$ versions,
we simply embed the $SU(3)_c$ spurion into the $SU(6)$ spurion,
namely, we define
$\tilde{\bA}_{L,\ba} = \bA_{L,a}$ for $\ba=a=1,2,3$,
and $\tilde{\bA}_{L,\ba} = 0$ for $\ba=4,5,6$.

At this point we have not made use of the last three components of
the $SU(6)$ multiplet, $\tB^A_{R,\ba}$, $\ba=4,5,6$.
These components occur in a difference place in $\cl_{EHC}$,
in the term that involve $\bB^A_L$,
and, thus, contains $\bj_{L,a} \sim \J_{a+3}$, as follows from Eq.~(\ref{SU6embed}).
This time, we ``lift'' the spurions to $SU(6)$ by letting
$\tA^i_{R,\ba}=0$ for $\ba=1,2,3$, and $\tA^i_{R,\ba}=A^i_{R,\ba-3}$ for $\ba=4,5,6$
and $i=1,2$.

These examples demonstrate that there is one-to-one correspondence
between the $SU(3)_c$-invariant and $SU(6)$-invariant forms of $\cl_{EHC}$.
The underlying reason is that the ``expectation values''
of the spurions are only constrained by SM symmetries.

\vspace{2ex}

\subsection{\label{su4yeff} Top Yukawa couplings}
Effective top-Yukawa couplings are generated by integrating out all the
states of the hypercolor theory except for the pNGBs.
These effective interactions are organized in a weak-coupling expansion
in the four-fermion couplings, as well as according to the
usual power counting of the chiral lagrangian.
To second order in the four-fermion couplings, and to leading order
in the chiral expansion, we find effective interactions that are
either linear or bilinear in $\S$ or $\S^*$.
Any effective interaction which is cubic or higher in the nonlinear field
must contain additional derivatives and/or mass insertions,\footnote{%
  See Sec.~\ref{su4disc} below for a discussion of explicit mass terms
  for the fermions of the hypercolor theory.
}
and therefore belongs to a higher order in the chiral expansion.

We begin with effective interactions that are linear in $\S$ or $\S^*$.
Each effective interaction contains one spurion and one anti-spurion,
one of which must be left-handed and the other right-handed.
The effective Yukawa interactions have the same symmetries as $\cl_{\rm EHC}$.
In order to form an $SU(4)$ singlet,
the spurion must belong to $A,A^c,S$ or $S^c$ and the anti-spurion
to $D$ or $N$, or the other way around, because the effective interaction
has to contain a $\S$ or a $\S^*$.  It follows that the spurion and
the anti-spurion must both come from $\cl_{\rm EHC,1}$, or both
from $\cl_{\rm EHC,2}$, which explains why we have grouped the
four-fermion interactions this way.
The list of possible top-Yukawa effective interactions is thus
\begin{eqnarray}
\label{su4yukawa}
  && 
  \F\tr(\bX_L\S (D_R^{1,2})^T)\ ,\quad
  \F^*\tr(\bX^c_L\S^* D_R^{1,2})\ ,\quad
  \F\tr(\bA_L\S N_R) \ ,
\\
  && 
  \F^*\tr(\bA^c_L\S^* N_R) \ ,\quad
  \F^*\tr((\bD_L^{1,2})^T \S^* X_R)\ ,\quad
  \F\tr(\bD_L^{1,2} \S X^c_R)\ ,
\nonumber
\end{eqnarray}
where the hermitian conjugate is to be added to each operator.
$X_L$ can be $A_L$ or $S_L$, and $X_R$ can be $A^{1,2}_R$ or $S_R$.
The explicit form of each effective interaction can be worked out
by assigning to each spurion its Standard-Model value
from Sec.~\ref{su4spurions}, and using Eqs.~(\ref{pions}) and~(\ref{Scsa})
for the $\S$ field.  Out of a total of 22 possible contributions
to the top-Yukawa coupling, we find that 8 of the possibilities
vanish identically, while the other 14 generate a non-zero
top-Yukawa coupling.

\begin{table}[t]
\vspace*{0ex}
\begin{center}
\begin{tabular}{cccccc} \hline \hline
$B_{R,L}^A$ & $B_{R,L}^{A^c}$ & $B_{R,L}^S$ & $B_{R,L}^{S^c}$ &
  $B_{R,L}^D$ & $B_{R,L}^N$ \\ \hline
 $1\pm q$ & $-1\pm q$ & $1\pm q$ & $-1\pm q$ & $\mp q$ & $\mp q$ \\
\hline\hline
\end{tabular}
\end{center}
\begin{quotation}
\floatcaption{TABzeta}{Axial charges of the hyperbaryons}
\end{quotation}
\vspace*{-4ex}
\end{table}

Each effective interaction in Eq.~(\ref{su4yukawa}) is (formally) invariant
under $SU(4)$ and $U(1)_A$.\footnote{%
  To maintain the invariance under $SU(6)$ we would have
  to reintroduce the corresponding nonlinear field.
}
The power of $\F$ is fixed by the axial charges of the spurions,
which, in turn, are determined by the axial charges of the hyperbaryons,
and the requirement
that the four-fermion lagrangian~(\ref{LEHCsu4}) will be invariant.
For example, the power of $\F$ in the first effective interaction
matches the axial charge of the product $B_R^A \bB_L^D$ (or $B_R^S \bB_L^D$).
See Eq.~(\ref{Axtrans}) for the axial transformations, and
Table~\ref{tabantihB} for the field content of the hyperbaryons.
The axial charges of the hyperbaryons are listed in Table~\ref{TABzeta}.
Notice that the dependence on the axial charge $q$ of the vector \irrep\
always cancels out in the effective Yukawa interactions.

Similar considerations give rise to the list of
effective interactions which are bilinear in $\S$ or $\S^\dagger$, given by
\begin{eqnarray}
  && \tr(\bA_L\S)\tr(A_R\S^*) \ ,\quad
     \tr(A^c_L\S)\tr(\bA^c_R\S^*) \ ,\quad
     \tr(\bD^T_L\S^* D_R\S) \ ,
\label{ytbil}\\
  && \F^2 \tr(A^c_{L,R}\S)\tr(\bA_{R,L}\S) \ ,\quad
     \F^2 \tr(\bS_{R,L}\S S^c_{L,R}\S) \ ,\quad
     \F^2 \tr(\bA_{R,L}\S A^c_{L,R}\S) \ ,
\nonumber
\end{eqnarray}
where again the hermitian conjugate is to be added to each operator.
This amounts to 18 additional possibilities, none of which vanish.

The coupling constant that multiplies a given effective top-Yukawa interaction
term is obtained using the procedure that we have discussed in detail
in Ref.~\cite{topsect}.  As an example, let us consider the term
$\F\tr(\bA_L\S N_R)$.
Denoting by $y_{A_L,N_R}$ the coupling constant that multiplies this term
in the effective theory, and using $\svev{\F}=1$, we have
\begin{eqnarray}
\label{yALNReff}
  \frac{\partial\ }{\partial N_{Ra\a}(y)}
  \frac{\partial\ }{\partial \bA_{Lijb\b}(x)}
  \log{Z_{\rm eff}}
  &=& -y_{A_L,N_R} \svev{\S_{ji}} \d_{ab} \d_{\a\b} \d(x-y) + \cdots
\\
  &=& y_{A_L,N_R}\,\e_{0,ij}\,\d_{ab} \d_{\a\b} \d(x-y) + \cdots \ ,
\nonumber
\end{eqnarray}
where $a,b$ are $SU(3)$-color indices, $\a,\b$ are Dirac indices,
and we have treated the singlet $N$ as an $SU(4)$ scalar
with no flavor indices.  In the microscopic theory,
\begin{eqnarray}
\label{yALNRmic}
  \frac{\partial\ }{\partial N_{Ra\a}(y)}
  \frac{\partial\ }{\partial \bA_{Lijb\b}(x)} \log{Z}
  &=& \l_1 \l_7 \svev{B^A_{Rjib\b}(x) \bB^N_{La\a}(y)} \ .
\end{eqnarray}
Demanding equality between the effective and microscopic theories,
and using that the right-hand side of Eq.~(\ref{yALNReff}) is the leading term
in a derivative expansion, we obtain
\begin{equation}
\label{yALNRmatch}
  y_{A_L,N_R} = \frac{\l_1 \l_7}{48} \e_{0,ij} \d_{ab} \d_{\a\b} S_{ijab\a\b}(0) \ ,
\end{equation}
where
\begin{equation}
\label{FT}
  \svev{B^A_{Rija\a}(x) \bB^N_{Lb\b}(y)}
  = \int \frac{d^4p}{(2\p)^4} e^{ip(x-y)} S_{ijab\a\b}(p) \ .
\end{equation}
In the absence of spontaneous symmetry breaking, this
two-point function would evidently vanish, because $B^A$ and $B^N$ belong to
two different \irreps\ of $SU(4)$.  But the antisymmetric \irrep\ of $SU(4)$
contains an $Sp(4)$ singlet (see App.~\ref{su4sp4coset}),
and so this two-point function is non-zero after symmetry breaking.
In terms of the elementary fermions of the hypercolor theory
we have, using Table~\ref{tabantihB},
\begin{eqnarray}
\label{FTmic}
  \svev{B^A_{Rija}(x) \bB^N_{Lb}(y)}
  &=& - \int \cd A\,\m(A)\, P_R \svev{\j_{Ia}(x) \bj_{Jb}(y) \g_\n} P_R
\\
  && \times \svev{(\bc_i(x) P_R \G_I \c_j(x))(\bc_k(y) P_L \g_\n \G_J \c_k(y))} \ ,
\nonumber
\end{eqnarray}
where $\cd A$ denotes the Haar measure for the gauge field,
and $\m(A)$ is the Boltzmann weight.
Inside the gauge-field integral, the expectation values denote
correlation functions of the elementary fermions in a fixed gauge-field
background.  There are three different ways to contract the four $\c$ fermions
into a product of two $\svev{\c\bc}$ propagators.
In every case we will have a $P_R$ applied to both sides
of one $\svev{\c\bc}$ propagator, which projects out
an order parameter for the $SU(4)\to Sp(4)$ symmetry breaking.
Expressions for all other contributions to the top-Yukawa coupling
can be worked out in a similar way.  It is clear that the experimental
value of the top-Yukawa coupling in the Standard Model provides only
one constraint on the many couplings present in $\cl_{\rm EHC}$.

\begin{boldmath}
\subsection{\label{su4Veff} $\Veff$}
\end{boldmath}
The effective potential for the pNGBs is generated by integrating out
all other states of the hypercolor theory, and, in addition,
the Standard-Model gauge and fermion fields.
Here we will calculate the effective potential $\Veff$ for the
$SU(4)/Sp(4)$ and $U(1)_A$ pNGBs (we keep disregarding the $SU(6)/SO(6)$ pNGBs),
which is obtained by integrating out the third-generation quarks.\footnote{%
For the gauge boson's contribution to the effective potential,
see Sec.~\ref{su4disc} below.}
To leading order in the four-fermion couplings,
the effective potential arises from correlation functions
of two four-fermion vertices, where every correlation function
is a convolution of a hyperbaryon two-point function with
a single massless quark propagator.  As a result, every term in the
effective potential will be quadratic in the spurions, and both spurions
will have the same handedness.

There is a large number of ways to generate an effective potential,
which we organize into twelve ``templates,''
\begin{align}
  \ct_1 &= \F^{1-2q} \tr(\bA\S N)+\hc\ ,
\allowdisplaybreaks
\label{V2su4}\\
  \ct_2 &= \F^{-1-2q} \tr(\bA^c\S^* N)+\hc\ ,
\allowdisplaybreaks
\nonumber\\
  \ct_3 &= \F^{1 \mp 2q}\tr(\bA\S D^T)+\mbox{h.c.}\ ,
\allowdisplaybreaks
\nonumber\\
  \ct_{4} &= \F^{1 \mp 2q}\tr(\bS\S D^T)+\mbox{h.c.}\ ,
\allowdisplaybreaks
\nonumber\\
  \ct_{5} &= \F^{-1 \mp 2q}\tr(\bA^c\S^* D)+\mbox{h.c.}\ ,
\allowdisplaybreaks
\nonumber\\
  \ct_{6} &= \F^{-1 \mp 2q}\tr(\bS^c\S^* D)+\mbox{h.c.}\ ,
\allowdisplaybreaks
\nonumber\\
  \ct_{7} &= \tr(\bA\S)\tr(A\S^*) \ ,
\allowdisplaybreaks
\nonumber\\
  \ct_{8} &= \tr(A^c\S)\tr(\bA^c\S^*) \ ,
\allowdisplaybreaks
\nonumber\\
  \ct_{9} &= \F^2 \tr(A^c\S)\tr(\bA\S) +\hc \ ,
\allowdisplaybreaks
\nonumber\\
  \ct_{10} &= \F^2 \tr(S^c\S \bS\S) +\hc \ ,
\allowdisplaybreaks
\nonumber\\
  \ct_{11} &= \F^2 \tr(A^c\S \bA\S) +\hc \ ,
\allowdisplaybreaks
\nonumber\\
  \ct_{12} &= \tr(D\S \bD^T\S^*) \ .
\nonumber
\end{align}
As in Sec.~\ref{su4yeff}, the power of $\F$ in each template matches
the axial charge of the associated product of hyperbaryons
in the microscopic theory.  The axial charge vanishes for templates
$\ct_{7}$,  $\ct_{8}$ and $\ct_{12}$.  For the other templates
it doesn't.
We have normalized the axial charge such that the $\c$'s give rise to
an integer power of $\F$.  Templates $\ct_1$ through $\ct_6$ are
sensitive also to $q$, the axial charge of the $\j$'s.
In templates $\ct_3$ through $\ct_6$, $\F^{-2q}$ ($\F^{+2q}$)
corresponds to right-handed (left-handed) spurions.
For $\ct_1$ and $\ct_2$ we always obtain $\F^{-2q}$ from the $\j$'s,
because only $t_R$ can be embedded into a neutral spurion (see below).

The alert reader will have noticed the similarity between
templates $\ct_1$ through $\ct_6$ and the effective Yukawa interactions in
Eq.~(\ref{su4yukawa}), and likewise, between templates $\ct_7$ through $\ct_{12}$
and the effective Yukawa interactions in Eq.~(\ref{ytbil}).
The underlying reason is the similar group theoretic structure,
as well as the power counting, which again allows for a maximum of two
non-linear fields ($\S$ or $\S^*$) in the leading-order effective potential.
While we will shortly explain in detail how the templates
encode the effective potential, already at this stage we point out
several important differences.  First, in the effective Yukawa interactions
the quark fields are present, whereas in the effective potential
they have been integrated out.  Second, the two spurions in the
effective Yukawa interactions are one right-handed and one left-handed,
whereas here both of them have the same handedness.
As a result, the pattern of axial charges in the effective Yukawa interactions
and in the effective potential is different as well.

Every template from Eq.~(\ref{V2su4}) will expand out
to several terms in $\Veff$.\footnote{%
  For an alternative, but ultimately equivalent, spurion technique,
  see for example Ref.~\cite{PW}.
}
We illustrate this using the example of $\ct_1$.
In this case the two spurions must be right-handed,
because $q_L$ cannot be embedded into a singlet of $SU(4)$.
As for $t_R$, it can be embedded into an antisymmetric spurion in two different ways.
Template $\ct_1$ thus gives rise to the following two terms
\begin{equation}
\label{tmpltN}
  C_{1R} \left( \l_7\tl_1 \svev{\F \tr(\bA_R^1\S N_R)+\hc}
         + \l_7\tl_2 \svev{\F \tr(\bA_R^2\S N_R)+\hc} \right) \ .
\end{equation}
Each term consists of the product of three elements:
a low-energy constant, a pair of coupling constants from $\cl_{EHC}$,
and an expression of the form $\svev{\ct_1}$, where we have made
a particular choice for the spurions in the template $\ct_1$.
The meaning of the notation $\svev{\cdot}$ here is the following.
For the right-handed case, this is the outcome of integrating out
the $t_R$ field, and the hyperbaryon fields to which it couples in
Eq.~(\ref{LEHCsu4}).
In practice, denoting the spurion and anti-spurion fields generically
as $X_R$ and $\bX_R$, they are traded inside the $\svev{\cdot}$ symbol
with the corresponding constant spurion matrices $\hX_{t_R}$ and $\hbX_{t_R}$,
see Eq.~(\ref{embedR}).  In the left-handed case,
we in addition sum over the contributions of $t_L$ and $b_L$ (Eq.~(\ref{embedL})).

As mentioned above, each correlation function that contributes to the
leading-order effective potential is built from two vertices from $\cl_{EHC}$,
and so it contains a hyperbaryon two-point function together with
a single quark propagator, which, in this approximation,
is a free massless propagator of a given chirality.
The spurion and the anti-spurion in each template must therefore
have the same chirality.  In the example of Eq.~(\ref{tmpltN}),
only $t_R$ can be embedded into a spurion belonging to the singlet \irrep.
Since there are two independent options for the embedding of $t_R$
into the antisymmetric \irrep, $A_R^1$ and $A_R^2$,
the template expands out to two terms in $\Veff$.

The four-fermion coupling constants together with
the low-energy constant are inferred from a matching procedure
that we have discussed in detail in Ref.~\cite{topsect}, and which is
similar to the one used in the previous subsection
for the case of the effective top-Yukawa couplings.
The four-fermion coupling constants are the two coupling constants
from $\cl_{EHC}$ associated with the spurion and the anti-spurion
that occur inside the $\svev{\cdot}$ symbol.
The remaining low-energy constant is expressed in terms of
a correlation function of the stand-alone hypercolor theory,
which does not depend on the particular embedding of the quark fields
into the spurion.  Therefore, for each template $\ct_i$ we have just two
low-energy constants $C_{iL}$ and $C_{iR}$, one for each chirality.
For the first two templates we only need $C_{iR}$,
because only $t_R$ can be embedded into an $N$ spurion.

Since both the hypercolor theory
and the four-fermion lagrangian are $CP$ invariant, so will be
the effective potential $\Veff$.  Using the assumed reality
of the four-fermion coupling constants, one can also verify directly
the $CP$ invariance of Eq.~(\ref{tmpltN}), and of the corresponding
expressions for all other templates.  Because $\Veff$ is always real,
it follows as a corollary that all the low-energy constants are real.
Similar statements apply to the low-energy constants that multiply the
effective Yukawa couplings discussed in the previous subsection.

We comment in passing that $CP$ is only an approximate symmetry of the
Standard Model, whose breaking is encoded in the Yukawa couplings.
In a similar spirit, one may relax the assumption that the coupling constants
in $\cl_{\rm EHC}$ are all real, and assume, instead, that any imaginary parts
of these coupling constants are parametrically small.
How the EHC theory would induce this small amount of $CP$ violation
goes beyond the scope of this paper.  To avoid confusion,
we stress that since we have defined the low-energy constants to be
independent of the four-fermion lagrangian, their reality is true
regardless of whether or not the coupling constants of $\cl_{EHC}$ are real.

In the rest of this subsection we list all the contributions to $\Veff$
for the twelve templates.  As explained above, the four-fermion couplings
that multiply each expression are easily read off from $\cl_{EHC}$.
The low-energy constants will be derived in the next subsection.
Thanks to the simplicity of the $SU(4)/Sp(4)$ coset,
it is possible to obtain the potential in closed form.
Because some of the templates depend on the
$U(1)$ field $\F$, in general an effective potential
will be generated for the $U(1)_A$ pNGB as well.

We begin with $\ct_1$, which gives rise to the two terms in Eq.~(\ref{tmpltN}).
Using Eqs.~(\ref{pions}) and~(\ref{Scsa}), we have
\begin{subequations}
\label{A12RNRu1}
\begin{eqnarray}
  \svev{\F^{1-2q}\tr(\bA_R^1\S N_R)+\hc}
  &=& 4\cos(\a)\cos((1-2q)\z)
\label{A1RNRu1}\\
  && -\frac{2\sqrt{2}\h\sin(\a)\sin((1-2q)\z)}{\a f} \ ,
\nonumber\\
  \svev{\F^{1-2q} \tr(\bA_R^2\S N_R)+\hc}
  &=& -4\cos(\a)\cos((1-2q)\z)
\label{A2RNRu1}\\
  && -\frac{2\sqrt{2}\h\sin(\a)\sin((1-2q)\z)}{\a f}\ ,
\nonumber
\end{eqnarray}
\end{subequations}
where $\a$ is given by (\seef\ App.~\ref{su4sp4coset})
\begin{equation}
\label{alpha}
\a=\frac{1}{f}\,\left(\half\eta^2+H^\dagger H\right)^{1/2}\ ,
\end{equation}
and we wrote
\begin{equation}
  \F = e^{i\z} \ .
\label{zeta}
\end{equation}
The field $\z$ is dimensionless, and is introduced here for the sake
of brevity.  For the chiral expansion, it is more natural to use
instead the expansion $\F=\exp(i\z/(\sqrt{2}f_\z))$, where the NGB field
has the appropriate canonical dimension, and $f_\z$ is the decay constant
of the $U(1)_A$ NGB \cite{tworeps}.

Because $t_R$ is embedded into the $A_R$ and $A_R^c$ spurions
in the same way, each result for $\ct_2$ may be obtained
from the corresponding result for $\ct_1$ by flipping the signs
of the $SU(4)/Sp(4)$ pNGBs, and multiplying $\z$ by
$-1-2q$ instead of $1-2q$.  The outcome is the same as just
replacing $1-2q$ by $1+2q$ everywhere.

Considering next templates $\ct_3$ through $\ct_6$, which also have
a single non-linear field, but a $D$ spurion instead of the $N$ spurion,
we find for $\ct_3$
\begin{eqnarray}
\label{T3}
\svev{\F^{1-2q}\tr(\bA^1_R\S D^{1T}_R)+\mbox{h.c.}}
&=&4\cos(\a)\cos((1-2q)\z)\\
&&-\frac{2\sqrt{2}\h\sin(\a)\sin((1-2q)\z)}{\a f}\ ,
\nonumber\\
\svev{\F^{1-2q}\tr(\bA^2_R\S D^{2T}_R)+\mbox{h.c.}}
&=&0\ ,\nonumber\\
\svev{\F^{1-2q}\tr(\bA^2_R\S D^{1T}_R)+\mbox{h.c.}}
&=&4\cos(\a)\cos((1-2q)\z)
\nonumber\\
&&+\frac{2\sqrt{2}\h\sin(\a)\sin((1-2q)\z)}{\a f}\ ,
\nonumber\\
\svev{\F^{1-2q}\tr(\bA^1_R\S D^{2T}_R)+\mbox{h.c.}}
&=&0\ ,\nonumber\\
\svev{\F^{1+2q}\tr(\bA_L\S D^{1T}_L)+\mbox{h.c.}}
&=& -4\cos(\a)\cos((1+2q)\z)
\nonumber\\
&& -\frac{2\sqrt{2}\h\sin(\a)\sin((1+2q)\z)}{\a f}\ ,
\nonumber\\
\svev{\F^{1+2q}\tr(\bA_L\S D^{2T}_L)+\mbox{h.c.}}
&=& 0 \ .
\nonumber
\end{eqnarray}
For $\ct_4$ we have
\begin{eqnarray}
\label{T4}
\svev{\F^{1-2q}\tr(\bS_R\S D^{1T}_R)+\mbox{h.c.}}
&=&0\, \\
\svev{\F^{1-2q}\tr(\bS_R\S D^{2T}_R)+\mbox{h.c.}}
&=&4\cos(\a)\cos((1-2q)\z)
\nonumber\\
&&+\frac{2\sqrt{2}\h\sin(\a)\sin((1-2q)\z)}{\a f}\ .
\nonumber\\
\svev{\F^{1+2q}\tr(\bS_L\S D^{1T}_L)+\mbox{h.c.}}
&=& 4\cos(\a)\cos((1+2q)\z)
\nonumber\\
&& +\frac{2\sqrt{2}\h\sin(\a)\sin((1+2q)\z)}{\a f}\ ,
\nonumber\\
\svev{\F^{1+2q}\tr(\bS_L\S D^{2T}_L)+\mbox{h.c.}}
&=& 0 \ ,
\nonumber
\end{eqnarray}
for $\ct_5$,
\begin{eqnarray}
\label{T5}
\svev{\F^{-1-2q}\tr(\bA^{c1}_R\S^* D^1_R)+\mbox{h.c.}}
&=&4\cos(\a)\cos((1+2q)\z)\\
&&-\frac{2\sqrt{2}\h\sin(\a)\sin((1+2q)\z)}{\a f}\ ,\nonumber\\
\svev{\F^{-1-2q}\tr(\bA^{c2}_R\S^* D^2_R)+\mbox{h.c.}}
&=&0\ ,\nonumber\\
\svev{\F^{-1-2q}\tr(\bA^{c2}_R\S^* D^1_R)+\mbox{h.c.}}
&=&4\cos(\a)\cos((1+2q)\z)\nonumber\\
&&+\frac{2\sqrt{2}\h\sin(\a)\sin((1+2q)\z)}{\a f}\ ,\nonumber\\
\svev{\F^{-1-2q}\tr(\bA^{c1}_R\S^* D^2_R)+\mbox{h.c.}}
&=&0\ ,\nonumber\\
\svev{\F^{-1+2q}\tr(\bA^c_L\S^* D^1_L)+\mbox{h.c.}}
&=&-4\cos(\a)\cos((1-2q)\z)
\nonumber\\
&&+\frac{2\sqrt{2}\h\sin(\a)\sin((1-2q)\z)}{\a f}\ ,\nonumber\\
\svev{\F^{-1+2q}\tr(\bA^c_L\S^* D^2_L)+\mbox{h.c.}}
&=&0\ ,\nonumber
\end{eqnarray}
and for $\ct_6$,
\begin{eqnarray}
\label{T6}
\svev{\F^{-1-2q}\tr(\bS^c_R\S^* D^1_R)+\mbox{h.c.}}
&=&0\ ,\nonumber\\
\svev{\F^{-1-2q}\tr(\bS^c_R\S^* D^2_R)+\mbox{h.c.}}
&=&-4\cos(\a)\cos((1+2q)\z)\\
&&-\frac{2\sqrt{2}\h\sin(\a)\sin((1+2q)\z)}{\a f}\ ,
\nonumber\\
\svev{\F^{-1+2q}\tr(\bS^c_L\S^* D^1_L)+\mbox{h.c.}}
&=&-4\cos(\a)\cos((1-2q)\z)
\nonumber\\
&&+\frac{2\sqrt{2}\h\sin(\a)\sin((1-2q)\z)}{\a f}\ ,\ ,\nonumber\\
\svev{\F^{-1+2q}\tr(\bS^c_L\S^* D^2_L)+\mbox{h.c.}}
&=&0\ .\nonumber
\end{eqnarray}

Turning to the templates with two non-linear fields,
for $\ct_{7}$ we have
\begin{subequations}
\begin{eqnarray}
\svev{\tr(\bA_R^1\S)\tr(A_R^1\S^*)} &=&
-4+\frac{4\sin^2{\a}}{\a^2}\,\frac{H^\dagger H}{f^2} \ ,
\label{A1RA1Rfull}\\
\svev{\tr(\bA_R^2\S)\tr(A_R^2\S^*)} &=&
-4+\frac{4\sin^2{\a}}{\a^2}\,\frac{H^\dagger H}{f^2}\ ,
\label{A2RA2Rfull}\\
\svev{\tr(\bA_R^1\S)\tr(A_R^2\S^*)+\hc} &=&
8-\frac{8\sin^2{\a}}{\a^2}\,\frac{(\eta^2+H^\dagger H)}{f^2}\ ,
\label{A1RA2Rfull}\\
\svev{\tr(\bA_L\S)\tr(A_L\S^*)} &=&
-\frac{\sin^2{\a}}{\a^2}\,\frac{4H^\dagger H}{f^2}\ .
\label{ALALfull}
\end{eqnarray}
\end{subequations}
Notice that $\vev{\tr(\bA_R^2\S)\tr(A_R^1\S^*)}$ is the hermitian conjugate
of $\vev{\tr(\bA_R^1\S)\tr(A_R^2\S^*)}$.  The results for $\ct_{8}$
are the same as for the corresponding results for $\ct_{7}$.
The last double-trace template is $\ct_{9}$, for which we obtain
\begin{subequations}
\begin{eqnarray}
  \svev{\F^2\tr(\bA_R^{1}\S)\tr(A_R^{c1}\S)+\hc}
  &=& 8\cos(2\z)\left(-1+\frac{\sin^2{\a}}{\a^2}\,
      \frac{(\eta^2+H^\dagger H)}{f^2}\right) \hspace{5ex}
\label{A1RAc1Ru1}\\
  && + \frac{4\sqrt{2}}{\a f}\,\h\sin(2\z)\sin(2\a)\ ,
\nonumber\\
  \svev{\F^2\tr(\bA_R^{2}\S)\tr(A_R^{c2}\S)+\hc}
  &=& 8\cos(2\z)\left(-1+\frac{\sin^2{\a}}{\a^2}\,
      \frac{(\eta^2+H^\dagger H)}{f^2}\right)
\label{A2RAc2Ru1}\\
  && - \frac{4\sqrt{2}}{\a f}\,\h\sin(2\z)\sin(2\a)\ ,
\nonumber\\
  \svev{\F^2\tr(\bA_R^{1}\S)\tr(A_R^{c2}\S)+\hc} &=&
  8\cos(2\z)\left(1-\frac{\sin^2{\a}}{\a^2}\,\frac{H^\dagger H}{f^2}\right)\ ,
\label{A1RAc2Ru1}\\
  \svev{\F^2\tr(\bA_R^{2}\S)\tr(A_R^{c1}\S)+\hc} &=&
  8\cos(2\z)\left(1-\frac{\sin^2{\a}}{\a^2}\,\frac{H^\dagger H}{f^2}\right)\ ,
\label{A2RAc1Ru1}\\
  \svev{\F^2\tr(\bA_L\S)\tr(A^c_L\S)+\hc} &=&
  8\cos(2\z)\,\frac{\sin^2{\a}}{\a^2}\,\frac{H^\dagger H}{f^2}\ .
\label{ALAcLu1}
\end{eqnarray}
\end{subequations}
Moving on to the single-trace templates, for $\ct_{10}$ we find
\begin{subequations}
\begin{eqnarray}
  \svev{\tr( \bS_R\S S^c_R\S) +\hc}
  &=& 4\cos(2\z) \left(-1+\frac{\sin^2{\a}}{\a^2}\,
      \frac{(\eta^2+H^\dagger H)}{f^2}\right)
\label{SRSRu1}\\
  &&  -\frac{2\sqrt{2}\h\sin(2\z)\sin(2\a)}{\a f} \ ,
\nonumber\\
  \svev{\tr( \bS_L\S S^c_L\S) +\hc}
  &=& 4\cos(2\z)\left(-2+\frac{3\sin^2{\a}}{\a^2}\,
      \frac{H^\dagger H}{f^2}\right) \ ,
\label{SLSLu1}
\end{eqnarray}
\end{subequations}
and for $\ct_{11}$,
\begin{subequations}
\begin{eqnarray}
  \svev{\tr( \bA^1_R\S A^{c1}_R\S) +\hc}
  &=& 4\cos(2\z)\left(-1+\frac{\sin^2{\a}}{\a^2}\,
      \frac{\eta^2+H^\dagger H}{f^2}\right)
\label{AR1AR1su1}\\
  &&  +\frac{2\sqrt{2}\h\sin(2\z)\sin(2\a)}{\a f} \ ,
\nonumber\\
  \svev{\tr( \bA^2_R\S A^{c2}_R\S) +\hc}
  &=& 4\cos(2\z)\left(-1+\frac{\sin^2{\a}}{\a^2}\,
      \frac{\eta^2+H^\dagger H}{f^2}\right)
\label{AR2AR2su1}\\
  &&  -\frac{2\sqrt{2}\h\sin(2\z)\sin(2\a)}{\a f} \ ,
\nonumber\\
  \svev{\tr( \bA^1_R\S A^{c2}_R\S) +\hc}
  &=& -4\cos(2\z)\, \frac{\sin^2{\a}}{\a^2}\,\frac{H^\dagger H}{f^2}\ ,
\label{AR1AR2csu1}\\
  \svev{\tr( \bA^2_R\S A^{c1}_R\S) +\hc}
  &=& -4\cos(2\z)\, \frac{\sin^2{\a}}{\a^2}\,\frac{H^\dagger H}{f^2}\ ,
\label{AR2AR1csu1}\\
  \svev{\tr( \bA_L\S A^c_L\S) +\hc}
  &=& 4\cos(2\z)\left(-2+\frac{\sin^2{\a}}{\a^2}\,
      \frac{H^\dagger H}{f^2}\right)\ .
\label{ALALsu1}
\end{eqnarray}
\end{subequations}
Finally, for $\ct_{12}$ the non-zero results are
\begin{subequations}
\begin{eqnarray}
  \svev{\tr((\bD_R^1)^T\S^*D_R^1\S)} &=&
  -4+8\frac{\sin^2{\a}}{\a^2}\,\frac{H^\dagger H}{f^2}\ ,
\label{DR1DR1full}\\
  \svev{\tr((\bD_R^2)^T\S^*D_R^2\S)} &=&
  2-2\frac{\sin^2{\a}}{\a^2}\,\frac{H^\dagger H}{f^2}\ ,
\label{DR2DR2full}\\
  \svev{\tr((\bD_L^1)^T\S^*D_L^1\S)} &=&
  \frac{\sin^2{\a}}{\a^2}\,\frac{H^\dagger H}{f^2}\ ,
\label{DL1DL1full}\\
  \svev{\tr((\bD_L^2)^T\S^*D_L^2\S)} &=&
  \frac{\sin^2{\a}}{\a^2}\,\frac{H^\dagger H}{f^2}\ .
\label{DL2DL2full}
\end{eqnarray}
\end{subequations}

\subsection{\label{su4LECs} Low-energy constants}
To complete the construction of the effective potential, we need
the low-energy constants.  In order to fully benefit from the $SU(4)$ symmetry
of the hypercolor theory, we now expand each spurion as
\begin{equation}
  X_{L,R}(x) = \h_{L,R}(x) \hX_{L,R}\ , \qquad
  \bX_{L,R}(x) = \bh_{L,R}(x) \hbX_{L,R}\ ,
\label{splitSSd}
\end{equation}
where $\h_{L,R}(x)$ is a free massless Weyl field.
Let us compare this with Eqs.~(\ref{embedL}) and~(\ref{embedR}).
In the latter case, the (hatted) matrices that carry the $SU(4)$ indices
are assigned a fixed numerical value that defines a particular embedding
of a quark field.
By contrast, we now treat $\hX_{L,R}$ and $\hbX_{L,R}$ as global spurions
that do not have any particular value, but, instead,
transform in an \irrep\ of $SU(4)$.  As a final preparatory step,
we eliminate from $\cl_{EHC}$ the information about any specific embedding
of the quark fields while keeping only the information about
the $SU(4)$ \irreps, by writing, e.g.,
$(\l_5 D_R^1(x) + \l_6 D_R^2(x))B_L^D(x) = \cd_R\, \h_R(x) B_L^D(x)$,
where $\cd_R$ is a global spurion in the adjoint \irrep.  In this process
we also deliberately suppress the information about the four-fermion
coupling constants.  As discussed above, this information can easily
be read off from the original definition~(\ref{LEHCsu4}).
We end up re-expressing $\cl_{EHC}$ in terms of the hyperbaryon fields,
the $\h_{L,R}(x)$ field, and a pair of global spurions for each \irrep:
singlet $\cn_{L,R}$, adjoint $\cd_{L,R}$, two-index antisymmetric $\ca_{L,R}$,
two-index symmetric $\cs_{L,R}$, and their complex conjugates
$\ca^c_{L,R}$ and $\cs^c_{L,R}$.

In the (templates for the) effective potential, Eq.~(\ref{V2su4}),
we simply trade every spurion field with the corresponding global spurion.
Each low-energy constant will be obtained by taking ordinary derivatives
with respect to the global spurions, and matching the results
between the microscopic and the effective theories.
This matching procedure will allow us to replace the $\S$ field in the
effective theory by its expectation value.  This, in turn,
simplifies considerably the calculation of the low-energy constants.
Indeed, by making use of the global symmetry,
we are able to extract the low-energy constants from correlation functions
of the microscopic theory that do not involve any NGB asymptotic states.

We start with $\ct_1$, whose contribution to $\Veff$ now reads
\begin{equation}
\label{tmplt1glbl}
  C_{1R} (\F^{1-2q} \tr(\bca_R\S \cn_R)+\hc) \ .
\end{equation}
We recall that we only need the right-handed low-energy constant $C_{1R}$,
because the left-handed quarks cannot be embedded into the singlet \irrep.
In the effective theory,
\begin{eqnarray}
\label{LECARNRalt}
  \e_{0,ij}\,\frac{\partial\ }{\partial \cn_R}
  \frac{\partial\ }{\partial \bca_{Rij}} \frac{\log{Z_{\rm eff}}}{V}
  &=& -C_{1R}\, \e_{0,ij} \svev{\S_{ji}} \
\\
  &=& -C_{1R}\, \e_{0,ij}\, \e_{0,ji} \ = \ 4C_{1R} \ ,
\nonumber
\end{eqnarray}
where we have used that $\svev{\S}=\e_0$ and $\svev{\F}=1$.
In the microscopic theory we have
\begin{eqnarray}
\label{LECARNRmic}
  \e_{0,ij}\,\frac{\partial\ }{\partial \cn_R}
  \frac{\partial\ }{\partial \bca_{Rij}} \log{Z}
  &=& \e_{0,ij}\int d^4x\int d^4y\svev{\bB_L^N(y)\h_R(y)\bh_R(x)B_{Lji}^A(x)}
\\
  &=& i\e_{0,ij}\int d^4x\int d^4y \int\frac{d^4p}{(2\p)^4}\frac{p_\m}{p^2}\,
  e^{ip(y-x)}\svev{\bB_L^N(y)\g_\m B_{Lji}^A(x)}\ .
\nonumber
\end{eqnarray}
Hence
\begin{equation}
\label{C1R}
  C_{1R} = \frac{i\e_{0,ij}}{4}
  \int d^4x \int\frac{d^4p}{(2\p)^4}\frac{p_\m}{p^2}\,
  e^{-ipx}\svev{\bB_L^N(0)\g_\m B_{Lji}^A(x)}\ .
\end{equation}
As in Sec.~\ref{su4yeff} we may express the hyperbaryon two-point function
in terms of the elementary fermions.
As can be seen from Table~\ref{tabantihB},
while  in the case of the $D$ and $N$ \irreps\ the hyperbaryon fields
have the same form for the $SO(5)$ and $SO(11)$ gauge theories,
their forms for the other \irreps\ are different in the two theories.
For definiteness, we will assume in this subsection that the microscopic theory
is the $SO(5)$ gauge theory,\footnote{%
  The reader can easily work out the minor changes for the $SO(11)$ case.
}
obtaining
\begin{eqnarray}
\label{ARNRhb2pt}
  \svev{\bB_L^N(y)\g_\m B_{Lji}^A(x)}
  &=& \half \int\cd A\,\m(A) \svev{ \bj_J(y)\g_\n \g_\m P_L \j_I(x)}
\\
  && \times \svev{\bc_k(y) \g_\n \G_J \c_k(y)}
     \svev{\bc_j(x) P_R \G_I \c_i(x)} + \cdots \ ,
\nonumber
\end{eqnarray}
where we have used that $\bc_k\G_J\g_5\g_\n\c_k=0$,
and the ellipses stand for
a term that vanishes when contracted with $\e_{0,ij}$ in Eq.~(\ref{C1R}).
As expected, the expectation value of $\bc_j P_R \G_I \c_i$ provides for
an order parameter for $SU(4)\to Sp(4)$ symmetry breaking.
Unlike the basic local order parameter (Eq.~(\ref{formal})),
because of the presence of the $SO(d)$ matrices $\G_I$ and $\G_J$
inside of the $\c$ bilinears, only the two-point function
as a whole is a gauge invariant (non-local) order parameter.
In addition, the factor $\svev{ \bj_J(y)\g_\n \g_\m P_L \j_I(x)}$
does not vanish because of the symmetry breaking $SU(6)\to SO(6)$,
so the non-vanishing of the correlator $\svev{\bB_L^N(y)\g_\m B_{Lji}^A(x)}$
requires both $SU(4)$ and $SU(6)$ to be spontaneously broken.

For template $\ct_2$, the only difference in the calculation of $C_{2R}$
is that the hyperbaryon $B^A_L$ is replaced by $B^{A^c}_L$.
This has the effect of replacing the $P_L$ projector inside
the $\bc_j(x) P_L \G_I \c_i(x)$ bilinear in Eq.~(\ref{ARNRhb2pt}) by a $P_R$.
For $\ct_3$, we need an adjoint hyperbaryon instead of the neutral one.
In this case both chiralities are needed, and by similar arguments we find
\begin{equation}
\label{C3final}
  C_{3R,L} = -\frac{i\e_{0,jk}}{4}
  \int d^4x\int \frac{d^4p}{(2\p)^4}\frac{p_\m}{p^2}\,e^{-ipx}
  \svev{\bB^D_{ki}(0)\g_\m P_{L,R} B^A_{ji}(x)}\ .
\end{equation}
The low-energy constants for templates $\ct_4$, $\ct_5$ and $\ct_6$
can be similarly obtained.

In the case of template $\ct_{7}$ we need to do a little more work,
because one can construct from the $\ca$ and $\bca$ spurions also a
symmetry-preserving term that does not depend on the $\S$ field,
$\tr(\bca\ca)$.  Considering the left-handed case
for definiteness, the relevant terms are
\begin{equation}
\label{C7Zeff}
  C_{7L} \tr(\bca_L\S)\tr(\ca_L\S^*) + C'_{7L} \tr(\bca_L\ca_L) \ ,
\end{equation}
and so
\begin{equation}
\label{LECAAdt}
  \frac{\partial}{\partial \ca_{Lij}}\frac{\partial}{\partial \bca_{Lk\ell}}
  \frac{\log Z_{\rm eff}}{V}
  = -C_{7L} \e_{0,ij}\e_{0,k\ell}
    -C'_{7L} \left(\d_{jk}\d_{i\ell}-\d_{ik}\d_{j\ell}\right)\ .
\end{equation}
We may now extract $C_{7L}$ by contracting this result with
the fully antisymmetric four-dimensional tensor $\e_{ijk\ell}$.
By applying the same differentiations to the microscopic theory, and
comparing the results, we find
\begin{equation}
\label{AAdtmic}
  C_{7L} = \frac{i\e_{ijk\ell}}{8}
  \int d^4x \int\frac{d^4p}{(2\p)^4}\,\frac{p_\m}{p^2}\,
  e^{-ipx} \svev{\bB^A_{ji}(0) \g_\m P_R B^A_{\ell k}(x)}\ .
\end{equation}
For $C_{7R}$, the chiral projector inside the hyperbaryon two-point function
is $P_L$.  In terms of the elementary fermions,
\begin{eqnarray}
\label{AAdtmichb2pt}
  \svev{\bB^A_{ji}(y) \g_\m P_{R,L} B^A_{\ell k}(x)}
  &=& \int \cd A\,\m(A) \svev{\bj_I(y) \g_\m P_{R,L} \j_J(x)}
\\
  && \times \svev{\bc_i(y)P_L\G_I\c_j(y)} \svev{\bc_\ell(x)P_R\G_J\c_k(x)}
  + \cdots \ ,
\nonumber
\end{eqnarray}
where again the ellipses denote terms that vanish when contracted with
$\e_{ijk\ell}$ in Eq.~(\ref{AAdtmic}).
We see that from each $\c$ propagator we pick up the
part proportional to $\e_0$ in flavor space, which is non-zero
in the broken phase.

For template $\ct_{8}$, the $\ca$ and $\bca$ spurions are replaced by
$\bca^c$ and $\ca^c$ spurions, respectively.
The result is similar, except that,
in Eq.~(\ref{AAdtmichb2pt}), the chiral projectors inside the $\c$ bilinears
get flipped.

For templates $\ct_{9}$, $\ct_{10}$ and $\ct_{11}$
there are no $\S$ independent terms.
For $\ct_{9}$ we find
\begin{equation}
\label{T9}
  C_{9L,R} = \frac{i\e_{ijk\ell}}{8}
  \int d^4x \int\frac{d^4p}{(2\p)^4}\,\frac{p_\m}{p^2}\,
  e^{-ipx} \svev{\bB^{A^c}_{ji}(0) \g_\m P_{R,L} B^A_{\ell k}(x)} \ ,
\end{equation}
where
\begin{eqnarray}
\label{B9}
  \svev{\bB^{A^c}_{ji}(y) \g_\m P_{R,L} B^A_{\ell k}(x)}
  &=& \int \cd A\,\m(A) \svev{\bj_I(y) \g_\m P_{R,L} \j_J(x)}
\\
  && \times \svev{(\bc_i(y)P_R\G_I\c_j(y))(\bc_\ell(x)P_R\G_J\c_k(x))} \ .
\nonumber
\end{eqnarray}
This time, the three possible contractions of the $\c$'s are all non-zero
in the broken phase, and contribute to the low-energy constants.
For template $\ct_{10}$,
\begin{equation}
\label{T10}
  C_{10L,R} = -\frac{i\e_{ijk\ell}}{8}
  \int d^4x \int\frac{d^4p}{(2\p)^4}\,\frac{p_\m}{p^2}\,
  e^{-ipx} \svev{\bB^{S^c}_{ji}(0) \g_\m P_{R,L} B^S_{\ell k}(x)} \ ,
\end{equation}
where
\begin{eqnarray}
\label{B10}
  \svev{\bB^{S^c}_{ji}(y) \g_\m P_{R,L} B^S_{\ell k}(x)}
  &=& \int \cd A\,\m(A) \svev{\bj_I(y) \s_{\s\r} \g_\m P_{R,L} \s_{\k\l} \j_J(x)}
\\
  && \times \svev{(\bc_i(y)P_R \s_{\s\r} \G_I\c_j(y))
                  (\bc_\ell(x)P_R\s_{\k\l} \G_J\c_k(x))} \ .
\nonumber
\end{eqnarray}
For template $\ct_{11}$ we find $C_{11L,R}=-C_{9L,R}$.

Finally, in the case of template $\ct_{12}$ we once more have
a symmetry preserving term, $C'_{12L,R}\tr(\bcd_{L,R}\cd_{L,R})$,
that we need to separate out.\footnote{%
  See Ref.~\cite{Wvac} for a similar calculation.
}
Expanding the adjoint fields on the basis
of $SU(4)$ generators $T_a$ we have in the effective theory
(omitting the chirality label)
\begin{equation}
\label{LECDD}
  \frac{\partial}{\partial \cd_a}\frac{\partial}{\partial \bcd_b}
  \frac{\log{Z_{\rm eff}}}{V}
  = -C_{12} \tr(T_b^T\e_0 T_a\e_0) - C'_{12} \tr(T_b T_a)\ .
\end{equation}
The right-hand side is proportional to $(\pm C_{12}+C'_{12})\d_{ab}$
when $T_a$ is an unbroken, respectively, broken generator.
By considering both cases we may extract the low-energy constant.
In the microscopic theory (considering the left-handed spurions for
definiteness)
\begin{equation}
\label{DDmic}
  \frac{\partial}{\partial \cd_{La}}\frac{\partial}{\partial \bcd_{Lb}}
  \frac{\log{Z}}{V}
  = i \int d^4x \int\frac{d^4p}{(2\p)^4}\,\frac{p_\m}{p^2}\,
      e^{-ipx}\svev{\bB^D_a(0)\g_\m P_R B^D_b(x)}\ ,
\end{equation}
where
\begin{equation}
\label{DDmichb2pt}
  \svev{\bB^D_a(y)\g_\m P_{L,R} B^D_b(x)}
  = \int \cd A\,\m(A)\, \cf_{IJ\n\r ab}(x,y)
  \svev{\bj_J(y) \g_\r \g_\m \g_\n P_{L,R} \j_I(x)} \ ,
\end{equation}
and
\begin{eqnarray}
\label{contraction}
  \cf_{IJ\n\r ab}(x,y) &=&
  \svev{(\bc(x) P_R\g_\n\G_I T_b\c(x))(\bc(y) P_R\g_\r\G_J T_a\c(y))}
\\
  &=& -\tr\Big(
  T_b \svev{\c(x)\bc(y)} P_R\g_\r\G_J T_a \svev{\c(y)\bc(x)} P_R\g_\n\G_I\Big)
\nonumber\\
  && -\tr\Big(
  T_b\svev{\c(x)\bc(y)}\g_\r P_R\G_J T_a^T\svev{\c(y)\bc(x)}P_R\g_\n\G_I\Big) \ .
\nonumber
\end{eqnarray}
The first term on the right-hand side of the second equality
picks up the kinetic part of
the $\c$ propagator, which is symmetry preserving and proportional
to $\d_{ij}$ in flavor space.  The flavor trace therefore collapses to
$\tr(T_b T_a)$, which corresponds to the $C'_{12}$ term in Eq.~(\ref{LECDD}).
The last term picks up the symmetry breaking part of the $\c$ propagator,
which is proportional to $\e_{0,ij}$.  This precisely corresponds
to the flavor trace multiplying the $C_{12}$ term in Eq.~(\ref{LECDD}),
and therefore the low-energy constants $C_{12L,R}$ are obtained by substituting
this term into Eq.~(\ref{DDmichb2pt}).  This completes the derivation
of the low-energy constants for this theory.

\subsection{\label{su4disc} Summary}
Collecting everything, we see that the effective potential
arising from integrating out the third-generation quarks takes the form
\begin{equation}
\label{Veff9}
  \Veff = c_0 + \sum_{i=1}^{9} c_i f_i \ ,
\end{equation}
with the following nine functions
\begin{eqnarray}
  f_{1,2} &=& \cos(\a)\cos((1\pm 2q)\z) \ , \qquad
  f_{3,4} \ = \ \frac{\h\sin(\a)\sin((1\pm 2q)\z)}{\a f} \ ,
\label{Vf}\\
  f_5 &=& \frac{\h\sin(2\a)\sin(2\z)}{\a f} \ ,
\nonumber\\
  f_6 &=& \frac{\sin^2 \a}{\a^2 f^2}\, H^\dagger H \ , \qquad
  f_7 \ = \ \cos(2\z) \frac{\sin^2 \a}{\a^2 f^2}\, H^\dagger H \ ,
\nonumber\\
  f_8 &=& \frac{\sin^2 \a}{\a^2 f^2}\, \h^2 \ , \qquad
  f_9 \ = \ \cos(2\z) \frac{\sin^2 \a}{\a^2 f^2}\, \h^2 \ ,
\nonumber
\end{eqnarray}
and where $\a$ is given by Eq.~(\ref{alpha}).
An interesting feature of this result is that, in general,
a potential is generated not only for the Higgs doublet and for $\h$,
which are the NGBs of the $SU(4)/Sp(4)$ coset, but also for the
singlet NGB $\z$.  (We recall that in this paper we disregard
the NGBs of the $SU(6)/SO(6)$ coset.)
The $c_i$'s of Eq.~(\ref{Veff9}) can be expressed in terms of the coupling constants
of $\cl_{EHC}$ and the low-energy constants that we have derived in the
previous subsection.  The low-energy constants can be determined
from a lattice calculation, which would then allow for a study
of the experimental constraints on the four-fermion coupling constants.
We note that experimental constraints on the effective potential alone
can, of course, be studied directly in terms of the $c_i$'s.
However, if one wants to incorporate the top Yukawa coupling into
this analysis, then it has to be done in terms of the four-fermion couplings,
and thus, it depends on the knowledge of the low-energy constants.

For completeness, we also give the gauge-boson contribution to
the effective potential, which is
\begin{equation}
\label{effpotweakPR}
  V_{EW} = -C_w\,\tr(\S Q_a\S^*Q_a^*)\ ,
\end{equation}
where $Q_a$ is to be summed over $gT_L^i$ and $g'Y=g'T_R^3$,
and where $C_w>0$ \cite{EW1983}.
The expression for the low-energy constant $C_w$ may be found in Ref.~\cite{Wvac}
for the case of a real \irrep.  The case of a pseudoreal \irrep\ defers
only by the overall sign.  However, relative to the definition of $V_{EW}$
given in Ref.~\cite{Wvac}, in Eq.~(\ref{effpotweakPR}) we have introduced an
extra a minus sign on the right-hand side.  This cancels out
against the sign that is encountered in the derivation,
so that now $C_w$ comes out
positive in the pseudoreal case as well.  With this, we find
\begin{equation}
\label{Vgb}
  V_{EW} = -\frac{C_w}{2} (3g^2+g'^2) (1-f_6) \ ,
\end{equation}
where $f_6$ is defined in Eq.~(\ref{Vf}).  The gauge bosons contribution
will therefore add up to the coefficient $c_6$.
As usual, taken by itself this contribution prefers
the trivial vacuum $\svev{H}=0$, a phenomenon that goes under the name
of vacuum alignment \cite{MP1980}.  But considering $\Veff$ as a whole,
there is ample room for a non-trivial minimum of the Higgs field.

A final contribution to the effective potential might come from mass terms
for the $\c$ fermions.  One can write down two mass terms which are
invariant under the Standard Model symmetries \cite{GELT}.
Introducing $\e_0^\pm = \pm(i/2)(1\pm \t_3)\times \t_2$ (where we are using
the notation of App.~\ref{su4sp4coset}), these mass terms are
\begin{eqnarray}
\label{Lmasspm}
  V_m^\pm &=&  Bm^\pm \tr(\F \S \e_0^\pm + \hc)
\\
  &=& Bm^\pm \left(-4\cos(\zeta) \cos(\alpha)
      \pm \frac{2\sqrt{2} \sin(\zeta)\sin(\a)}{\a f} \right)\ ,
\nonumber
\end{eqnarray}
where we have used Eqs.~(\ref{formal}),~(\ref{zeta}) and~(\ref{Scsa}),
and $B$ is a low-energy constant.
For $m^+=m^-=m$, the mass term simplifies to
\begin{eqnarray}
V_m &=&  Bm \tr(\F \S \e_0 + \hc)
\label{Lmass}\\
  &=& -8Bm \cos(\zeta) \cos(\alpha) \ .
\nonumber
\end{eqnarray}
The mass term~(\ref{Lmass}) breaks the global $SU(4)$ symmetry
explicitly to $Sp(4)$, and the individual mass terms~(\ref{Lmasspm})
further break it explicitly to the Standard Model symmetry
$SU(2)_L\times SU(2)_R$.  From the point of view of the stand-alone
hypercolor theory it may be more natural to avoid any mass terms,
since this keeps the full $SU(4)$ global symmetry intact.
Having said this, we observe that explicit breaking of the flavor symmetry
of the hypercolor theory, encoded in the four-fermion lagrangian~(\ref{LEHCsu4}),
must originate from the EHC theory.  Since we do not know the details
of this EHC theory, we cannot rule out that it might also induce
some of the mass terms discussed above.
Similar statements apply to a Dirac mass term $\propto \bj \j$
for the vector-\irrep\ fermions,
which breaks the $SU(6)$ symmetry explicitly to $SU(3)_c$.

The structure of the total potential is complicated. Its minimum will depend
on the values of the low-energy constants, which can be determined within
the hypercolor theory, and on the four-fermion couplings $\l_i$ and $\tl_i$,
which arise from integrating out heavy degrees of freedom of the EHC theory.
In addition, the potential depends on the electroweak couplings
through Eq.~(\ref{Vgb}), and possibly, on the mass term~(\ref{Lmasspm})
or~(\ref{Lmass}).
Here we will be content with an example of a phenomenologically viable
potential obtained by setting to zero by hand most of
the four-fermion couplings.

Our example consists of turning on the following couplings:
$\l_2$, $\l_7$, and $\tl_1=-\tl_2$, setting to zero the rest of the
four-fermion couplings and the mass terms.
Notice that $\tl_1$ and $\tl_2$ involve the same hyperbaryon, $B_L^A$,
hence the notion of a fixed ratio $\tl_1/\tl_2$ is invariant under
renormalization-group evolution.  Also, $\tl_1=-\tl_2$ implies that
the spurions $A_R^1$ and $A_R^2$ always occur
as the linear combination $A_R^1-A_R^2 \propto \e_0$.

With this choice, the only contribution that depends on $\z$
arises from template $\ct_1$ (see Eq.~(\ref{A12RNRu1})), and is given by
\begin{equation}
  8\,C_{1R}\,\l_7\tl_1 \cos(\a)\cos((1-2q)\z) \ .
\label{VT1}
\end{equation}
We will demand that the minimum of the potential occurs for $|\a|<\p/2$,
as is required for a phenomenologically viable solution.  Further
assuming that
\begin{equation}
  C_{1R}\,\l_7\tl_1<0 \ ,
\label{C1Rineq}
\end{equation}
then implies that $\svev\z=0$ at the minimum of the potential.
(Alternatively, we may set $\l_7=0$ and achieve a similar result
by turning on the mass term~(\ref{Lmass}) with $m>0$.)
Setting $\z=0$, the complete potential is then give by
\begin{equation}
\label{potMG}
  V(H,\eta) = -a_1 \cos\a + a_2 \sin^2\a
  + a_3\,\frac{\sin^2\a}{\a^2f^2}H^\dagger H\ ,
\end{equation}
where $a_1=-8C_{1R}\,\l_7\tl_1$, $a_2=16\,C_{3R}\,\tl_1^2$, and
\begin{equation}
  a_3 = C_w(3g^2+g'^2)/2 - 4\,C_{8L}\,\l_2^2 \ .
\label{ggAL}
\end{equation}
The $a_2$ term arises from the contributions of right-handed spurions
to $\ct_{7}$, while the $a_3$ term arises from the gauge-bosons contribution
as well as from the left-handed spurions in $\ct_{7}$.
The $a_1$ and $a_2$ terms have full $Sp(4)$ invariance since they depend
on $H$ and $\h$ only through $\a$.  It follows that, if the minimum of
the potential occurs for non-zero $\a$, it will point in the $H$ direction
(\ie, $\svev{H}\ne 0$ and $\svev{\h}=0$) when $a_3<0$,
and in the $\h$ direction when $a_3>0$.  This conclusion is confirmed
by studying the saddle-point equations.  Thus, to be phenomenologically viable,
the top-sector contribution to $a_3$ must be (negative and) large enough
to overcome the positive contribution of the gauge bosons.
A sufficient set of conditions to ensure a vacuum with
$\svev\z=\svev\h=0$ and $\svev{H}\ne 0$ is $a_1>0$, $a_3<0$, and
\begin{equation}
  -C_{1R}\,\l_7\tl_1 + 4\,C_{7R}\,\tl_1^2 - C_{8L}\,\l_2^2
  +C_w(3g^2+g'^2)/8 < 0 \ ,
\label{C3Lineq}
\end{equation}
where this last condition implies that the curvature in the $H$ direction
is negative at the origin, and thus that the minimum of the potential
cannot occur for $\a=0$.  Once $\svev\z=\svev\h=0$, the potential
further simplifies.  We defer further discussion of the resulting potential
to the concluding section.

Returning momentarily to the EHC theory, we observe that if
the four-fermion couplings arise from integrating out heavy gauge bosons,
then each four-fermion term must take the form of
a current-current interaction (possibly up to a Fierz rearrangement).
Checking Table~\ref{tabantihB} shows that this condition is satisfied
for all the four-fermion couplings that contribute to our example potential.
Some other four-fermion couplings, such as, for example, the $\l_1$ term,
cannot be brought to the form of a current-current interaction,
and would thus vanish.  However, if the heavy EHC degrees of freedom
that have been integrated out include not only gauge bosons but also fermions
(whose mass could have either an explicit or a dynamical origin),
or scalars, then none of the four-fermion operators in Eq.~(\ref{LEHCsu4})
is ruled out.  In that case we could, for example, turn off $\l_2$
and turn on $\l_1$ instead.  The only change in the potential would be
that $C_{8L}\,\l_2^2$ gets replaced by $C_{7L}\,\l_1^2$.

\begin{table}[t]
\vspace*{0ex}
\begin{center}
\begin{tabular}{r|l|l} \hline \hline
 +1 & $D_L^1$ & $A_R^1$, $A_R^{c1}$ \\
  0 & $A_L$, $A_L^c$, $S_L$, $S_L^c$ & $N_R$, $D_R^1$, $D_R^2$ \\
 -1 & $D_L^2$ & $A_R^2$, $A_R^{c2}$, $S_R$, $S_R^c$ \\ \hline\hline
\end{tabular}
\end{center}
\begin{quotation}
\floatcaption{TABeta}{Values $n=-1,0,+1$ of the phase transformation
$\exp(-in\tileta_0)$, which is to be applied to a Standard Model field,
together with the spurion embeddings of $q_L$ (2nd column)
and $t_R$ (3rd column) for which, for this $n$,
the corresponding term in $\cl_{EHC}$ remains invariant when
the $SU(4)$ transformation $U_0$ is applied to the $\c$ fields.}
\end{quotation}
\vspace*{-4ex}
\end{table}

\begin{boldmath}
\subsection{\label{SCPB} Spontaneous $CP$ breaking}
\end{boldmath}
The Standard-Model neutral fields $\eta$ and $\zeta$ are pseudoscalars, and so,
at face value, their expectation values break $CP$ spontaneously.
(We are assuming that all the four-fermion couplings are real,
so that $CP$ is not broken explicitly.)  Recently, it has been pointed out
in Ref.~\cite{Sannino} that this is not necessarily true,
because it might be possible to shift the expectation value to zero
through field redefinitions.\footnote{
  However, in our opinion the discussion of Ref.~\cite{Sannino} is incomplete.
}
Here we address this question, first for $\svev\eta$,
and then for $\svev\zeta$.

Assume that at the minimum of the effective potential, $\svev\eta=\eta_0\ne0$.
In order to ``rotate away'' this expectation value we need to apply
to the $\c$ fields of the hypercolor theory the $SU(4)$ transformation
$U_0=\exp(-i\tileta_0 X/2)$, where we have introduced the dimensionless
quantity $\tileta_0=\eta_0/(\sqrt{2}f)$, and $X = \t_3 \times 1$ is the
generator associated with $\eta$ (see Eq.~(\ref{pions})).  Indeed,
if $\svev\S=\exp(i\tileta_0 X)\e_0$, then $U_0 \svev\S U_0^T = \e_0$.
If initially both $\eta$ and $H$ have non-zero expectation values,
then the $U_0$ transformation will set $\svev\eta=0$ while in general
changing the expectation value of $H$ as well.

The question now is whether we can find a matching transformation
of the Standard Model fields $q_L$ and $t_R$, such that, together with the
transformation $\c\to U_0\c$, the total lagrangian $\cl_{HC}+\cl_{EHC}$
will be invariant.  If the answer is Yes, then we have achieved
$\svev\eta=0$ via the field redefinitions, which implies that
$\svev\eta$ was indeed unphysical.

In order to keep a particular term in $\cl_{EHC}$ invariant,
the transformation needed for a given Standard Model field
depends on its spurion embedding.  Using the $SU(4)$ transformation rules
of the spurions, and applying the transformation to each
spurion embedding in turn, we find that this transformation
can always be realized via the multiplication of the Standard Model field by
a $U(1)$ phase $\exp(-in\tileta_0)$, where the possible values
of $n$ are $-1,0,+1$.  We list the values of $n$ for all spurion
embeddings of $q_L$ and $t_R$ in Table~\ref{TABeta}.

The answer to the question is now clear.  Consider the set of
non-zero couplings in $\cl_{EHC}$.  If all of the spurion embeddings
of $q_L$ belong to the same row of Table~\ref{TABeta}, and the same is true
also for the embeddings of $t_R$, then invariance of $\cl_{EHC}$ will be
achieved by applying the corresponding phase transformations to $q_L$
and to $t_R$.  In this case the expectation value of $\eta$
can indeed be rotated away, and is thus unphysical.
But if the spurion embeddings of $q_L$ and/or $t_R$
belong to more than one row of the table, then it is not possible
to maintain the invariance of $\cl_{EHC}$.  In this case $\svev\eta$
is physical, and $\svev\eta\ne0$ signifies the spontaneous breaking of $CP$
(for an exception, see below).

A similar argument applies to $\svev\zeta$.  The phase transformation
of a Standard Model field that we now need for a particular term
in $\cl_{EHC}$ is determined by the axial charge of the hyperbaryon
to which it couples (see Table~\ref{TABzeta}).
Once again, in order to be able to rotate $\svev\zeta$
away, the necessary and sufficient condition is that $q_L$ couples
to hyperbaryons that all have the same axial charge, and that the same
is true for $t_R$.

For the example potential discussed in the previous subsection
we have turned on the couplings $\l_2$, $\l_7$,
$\tl_1$ and $\tl_2$.  Only the $\l_2$ term is a spurion embedding
of $q_L$, so this poses no difficulty.  However, the three spurion embeddings
of $t_R$ associated with the remaining three couplings populate
all three lines of Table~\ref{TABeta}.  Therefore, the invariance
of $\cl_{EHC}$ under the field redefinition $\c\to U_0\c$
cannot be maintained, which implies that
$\svev\eta$ is physical.  The same is true for $\svev\zeta$
since the axial charges of the relevant hyperbaryons are all different
from each other.  As a result, for $\svev\eta\ne0$ and/or $\svev\zeta\ne0$,
$CP$ is broken spontaneously.

An exception is the special case $\svev\S=(\t_3\times 1)\e_0$, which
corresponds to specific non-zero values of both $\svev\eta$ and $\svev\zeta$.
Even if both expectation values are physical,
in this special case $CP$ is not broken spontaneously, because
$\svev\S$ is real, and so it remains invariant under the combined
sign flip of $\eta$ and $\zeta$.

Finally, we comment that an advantage of the $SU(4)/Sp(4)$ coset is
that it does not contain any isospin-triplet fields, and, as a result,
the difficulties with triplet expectation values and their potential
influence on the $\r$-parameter do not arise.

\vspace{2ex}

\begin{boldmath}
\section{\label{su5so5} The $SU(5)/SO(5)$ coset}
\end{boldmath}
The list of Ref.~\cite{ferretti16} includes two models in which
the spinor \irrep\ is real, based on the gauge groups $SO(7)$ and $SO(9)$.
These models are the subject of this section.  While the vector-\irrep\ fermions
$\j_{Ia}$ are the same as before, $\c_i$ will now denote 5 Majorana
fermions in the real spinor \irrep\ (the relation between $\c_i$ and $\bc_i$
is still given by Eq.~(\ref{maj})).
In comparison with the $SU(4)/Sp(4)$ coset we have studied in
the previous section, the $SU(5)/SO(5)$ coset is larger.
Apart from the Higgs field and the singlet $\eta$,
it contains nine additional NGBs that fill up the $(3,3)$ representation
of $SU(2)_L\times SU(2)_R$.  For the basic features
of the $SU(5)/SO(5)$ coset, and the embedding of the 14 NGBs
into the pion field, see App.~\ref{su5so5coset}.

The order parameter $\svev{\bc_i\c_j}$ is symmetric on its indices
for a real \irrep.  We will assume that the vacuum state has
$\svev{\bc_i\c_j}\propto\d_{ij}$.  Applying the infinitesimal
flavor transformation, Eq.~(\ref{flavormaj}),
we see that the NGB fields are all pseudoscalars,
\begin{equation}
\label{pireal}
  \d_a (\bc\c) = i\bc\g_5(T_a+T_a^T)\c \ .
\end{equation}
The NGBs correspond to the 14 real symmetric generators of $SU(5)$.
For the 10 antisymmetric, imaginary generators of $SU(5)$, we have
$\d_a (\bc\c)=0$, showing that the unbroken group is $SO(5)$.

These features of the NGBs resembles QCD, and are different
from what we saw in the previous section for the case of a pseudoreal \irrep.
As in QCD, it is easy to check that all the NGB fields flip sign
under the $CP$ transformation of the hypercolor theory, Eq.~(\ref{CPc}).
This creates a phenomenological problem concerning the Higgs field.
The Standard Model's $CP$ transformation, which we will denote
as $\CPSM$, must be different from the original $CP$ transformation
of the hypercolor theory, because the real components of $H_0$ and $H_+$
are even under $\CPSM$, but, like all NGBs, they are odd under
the $CP$ transformation of the hypercolor theory.
As it turns out, $\CPSM$ may be obtained as the product of
the original $CP$ and a diagonal $SO(5)$ transformation.\footnote{%
  For a similar situation, see Ref.~\cite{LV}.
}
Explicitly,
\begin{equation}
\label{CPSM}
  \CPSM = \cq \,\circ\, CP \ , \qquad
  \cq = diag(1,-1,1,-1,1) \ .
\end{equation}

The formal correspondence of the effective fields with the microscopic theory
takes a similar form to Eq.~(\ref{formal}), except that now the
non-linear coset field $\S$ is a symmetric unitary $5\times 5$ matrix.
The pion field $\P$ is real, symmetric, and traceless (see Eq.~(\ref{Sigma})).  Using the
embedding of the Higgs field into the pion field, given in App.~\ref{su5so5coset},
it is straightforward to check that Eq.~(\ref{CPSM}) correctly reproduces the
Standard-Model transformation rules of all components of the Higgs field.

The organization of this section is as follows.
Since the methodology is the same as in the previous section, we will be brief,
and focus on those features of the $SU(5)/SO(5)$ coset that are
different from the $SU(4)/Sp(4)$ coset.
As before we begin with the spurions in Sec.~\ref{su5spurions},
and write down the four-fermion lagrangian in Sec.~\ref{su5LEHC},
which is then followed by the list of top Yukawa effective couplings
in Sec.~\ref{su5yeff}. Turning to the effective potential for the pNGBs,
we begin in Sec.~\ref{su5LECs} with the templates,
which are followed by the list of low-energy constants.
Because of the complexity of the $SU(5)/SO(5)$ coset we were unable
to obtain the effective potential in closed form.
The expansion of $\Veff$ to second order in the pNGB fields
is relegated to App.~\ref{Veff2nd},
while in Sec.~\ref{su5Veff} and App.~\ref{Veffphi} we focus on the third order terms
and their phenomenological role.

\vspace{2ex}

\subsection{\label{su5spurions} Spurions}
As usual, we assume that the third-generation quark fields
couple linearly to three-constituent baryons of the hypercolor theory,
via four-fermion interactions that originate from an extended hypercolor theory
which is operative at an as yet much higher energy scale.
In view of our ignorance of the EHC theory, we must allow for
the most general form of the four-fermion lagrangian 
which is compatible with the symmetries of the Standard Model:
the continuous symmetries $SU(3)_c$, $SU(2)_L$, $T_R^3$ and $B$,
and the discrete symmetry $\CPSM$.
Analogous to Sec.~\ref{su4sp4}, we do this by looking for
all the embeddings of $q_L$ and $t_R$ into $SU(5)$ spurions.
Demanding consistency with the assignment of Standard-Model quantum numbers
then yields the most general coupling between the third-generation quarks
and the hyperbaryons.

We begin with the left-handed doublet $q_L=(t_L,b_L)$.  Introducing
the $5\times 5$ matrices
\begin{equation}
\label{Thetaq}
  \Theta_q = -i \left(\begin{array}{ccccc}
    0&0&0&0& ib_L \\
    0&0&0&0& b_L  \\
    0&0&0&0& it_L \\
    0&0&0&0& -t_L \\
    0&0&0&0&0
  \end{array}\right)\ , \qquad
  \bTheta_q = i \left(\begin{array}{ccccc}
    0&0&0&0&0 \\
    0&0&0&0&0 \\
    0&0&0&0&0 \\
    0&0&0&0&0 \\
    -i\bb_L & \bb_L & -i\bt_L & -\bt_L & 0
  \end{array}\right)\ ,
\end{equation}
all the spurion embeddings of $q_L$ may be constructed using
$\Theta_q$ and $\Theta_q^T$, and all the embeddings of $\bq_L$
may be constructed using $\bTheta_q$ and $\bTheta_q^T$.
For the adjoint \irrep\ we have two independent embeddings,
$D_L^1=\Theta_q$ and $D_L^2=\Theta_q^T$.
For the symmetric \irrep\ there is only one embedding
$S_L=S^c_L=\Theta_q+\Theta_q^T$,
and similarly for the anti-symmetric \irrep\ we have
$A_L=A^c_L=\Theta_q-\Theta_q^T$.
Notice that while the quark content of the spurions $S_L$ and $S^c_L$
is the same (and similarly for $A_L$ and $A^c_L$),
they are nevertheless different spurions,
because their $SU(5)$ transformation rules are different.
For $g\in SO(5)$, the transformation rules of all the two-index
$SU(5)$ \irreps\ collapse to the common rule $X\to gXg^T$.
The relative phases of different entries of
$\Theta_q$ and $\bTheta_q$ are fixed by the embedding of $SU(2)_L$
and $SU(2)_R$ as subgroups of $SO(5)$ (see Eq.~(\ref{tensorprod})).
Our choice of the overall phase of $\Theta_q$ will be explained shortly.

Being an $SU(2)_L$ singlet with $T_R^3=0$,
the right-handed quark field $t_R$ can be embedded into a $(1,1)$
or into a $(1,3)$ of $SU(2)_L\times SU(2)_R$.\footnote{%
  The basis elements that span the $(3,1)$ and $(1,3)$ \irreps\
  are the generators $T_L^i$, respectively $T_R^i$, themselves.
}
The simplest possibility is the $SU(5)$ singlet $N_R=diag(1,1,1,1,1)t_R$.
For the adjoint \irrep\ we again have two embeddings,
$D_R^1 = T_R^3 t_R$ and $D_R^2 = diag(1,1,1,1,-4)t_R$,
which correspond to the $(1,3)$ and $(1,1)$ cases, respectively.
There are two more possibilities for the symmetric \irrep,
$S_R^1=S_R^{1c}= diag(1,1,1,1,0)t_R$ and $S_R^2=S_R^{2c}= diag(0,0,0,0,1)t_R$,
both of which correspond to the $(1,1)$ case.
Finally, there is a single embedding for antisymmetric \irrep,
$A_R=A^c_R=T_R^3 t_R$, which belongs to $(1,3)$.
As for $q_L$, we sometimes encounter the same embedding of $t_R$
for different $SU(5)$ \irreps.
For example, in each of the spurions $D_R^1$, $A_R$ and $A^c_R$,
the quark field $t_R$ is multiplied by the same constant matrix, $T_R^3$.
Again, these are nevertheless different spurions,
because of their different $SU(5)$ transformation properties.

The $c$-number matrices that define the anti-spurion embeddings
(recall Eqs.~(\ref{embedL}) and~(\ref{embedR})) are always given by
\begin{equation}
  \hbX \equiv \hX^\dagger = \cq \hX^T \cq \ ,
\label{hX}
\end{equation}
where the $SO(5)$ matrix $\cq$ is defined in Eq.~(\ref{CPSM}).
The last equality, which can be verified on a case-by-case basis,
depends on the fact that all the right-handed spurion matrices $\hX_R$
are real, and all the left-handed spurion matrices $\hX_L$ were constructed
using $\Theta_q$ (and its transpose), which implies that
$\hX_{L,ij}$ is always real for even $i+j$, and imaginary for odd $i+j$.
Of course, choosing to multiply any spurion matrix by some arbitrary phase
would spoil these features.  As already explained in Sec.~\ref{su4LEHC},
we refrain from doing this because we are after the most general
four-fermion lagrangian which is consistent with the Standard Model's
symmetries, including, in particular, $\CPSM$.

\begin{boldmath}
\subsection{\label{su5LEHC} $\cl_{EHC}$}
\end{boldmath}
\hspace{-2ex}
With all the spurion embeddings at hand, the four-fermion lagrangian is
\begin{subequations}
\label{LEHC}
\begin{eqnarray}
  \cl_{\rm EHC} &=& \cl_{\rm EHC,1} + \cl_{\rm EHC,2} \ ,
\label{LEHCa}\\
  \cl_{\rm EHC,1} &=&
  \tr\Big(\l_1\bS_L B_R^{S} + \l_2\bS^c_L B_R^{S^c}
  + \l_3\bA_L B_R^{A} + \l_4\bA^c_L B_R^{A^c}
\label{LEHCb}\\
  &&  + (\l_5 \bD_R^1 + \l_6 \bD_R^2) B_L^D
      + \l_7 \bN_R B_L^N +\hc \Big) \ ,
\nonumber\\
  \cl_{\rm EHC,2} &=& \tr \Big(
  (\tl_1\bS_R^1 + \tl_2\bS_R^2) B_L^{S}
  + (\tl_3\bS_R^{1c} + \tl_4\bS_R^{2c}) B_L^{S^c}
\label{LEHCc}\\
  && + \tl_5\bA_R B_L^A + \tl_6\bA^c_R B_L^{A^c}
     + (\tl_7 \bD_L^1 + \tl_8 \bD_L^2) B_R^D +\hc \Big) \ ,
\nonumber
\end{eqnarray}
\end{subequations}
where now the trace is over $SU(5)$ indices.  As usual, the invariance of
$\cl_{\rm EHC}$ under Standard-Model continuous symmetries
follows from the consistency of the spurion embeddings with those symmetries.
Assuming again that all the coupling constants are real,
and using that all the $c$-number spurion matrices
satisfy the algebraic property~(\ref{hX}),
one can verify that $\cl_{\rm EHC}$ is also invariant under $\CPSM$.
As discussed above, our spurion construction ensures that $\cl_{\rm EHC}$
is in fact the most general four-fermion lagrangian
that enjoys these symmetries.
As in Sec.~\ref{su4sp4}, one can then infer that all the low-energy constants
occurring in the effective top Yukawa interactions and in the effective
Higgs potential are real.

\vspace{2ex}

\subsection{\label{su5yeff} Top Yukawa couplings}
As in Sec.~\ref{su4yeff}, the leading effective top Yukawa couplings
are either linear or bilinear in $\S$ and $\S^*$.
For the same reason as before, those interactions
that are linear in $\S$ or $\S^*$
must involve a spurion and an anti-spurion that both come from
$\cl_{\rm EHC,1}$ or both from $\cl_{\rm EHC,2}$.  In the former case
we obtain 10 effective interactions
\begin{eqnarray}
\label{tY1}
  &&\hspace{-1cm}  \F\tr(\bD_R^i \S S^c_L)\ , \quad  \F\tr(\bD_R^i \S A^c_L)\ ,
  \quad  \F^*\tr(\bD_R^{iT} \S^* S_L)\ , \quad \F^*\tr(\bD_R^{iT} \S^* A_L)\ ,
\\
  && \hspace{2cm} \F\bN_R\tr(\S S^c_L)\ ,\quad \F^*\bN_R\tr(\S^*S_L)\ ,
\nonumber
\end{eqnarray}
where $i=1,2$, and in the latter case we obtain 12 more,
\begin{equation}
\label{tY2}
  \F\tr(\bD_L^i\S S_R^{jc})\ , \quad  \F\tr(\bD_L^i \S A^c_R)\ , \quad
  \F^*\tr(\bD_L^{iT} \S^* S_R^j)\ , \quad  \F^*\tr(\bD_L^{iT} \S^* A_R)\ ,
\end{equation}
where $i,j=1,2$.  The extraction of the associated low-energy constants
can be done following the example we have given in Sec.~\ref{su4yeff}.

The effective Yukawa couplings that are bilinear in $\S$ and $\S^*$
may be read off from templates $\ct_{7}$ thru $\ct_{12}$ in Eq.~(\ref{V2}) below,
in the same way that the effective interactions in Eq.~(\ref{ytbil})
are related to templates $\ct_{7}$ through $\ct_{12}$ of Eq.~(\ref{V2su4}).

\subsection{\label{su5LECs} Low-energy constants}
We now move on to the effective potential for the pNGBs,
and begin by listing the templates for $\Veff$.  This time, they are given by
\begin{align}
  \ct_1 &= \F^{1-2q} \tr(\bS\S N) +\hc \ ,
\allowdisplaybreaks
\label{V2}\\
  \ct_2 &= \F^{-1-2q} \tr(\bS^c \S^* N) +\hc \ ,
\allowdisplaybreaks
\nonumber\\
  \ct_3 &= \F^{1 \mp 2q}\tr(\bA\S D^T)+\mbox{h.c.}\ ,
\allowdisplaybreaks
\nonumber\\
  \ct_{4} &= \F^{1 \mp 2q}\tr(\bS\S D^T)+\mbox{h.c.}\ ,
\allowdisplaybreaks
\nonumber\\
  \ct_{5} &= \F^{-1 \mp 2q}\tr(\bA^c\S^* D)+\mbox{h.c.}\ ,
\allowdisplaybreaks
\nonumber\\
  \ct_{6} &= \F^{-1 \mp 2q}\tr(\bS^c\S^* D)+\mbox{h.c.}\ ,
\allowdisplaybreaks
\nonumber\\
  \ct_{7} &= \tr(\bS\S)\tr(S\S^*) \ ,
\allowdisplaybreaks
\nonumber\\
  \ct_{8} &= \tr(S^c\S)\tr(\bS^c\S^*) \ ,
\allowdisplaybreaks
\nonumber\\
  \ct_{9} &= \F^2 \tr(S^c\S)\tr(\bS\S) +\hc \ ,
\allowdisplaybreaks
\nonumber\\
  \ct_{10} &= \F^2 \tr(S^c\S \bS\S) +\hc \ ,
\allowdisplaybreaks
\nonumber\\
  \ct_{11} &= \F^2 \tr(A^c\S \bA\S) +\hc \ ,
\allowdisplaybreaks
\nonumber\\
  \ct_{12} &= \tr(D\S \bD^T\S^*) \ .
\nonumber
\end{align}
The main difference compared to the previous case (Eq.~(\ref{V2su4})),
is that the roles of the $A$ and $S$ \irreps\ have been interchanged,
because $\S$ is now symmetric instead of antisymmetric.

For completeness, we note that one can write down two mass terms
which are invariant under the Standard model symmetries, given by
$B\tr((m_1M_1+m_2M_2)\S+\hc)$, where the mass matrices are
$M_1=diag(1,1,1,1,0)$ and $M_2=diag(0,0,0,0,1)$.  For $m_1=m_2$,
the mass term is invariant under $SO(5)$.  Because of the similarity
between the mass matrices $M_{1,2}$ and the symmetric right-handed
spurions $S_R^{1,2}$, the explicit form of the mass terms bears
resemblance to the effective potential for template $\ct_1$.
We leave the details to the reader.

The derivation of the low-energy constants is very similar to the
previous section, and so we will only give the results.
Also, except for $\ct_{12}$,
we leave it to the reader to work out the explicit expressions
for the hyperbaryon two-point functions, using Table~\ref{tabantihB}.
In all cases it can be verified that $SU(5)$ must break spontaneously
to $SO(5)$ for the relevant two-point function not to vanish.
In some cases, $SU(6)$ must be broken to $SO(6)$ as well.

As in the previous section,
for $\ct_1$ we only need the right-handed low-energy constant,
\begin{equation}
\label{calcSN}
  C_{1R} = -\frac{i}{5} \int d^4x \int\frac{d^4p}{(2\p)^4}
  \frac{p_\m}{p^2}\,e^{-ipx} \svev{\bB^N(0)\g_\m P_L B_{ii}^S(x)}\ .
\end{equation}
For $\ct_2$, $B^S$ gets replaced by $B^{S^c}$.
For $\ct_3$ both chiralities occur in $\Veff$, and
\begin{equation}
\label{LECDA}
  C_{3R,L} = -\frac{i}{5}
  \int d^4x\int \frac{d^4p}{(2\p)^4}\frac{p_\m}{p^2}\,e^{-ipx}
  \svev{\bB^D_{ji}(0)\g_\m P_{L,R} B^A_{ji}(x)}\ .
\end{equation}
Again the low-energy constants for templates $\ct_4$, $\ct_5$ and $\ct_6$
can be similarly obtained.  For $\ct_7$ we find
\begin{equation}
\label{calcSSa}
  C_{7L,R} = -i \int d^4x \int \frac{d^4p}{(2\p)^4} \frac{p_\m}{p^2}\,
  e^{-ipx} \svev{\bB_{k\ell}^S(0) \g_\m P_{R,L} B_{ij}^S(x)}
  \bigg|_{i=j\ne k=\ell} \ .
\end{equation}
The special choice of flavor indices we have made separates out the coefficient
of $\tr(\bcs\S)\tr(\cs\S^*)$, which is what we need for $\Veff$,
from the coefficient of $\tr(\bcs \cs)$,
which is a $\S$-independent effective term
(for the spurion notation we use here, see Sec.~\ref{su4LECs}).
For $\ct_{8}$, we replace $B^S$ by $B^{S^c}$ and $\bB^S$ by $\bB^{S^c}$
in Eq.~(\ref{calcSSa}).  Next,
the low-energy constants for $\ct_{9}$ and $\ct_{10}$ are obtained from the same
hyperbaryon two-point function,
\begin{equation}
\label{calcSSc}
  -i \int d^4x \int \frac{d^4p}{(2\p)^4} \frac{p_\m}{p^2}\,
  e^{-ipx} \svev{\bB_{k\ell}^{S^c}(0) \g_\m P_{R,L} B_{ij}^S(x)} \ ,
\end{equation}
and differ only by the choice of flavor indices needed to project them out.
For $C_{9L,R}$ we set $i=j\ne k=\ell$ in Eq.~(\ref{calcSSc}), whereas for $C_{10L,R}$
we set $j=k\ne\ell=i$.  For $\ct_{11}$ we use the same choice of
flavor indices as for $\ct_{10}$, so that
\begin{equation}
\label{calcAAc}
  C_{11L,R} = -i \int d^4x \int \frac{d^4p}{(2\p)^4} \frac{p_\m}{p^2}\,
  e^{-ipx} \svev{\bB_{k\ell}^{A^c}(0) \g_\m P_{R,L} B_{ij}^A(x)}
  \bigg|_{j=k\ne\ell=i} \ .
\end{equation}

We finally consider $\ct_{12}$, where, just like in Sec.~\ref{su4LECs}, we need
to separate out the low-energy constant of interest from the
$\S$-independent effective term $C'_{12L,R}\tr(\bcd_{L,R}\cd_{L,R})$.
Instead of Eq.~(\ref{LECDD}), in the effective theory we now have
(again omitting the common chirality index)
\begin{equation}
\label{LECDDsu5}
  \frac{\partial}{\partial \cd_a}\frac{\partial}{\partial \bcd_b}
  \frac{\log{Z_{\rm eff}}}{V}
  = -C_{12} \tr(T_b^T T_a) - C'_{12} \tr(T_b T_a)\ .
\end{equation}
In the microscopic theory, the hyperbaryon two-point function
is given by Eqs.~(\ref{DDmic}) through~(\ref{contraction}) as before.
But the symmetry-breaking part of $\vev{\c_i \bc_j}$ is now
proportional to $\d_{ij}$, instead of to $\e_{0,ij}$, as it was in
Sec.~\ref{su4sp4}.  The upshot is that $C_{12L,R}$ can be expressed in terms of
the contraction on the last line of Eq.~(\ref{contraction}) in the same way
as in the previous section.

\begin{boldmath}
\subsection{\label{su5Veff} $\Veff$}
\end{boldmath}
With its nine additional NGBs, the structure of the $SU(5)/SO(5)$ coset
is richer than that of $SU(4)/Sp(4)$, and the calculation of $\Veff$
is more difficult.  We have not been able to obtain $\Veff$ in closed form.
As a first step, we have worked it out to second order in the pNGBs.
The results may be found in App.~\ref{Veff2nd}.

One way to understand the extra complexity of the $SU(5)/SO(5)$ coset
is to consider the invariants of the Standard Model symmetries
$SU(2)_L$ and $T_R^3$ that can be constructed from the pNGB fields.
If, in addition, such an operator (possibly together with its
hermitian conjugate) is invariant also under $\CPSM$,
it can occur as a separate term in the effective potential.
In the case of the $SU(4)/Sp(4)$ coset,
the simplest invariants that can occur in $\Veff$ were the bilinear
$H^\dagger H$ and powers of the inert pNGBs $\h$ and $\z$.  Moreover,
the $SU(4)/Sp(4)$ non-linear field $\S$ can be expressed
as a linear function of the pion field $\P$, with coefficients
that depend on the bilinears
$\h^2$ and $H^\dagger H$ (see Eq.~(\ref{Scsa})).  This has enabled us
to obtain the effective potential in closed form.
By contrast, in the case of the
$SU(5)/SO(5)$ coset we also have a $(3,3)$-plet of $SU(2)_L\times SU(2)_R$
at our disposal.  There are two new invariant bilinears, given by
$\tr(\hF_0^2)$ and $\tr(\hF_+\hF_-)$ in the notation of App.~\ref{su5so5coset}.
At third order there are new invariants that
depend only on the triplet fields: $\tr(\hF_0^3)$ and $\tr(\hF_0\hF_+\hF_-)$,
as well as mixed invariants that depend on both the Higgs
and the triplet fields:
$H^\dagger \hF_0 H$, $H^T \e \hF_- H$ and $H^\dagger \hF_+\e H^*$.

The mixed invariants are particularly important for phenomenology.
This is best illustrated through an example.  We consider
the contribution of $q_L$ to template $\ct_7$, whose third-order term is
(see Sec.~\ref{su4Veff} for the $\svev{\cdot}$ notation)
\begin{eqnarray}
  \svev{\tr(\bS_L\S)\tr(S_L\S^*)}\Big|_{\mbox{3rd order}}
  &=& \frac{32}{f^3}\,( H^T \e \hF_- H + \hc )
\label{SMaL3rd}\\
  &=& \frac{32}{f^3}\,(2 H_0 H_+ \phi_-^0
      -i\sqrt{2} H_+^2 \phi_-^- +i\sqrt{2} H_0^2 \phi_+^-  + \hc ) \ .
\nonumber
\end{eqnarray}
We see that once the Higgs field acquires an expectation value,
$\svev{H_0}=h/\sqrt{2}\ne 0$, this induces a linear potential for
$\Im \phi_+^-$ (see Eq.~(\ref{Vtriplet})).  As a result, the expectation value
$\vev{\Im \phi_+^-}=\varphi/\sqrt{2}$ will necessarily move away
from zero \cite{ferretti16},
while the expectation values of all the remaining components of
the $(3,3)$-plet remain zero at this order.
As explained in the introduction, this is undesirable, because
$\vev{\Im \phi_-^+}$ does not preserve the diagonal subgroup
of $SU(2)_L\times SU(2)_R$ (the custodial symmetry) \cite{GM}.
Therefore, this expectation value will drive
the $\r$-parameter away from unity.

Let us investigate this issue in more detail.  While we have not been able
to obtain the effective potential in closed form for arbitrary values
of the pNGB fields, this can be done when only $h$ and $\varphi$ are turned on.
The results may be found in App.~\ref{Veffphi}.
Examining these results, we see that odd-order terms,
and, in particular, the cubic term $h^2\varphi$, are present in several cases.
These include the contribution of $q_L$
to templates $\ct_{7}$ (Eq.~(\ref{phiSMaL})) and $\ct_{8}$ (Eq.~(\ref{phiSMbL})).
Similar terms are obtained for template $\ct_{12}$, see
Eqs.~(\ref{phiDMvv}),~(\ref{phiDMhh}),~(\ref{phiDMad}) and~(\ref{phiDMda}).

The question arises whether these undesirable contributions can be avoided.
A simple observation is that odd-order terms would be absent if
one could show that the effective potential is invariant under
an ``intrinsic parity'' transformation that takes $\S\to\S^*$ and $\F\to\F^*$,
while leaving the Standard-Model quark fields unchanged.  The obvious reason
is that this transformation flips the sign of all the pNGB fields.\footnote{
  The transformation $\S\to\S^*$ is physically equivalent
  to the transformation $P_\p$ considered in Ref.~\cite{ferretti16},
  because the difference between them is an $SO(5)$ transformation.
}
A case-by-case check, using the explicit forms of the spurions
(and assuming the general form of the pion field, Eq.~(\ref{Pi})),
reveals that the individual contributions to $\Veff$ are each invariant
under the intrinsic parity transformation,
except for the six cases we have listed above,
where the cubic term $h^2\varphi$ is actually present.

Individual odd-order contributions can be avoided by imposing suitable
constraints on the coupling constants of $\cl_{EHC}$.
For example, the contributions of Eqs.~(\ref{phiDMvv}) and~(\ref{phiDMhh})
cancel each other if $\tl_7=\pm\tl_8$ \cite{ferretti16}.
The contributions from Eqs.~(\ref{phiDMad}) and~(\ref{phiDMda}) are absent
if $\l_5$ and/or $\l_6$ vanish.
Similarly, Eq.~(\ref{phiSMaL}) is absent when $\l_1$ vanishes,
and Eq.~(\ref{phiSMbL}) when $\l_2$ does.
Interestingly, for the parametrization~(\ref{Sigphi}) all the odd-order
contributions happen to involve the same function of $h$ and $\varphi$.
In $\Veff$, every term from App.~\ref{Veffphi} comes multiplied
by two coupling constants from $\cl_{EHC}$, and a low-energy constant
(see Eq.~(\ref{tmpltN})).
Therefore, mathematically, the minimal requirement that would
eliminate all the odd-order terms for the parametrization~(\ref{Sigphi})
is a single constraint,
which is bilinear in the coupling constants of $\cl_{EHC}$,
and linear in the low-energy constants.

Physically, the four-fermion couplings and the low-energy constants
have an entirely different origin.  The former arise from integrating out
heavy gauge bosons of the EHC theory, whereas the latter only depend on
correlation functions of the hypercolor theory.  Therefore,
it is unlikely that they will satisfy a constraint of the kind
described above.  Intuitively, what makes more sense is that
the odd-order terms in $\Veff$ might vanish thanks to the vanishing
of sufficiently many four-fermion couplings.
Some new constraint in the EHC theory would have to set the
proper linear combinations of the couplings $\l_i$ and $\tl_i$ equal to zero.
One way this might happen is if the intrinsic parity symmetry discussed above
would arise from some discrete symmetry of the EHC theory.
Unfortunately, we have not been able to identify such a symmetry.
Having said this, it remains a possibility that integrating out
the heavy gauge bosons of the EHC theory would give rise to a small set
of four-fermion couplings, that happens to satisfy the needed constraints
on the couplings of $\cl_{EHC}$, at least when the heavy gauge bosons
exchange is considered at tree level.

\vspace{2ex}

\begin{boldmath}
\section{\label{fix} Revisiting the $SU(4)$ composite-Higgs model}
\end{boldmath}

Another composite Higgs model whose low-energy sector yields
the $SU(5)/SO(5)$ coset was first studied in detail
by Ferretti in Ref.~\cite{ferretti},
and later by us in Ref.~\cite{topsect}.  In this section we revisit
the effective potential induced by the coupling to third generation
quarks in this model.  We begin with a brief summary.
The model is an $SU(4)$ gauge theory.
The matter content includes 5 Majorana fermions $\c_i$
in the 2-index antisymmetric (sextet) \irrep, together with
3 Dirac fermions $\j_a$ in the fundamental \irrep.\footnote{%
  For a lattice study of a closely related $SU(4)$ gauge theory,
  see Ref.~\cite{TACO}.
}
The global symmetry is\footnote{%
  In Ref.~\cite{topsect}, $U(1)_A$ is denoted $U(1)'$.
}
\begin{equation}
\label{flavor}
  G = SU(5)\times SU(3)\times SU(3)'\times U(1)_X\times U(1)_A\ ,
\end{equation}
where $\c_R$ transforms as $\five$ of $SU(5)$, $\j_R$ as $\threebar$
of $SU(3)$, and $\j_L$ as $\threebar$ of $SU(3)'$.  The embedding
of $SU(2)_L\times SU(2)_R\subset SO(5)\subset SU(5)$ is the same as
in Sec.~\ref{su5so5} (see App.~\ref{su5so5coset}), while $SU(3)_c$
is the vector subgroup of $SU(3)\times SU(3)'$.  $U(1)_X$ is the
conserved fermion number of the $\j$'s.  If we take the $U(1)_X$ charge
of $\j$ to be $1/6$,\footnote{%
  This normalization is different by a factor two from that of Ref.~\cite{topsect}.
}
it will coincide with ordinary baryon number.  $U(1)_A$ is the conserved
axial current.  As in Ref.~\cite{topsect}, we take the axial charge
of $\c_R$ to be $-1$.  The axial charges are then
$5/3$ for $\j_R$ and $-5/3$ for $\j_L$.

In Ref.~\cite{topsect} we studied the top-induced effective potential.
Making rather restrictive assumptions, we found that the potential
is quartic in the spurions (equivalently, in the four-fermion couplings),
and we discussed it in some detail.  In this section, as in the rest
of this paper, we will instead make minimal assumptions about the four-fermion
lagrangian.  We begin by reconsidering the dimension-9/2 hyperbaryons
that can serve as top partners, finding two more operators
that can play this role, in addition to the four operators already considered
in Ref.~\cite{topsect}.  The most general four-fermion lagrangian thus
contain six independent couplings.  Using this lagrangian
we find that, in general, an effective potential is induced already
at second order in the four-fermion couplings.
We also reconsider the potential that is induced by the same four-fermion
lagrangian as in Ref.~\cite{topsect}, and find that it contains two additional
terms that we overlooked.  We conclude with a short
discussion of the phenomenological implications of our findings.

\begin{table}[t]
\vspace*{3ex}
\begin{center}
\begin{tabular}{l|l|cccc} \hline
\multicolumn{2}{c|}{} & $SU(5)$ & $SU(3)\times SU(3)'$
  & $SU(3)_c$ & $U(1)_A$
\\ \hline\hline
$B^{(\sfive,\sthree,\sone)}_R$ &    $B_R$ & \five
  & $(\threebar,\one)\times(\threebar,\one)\to(\three,\one)$
  & $\three$ & 7/3 \\
$B^{(\sfivebar,\sthree,\sone)}_L$ & $B_L$ & \fivebar
  & $(\threebar,\one)\times(\threebar,\one)\to(\three,\one)$
  & $\three$ &  13/3 \\
$B^{(\sfive,\sone,\sthree)}_R$ & $B'_R$ & \five
  & $(\one,\threebar)\times(\one,\threebar)\to(\one,\three)$
  & $\three$ & -13/3 \\
$B^{(\sfivebar,\sone,\sthree)}_L$ & $B'_L$ & \fivebar
  & $(\one,\threebar)\times(\one,\threebar)\to(\one,\three)$
  & $\three$ &  -7/3 \\
$B^{(\sfivebar,\sthreebar,\sthreebar)}_R$ & & \fivebar
  & $(\threebar,\one)\times(\one,\threebar)=(\threebar,\threebar)$
  & $\three+\sixbar$ &  1 \\
$B^{(\sfive,\sthreebar,\sthreebar)}_L$ & & \five
  & $(\threebar,\one)\times(\one,\threebar)=(\threebar,\threebar)$
  & $\three+\sixbar$ & -1 \\
\hline\hline
\end{tabular}
\end{center}
\vspace*{-3ex}
\begin{quotation}
\floatcaption{tabHC}{%
Hyperbaryon operators.  The first four lines correspond to Table~1 of
Ref.~\cite{topsect} (omitting the anti-hyperbaryons), and the last
two lines to Eq.~(\ref{Bbc}).  The left column is the name of the hyperbaryon
in the notation used in this section.  When relevant, we give for comparison
the name we used for the same operator in Ref.~\cite{topsect} in the second column.
The remaining columns list the quantum numbers.
The (ordinary) baryon number of all these hyperbaryons is 1/3.
}
\end{quotation}
\vspace*{-4.5ex}
\end{table}

The top-partners we consider are limited to three-fermion operators
of the minimal dimension, 9/2.
They must transform as $\three$ under $SU(3)_c$, and can belong
to $\five$ or $\fivebar$ of $SU(5)$.  Since $SU(3)_c$ is the diagonal
subgroup of $SU(3)\times SU(3)'$, this allows for several possibilities
for the $SU(3)\times SU(3)'$ quantum numbers of the hyperbaryons.
Altogether, we can construct 3 right-handed and 3 left-handed hyperbaryons
that satisfy the requirements.  We list them in Table~\ref{tabHC}.
The first four were already introduced in Ref.~\cite{topsect}.
The last two are given by
\begin{eqnarray}
\label{Bbc}
  B^{(\sfivebar,\sthreebar,\sthreebar)}_{R,bc}
  &=& \e_{ABCD} (\g_\m \c_{L,AB}) (\j^T_{L,Cb} C \g_\m \j_{R,Dc}) \ ,
\\
  B^{(\sfive,\sthreebar,\sthreebar)}_{L,bc}
  &=& \e_{ABCD} (\g_\m \c_{R,AB}) (\j^T_{L,Cb} C \g_\m \j_{R,Dc}) \ ,
\nonumber
\end{eqnarray}
where the subscripts $A,B,\ldots,$ are $SU(4)$-hypercolor indices.
In this section we label the hyperbaryons by a superscript
that specifies the $SU(5)\times SU(3)\times SU(3)'$ quantum numbers.\footnote{%
  We label a hyperbaryon and its anti-hyperbaryon by the same superscript.
}
Under $SU(3)_c$, the operators in Eq.~(\ref{Bbc}) describe a $\three$ and
a $\sixbar$, but only the $\three$ will couple to Standard-Model fields.

The most general four-fermion lagrangian that we can construct
using these hyperbaryons is
\begin{eqnarray}
\cl_{\rm EHC} &=&
    \l_1 \bT^{(\sfive,\sthree,\sone)}_L B^{(\sfive,\sthree,\sone)}_R
  + \l_2 \bT^{(\sfivebar,\sthree,\sone)}_R B^{(\sfivebar,\sthree,\sone)}_L
  + \l_3 \bT^{(\sfive,\sone,\sthree)}_L B^{(\sfive,\sone,\sthree)}_R
\label{lagSU5all}\\
  && + \l_4 \bT^{(\sfivebar,\sone,\sthree)}_R B^{(\sfivebar,\sone,\sthree)}_L
  + \l_5 \bT^{(\sfivebar,\sthreebar,\sthreebar)}_L B^{(\sfivebar,\sthreebar,\sthreebar)}_R
  + \l_6 \bT^{(\sfive,\sthreebar,\sthreebar)}_R B^{(\sfive,\sthreebar,\sthreebar)}_L
  + \hc \ .
\nonumber
\end{eqnarray}
The embeddings of the Standard Model fields into the spurions are
$T^{(\sfive,\sthree,\sone)}_{L,a} = T^{(\sfive,\sone,\sthree)}_{L,a}=T_{L,a}$,
$T^{(\sfivebar,\sthree,\sone)}_{R,a} = T^{(\sfivebar,\sone,\sthree)}_{R,a} = T_{R,a}$,
$T^{(\sfivebar,\sthreebar,\sthreebar)}_{L,bc} = \e_{abc}T_{L,a}$ and
$T^{(\sfive,\sthreebar,\sthreebar)}_{R,bc} = \e_{abc}T_{R,a}$.  Here
\begin{eqnarray}
  T_{L,a}(x) &=& t_{L,a}(x)\hatt_L + b_{L,a}(x)\hatb_L \ ,
\label{SMspurL}\\
  T_{R,a}(x) &=& t_{R,a}(x)\hatt_R \ ,
\label{SMspurR}
\end{eqnarray}
where the constant 5-vectors are
\begin{equation}
  \hatt_L = \frac{1}{\sqrt{2}}\left(\begin{array}{c}
    0 \\ 0 \\ i \\ -1 \\0
  \end{array}\right)\ ,\qquad
  \hatb_L = \frac{1}{\sqrt{2}}\left(\begin{array}{c}
    i \\ 1 \\ 0 \\ 0 \\0
  \end{array}\right)\ ,\qquad\qquad
  \hatt_R = \left(\begin{array}{c}
    0 \\ 0 \\ 0 \\ 0 \\ 1
  \end{array}\right)\ .
\label{tspur}
\end{equation}

In Ref.~\cite{topsect}, the effective potential was $O(\l^4)$, \ie,
it was quartic in the coupling constants of $\cl_{\rm EHC}$.
Correspondingly, the low-energy constant discussed in Ref.~\cite{topsect}
was determined in terms of a hyperbaryon 4-point function.
The additional terms proportional to $\l_5$ and $\l_6$
present in Eq.~(\ref{lagSU5all}) allow for the
generation of an effective potential already at $O(\l^2)$,
with low-energy constants that depend on hyperbaryon two-point functions.
The $O(\l^2)$ potential is given by
\begin{eqnarray}
\label{Vall}
  \Vefftop &=& \half\l_1\l_5 C_L\, \e_{abc} \O^*_{cd} \,\F^{4/3}\,
    v^{(\sfive,\sthree,\sone)\dagger}_{L,a} \S v^{(\sfivebar,\sthreebar,\sthreebar)}_{L,bd}
\\
  && + \half\l_3\l_5 C'_L\, \e_{abc} \O_{dc} \,\F^{-16/3}\,
    v^{(\sfive,\sone,\sthree)\dagger}_{L,a} \S v^{(\sfivebar,\sthreebar,\sthreebar)}_{L,db}
\nonumber\\
  && + \half\l_2\l_6 C_R\, \e_{abc} \O^*_{cd} \,\F^{16/3}\,
    v^{(\sfivebar,\sthree,\sone)\dagger}_{R,a} \S^* v^{(\sfive,\sthreebar,\sthreebar)}_{R,bd}
\nonumber\\
  && + \half\l_4\l_6 C'_R\, \e_{abc} \O_{dc} \,\F^{-4/3}\,
    v^{(\sfivebar,\sone,\sthree)\dagger}_{R,a} \S^* v^{(\sfive,\sthreebar,\sthreebar)}_{R,db}
    + \hc \ .
\nonumber
\end{eqnarray}
As in the previous sections, the global spurions (the $v$'s)
result from integrating over the quark fields $q_L$ and $t_R$.
The dependence of each spurion field on the relevant global spurions
is similar to Eqs.~(\ref{embedL}) and~(\ref{embedR}).  It follows that
in the right-handed case we simply need to substitute $\hatt_R$ for $v_R$.
For the left-handed case, we have to sum over $\hatt_L$ and $\hatb_L$,
paying attention to the possible presence of the $SU(3)_c$
invariant tensor $\e_{abc}$ in the embedding of the Standard-Model
fields into the spurions.

Since in this section we keep track of the $SU(3)\times SU(3)'$ symmetry,
we show in Eq.~(\ref{Vall}) the dependence of the potential on $\O$, the
nonlinear field for $SU(3)\times SU(3)'\to SU(3)_c$ symmetry breaking.
$\O$ transforms as $\O\to g\O g'^\dagger$, with $g\in SU(3)$ and $g'\in SU(3)'$,
\ie, it belongs to $(\one,\three,\threebar)$.
As for the dependence on the $U(1)_A$ nonlinear field $\F$, its power
in each term is given by the axial charge of the hyperbaryon
two-point function occurring in the calculation of the low-energy constant.
(The actual calculation of the low-energy constants
is similar to Ref.~\cite{topsect}, and is left for the reader.)

In order to proceed, we will for simplicity set $\O_{ab}=\d_{ab}$.
This means that, as in the previous sections, we do not calculate
the effective potential for the colored pNGBs.  The result is
\begin{eqnarray}
\label{Vtrall}
  \Vefftop &=&
  \l_5 \left(\l_1 C_L \F^{4/3} + \l_3 C'_L \F^{-16/3} \right) \tr(\S P_1)
\\
  && + \l_6 \left(\l_2 C_R \F^{16/3} + \l_4 C'_R \F^{-4/3} \right) \tr(\S^* P_2)
  + \hc \ .
\nonumber
\end{eqnarray}
Here we introduced the orthogonal projectors
\begin{subequations}
\label{proj}
\begin{eqnarray}
  P_1 &=& \sum_{v_L=\hatt_L,\hatb_L} v_L \times v_L^\dagger \ ,
\label{proj1}\\
  P_2 &=& v_R \times v_R^\dagger \ ,
\label{proj2}\\
  P_3 &=& \sum_{v_L=\hatt_L,\hatb_L} v_L^* \times v_L^T \ ,
\label{proj3}
\end{eqnarray}
\end{subequations}
whose sum $P_1+P_2+P_3$ is equal to the $5\times 5$ identity matrix.

As in the previous section, we were unable to work out the dependence
of $\Vefftop$ on all the pNGBs in closed form.  But, as before, we can obtain
the potential in some special cases.  First, expanding the potential
to second order in all the pNGBs gives
\begin{eqnarray}
\label{Vvvc2nd}
  \Vefftop &=& -(\l_1\l_5 C_L + \l_3\l_5 C'_L)
  \frac{(2/5)\h^2 + 4H^\dagger H + 4\F_0^2 + 8\F_+\F_-}{f^2}
\hspace{5ex}
\\
  && -(\l_2\l_6 C_R + \l_4\l_6 C'_R)
  \frac{(16/5)\h^2 + 8H^\dagger H}{f^2}
\nonumber\\
  && + \left( -\l_1\l_5 C_L + 4\l_3\l_5 C'_L
       - 8\l_2\l_6 C_R + 2\l_4\l_6 C'_R \right)\frac{16\h\z}{3\sqrt{5}f}
\nonumber\\
  && - \left( 2\l_1\l_5 C_L + 32\l_3\l_5 C'_L
       + 16\l_2\l_6 C_R + \l_4\l_6 C'_R \right) \frac{16\z^2}{9} + \cdots \ .
\nonumber
\end{eqnarray}
If we use the parametrization~(\ref{Sigphi}), \ie, we retain only
the $h$ and $\varphi$ fields of Eq.~(\ref{Vdef}), the potential is given by
\begin{eqnarray}
\label{Vvvchphi}
  \Vefftop &=&
  4\Big( \l_1\l_5 C_L \cos(4\z/3) +  \l_3\l_5 C'_L \cos(16\z/3) \Big)
  \left( 1 - \hatc\,\frac{2\varphi^2+h^2}{f^2} \right)
  \hspace{5ex}
\\
  && + 2 \Big( \l_2\l_6 C_R \cos(16\z/3) + \l_4\l_6 C'_R \cos(4\z/3) \Big)
  \left( 1 - 4\hatc\,\frac{h^2}{f^2} \right) \ .
\nonumber
\end{eqnarray}
We observe that there are no odd-order terms.
Indeed, it is easy to check that the potential~(\ref{Vtrall}) is invariant
under the intrinsic parity transformation of Sec.~\ref{su5Veff}.
The gauge bosons contribution for this parametrization is the same as
in Sec.~\ref{su5so5}, see Eq.~(\ref{Vggphi}).

In this section we have allowed for spurions with all possible
$SU(3)\times SU(3)'$ quantum numbers, resulting in the four-fermion
lagrangian~(\ref{lagSU5all}).  By contrast, in Ref.~\cite{topsect}
we only considered top spurions with
particular $SU(3)\times SU(3)'$ quantum numbers.  This corresponds
to retaining only the $\l_1$ and $\l_2$ terms in Eq.~(\ref{lagSU5all}),
while setting $\l_3=\l_4=\l_5=\l_6=0$.
In this case the $O(\l^2)$ potential vanishes, and the leading potential
is $O(\l^4)$.  Explicitly,
\begin{eqnarray}
  \Vefftop &=&
  \l_1^2\l_2^2\,\Ctop_{LR}
  \sum_{v_L=\hatt_L,\hatb_L} \left| v_L^\dagger \S \hatt_R \right|^2 +
  \l_2^4\,\Ctop_{RR} \left| \hatt_R^T \S \hatt_R \right|^2
\label{V4}\\
  && + \l_1^4\,\Ctop_{LL}
  \sum_{v_L,u_L=\hatt_L,\hatb_L} \left| u_L^\dagger \S v_L^* \right|^2
\nonumber\\
  &=& \l_1^2\l_2^2\,\Ctop_{LR} \tr(P_1 \S P_2 \S^*)
  + \l_2^4\,\Ctop_{RR} \tr(P_2 \S P_2 \S^*)
\nonumber\\
  && + \l_1^4\,\Ctop_{LL} \tr(P_1 \S P_3 \S^*) \ , \rule{0ex}{3ex}
\nonumber
\end{eqnarray}
where we have used Eq.~(\ref{proj}).
The $\Ctop_{LR}$ term was discussed in Ref.~\cite{topsect},
whereas the other two terms were overlooked.\footnote{%
  In the conventions of Ref.~\cite{topsect},
  $\l_1^2\l_2^2\,\Ctop_{LR}$ corresponds to $y^2\,C_{\rm top}$.
  In Ref.~\cite{topsect} we argued that $\Ctop_{LR}$
  dominates over the gauge bosons contribution in a certain
  large-$N$ framework.  Unfortunately, it is not possible
  to incorporate $\Ctop_{RR}$ and $\Ctop_{LL}$ into the same
  large-$N$ framework in a meaningful way.
}
As in the rest of this paper, the low-energy constants introduced
in this section are always determined by the stand-alone hypercolor theory.
Expanding this potential to second order in the pNGB fields gives
\begin{equation}
\label{Vefftop2nd}
  \Vefftop = \left( (4\l_1^2\l_2^2\,\Ctop_{LR} -8\l_2^4\,\Ctop_{RR})H^\dagger H
  +8\l_1^4\,\Ctop_{LL} \F_+\F_- \right) + \cdots \ ,
\end{equation}
while for the parametrization~(\ref{Sigphi}) we obtain
\begin{eqnarray}
  \Vefftop &=&
  \l_1^2\l_2^2 \Ctop_{LR}\,
  \frac{2h^2}{f^2}\left( \hs - \frac{2\hatc \varphi}{f} \right)^2
  + \l_2^4 \Ctop_{RR} \left( 1 - \frac{4\hatc h^2}{f^2} \right)^2
\label{Vtophphi}\\
  && + 4\l_1^4 \Ctop_{LL}
  \left(\frac{\hatc h^2}{f^2}+\frac{\hs\varphi}{f}\right)^2 \ .
\nonumber
\end{eqnarray}
This result shows that there are no odd-order terms
associated with $\Ctop_{RR}$, consistent with the invariance
of the corresponding term in Eq.~(\ref{V4})
under the intrinsic parity transformation of Sec.~\ref{su5Veff}.
Cubic terms arise from the contributions associated with $\Ctop_{LR}$
and $\Ctop_{LL}$.  These contributions will be absent if $\l_1=0$.\footnote{%
  Notice, however, that in order to generate a mass for the top quark,
  at least two four-fermion couplings must be non-zero,
  \eg, $\l_1$ and $\l_2$ \cite{topsect}.
}
If both $\l_1$ and $\l_2$ are non-zero, then the cubic terms will be present
except in the (unlikely, because arbitrarily fine-tuned) case that
$\l_2^2\Ctop_{LR}=\l_1^2\Ctop_{LL}$.  In this case the sum of the two terms
is proportional to
\begin{equation}
\label{P2plus3}
  \tr(P_1 \S P_2 \S^*) + \tr(P_1 \S P_3 \S^*)
  = \tr(P_1 \S (1-P_1) \S^*) \ ,
\end{equation}
which is again invariant under the intrinsic parity transformation.

The main phenomenological implications of the results of this section
are discussed in the concluding section.

\vspace{2ex}

\section{\label{discconc} Discussion and conclusions}

The composite Higgs approach is often discussed taking the
low-energy, non-linear sigma model as a starting point.
In this paper we studied in detail several concrete realizations
(ultraviolet completions)
of this approach as an asymptotically free gauge theory with fermionic matter.
In this concluding section, we discuss the lessons
that can be drawn from our findings.

We begin with a simple technical observation about the Higgs potential.
It is a generic feature of composite Higgs models that,
if we turn off all the pNGBs except for $h=\sqrt{2}\Re H_0$,
then the coset field $\S$ describes a rotation matrix by an angle $\a\propto h$
in some generalized space.
In other words, the non-zero entries of $\S$
depend linearly on $\cos(\a)$ or $\sin(\a)$.
This is true in particular for the two cosets discussed in this paper.\footnote{%
  For the $SU(4)/Sp(4)$ coset $\a=h/(\sqrt{2}f)$ by Eq.~(\ref{alpha}).
  For the $SU(5)/SO(5)$ coset $\a=2h/f$, see App.~\ref{Veffphi}.
}
For an effective potential that is at most quadratic in $\S$ and/or $\S^*$,
it follows that the effective potential is then a second-order polynomial in
$\cos(\a)$ and $\sin(\a)$.  Furthermore, $SU(2)_L$ invariance requires
that, when all triplet fields are turned off,
the potential must be an even function of the Higgs field $H$, and
this remains true when we retain $h=\sqrt{2}\Re H_0$ only.
The form of the resulting effective potential is very restricted.
It depends on just two trigonometric functions of $\a$,
and we may take it to be \cite{PW,topsect,DDEZ}
\begin{equation}
\label{Vh}
  \Veff = const. - A \cos(\a) + B \cos^2(\a) \ .
\end{equation}
The solutions of the saddle-point equation are $\sin(\a)=0$ or
\begin{equation}
\label{Vsaddle}
  \frac{A}{2B} = \cos(\a) \ ,
\end{equation}
which is the symmetry-breaking solution of interest.\footnote{%
  For Eq.~(\ref{Vsaddle}) to describe a global minimum at small $\a$
  we must have $0<A<2B$.
}
We may rewrite this solution as
\begin{equation}
\label{Vapprox}
  1 -\frac{A}{2B} = \frac{\a^2}{2} + O(\a^4) \ .
\end{equation}
Current experimental constraints suggest $h^2/f^2\leqx 0.1$
as a figure of merit \cite{BCS,PW,ATLAS}.\footnote{%
  We may expect this bound to become tighter in the future if
  no new particles are found at the LHC.
}
Thus, for the right-hand side of Eq.~(\ref{Vapprox}) to be small,
an ``irreducible fine-tuning'' at a similar level
of the coefficients $A$ and $B$ is needed.

The effective potential receives contributions from two different sources.
First, there are $O(g^2,g'^2)$ terms, arising from the interaction between
the electro-weak gauge bosons and the pNGBs.  The form of these terms
is constrained by gauge invariance, and they depend on
a single low-energy constant $C_{LR}$.
The other source of an effective potential arises from integrating out
the third generation quark fields.  This is the prime focus of this paper.
In order to explain
the four-fermion lagrangian that couples the quark fields to three-fermion
states of the hypercolor theory, we have to postulate the existence of an
``extended hypercolor'' theory.  This new dynamics is operative at a yet higher
energy scale, $\L_{EHC}$, and requires the existence of new heavy gauge bosons
that can transform an ordinary quark into one of the fermion species
of the hypercolor theory.\footnote{
  Here we assume that also the EHC theory, which otherwise remains unspecified,
  is a renormalizable gauge theory.
}
The leading contributions to the effective potential
from this sector are $O(\l^2)$, where we use $\l$ as a generic name
for a four-fermion coupling.  In the case of the model of Sec.~\ref{fix},
for reasons that we explain below,
we are also interested in $O(\l^4)$ contributions.

Having a minimum of the effective potential with $h^2/f^2 \leqx 0.1$
thus requires balancing between $O(g^2,g'^2)$ effects,
which depend on the gauge couplings of the Standard Model,
and $O(\l^2)$ effects (or, in special circumstances, $O(\l^4)$ effects),
which depend on the dynamics of the EHC theory,
and can generically arise from several distinct four-fermion couplings.
If the effects of the third-generation quarks dominates over the
gauge bosons, then the balancing has to happen between
the contributions coming from different four-fermion couplings.
We have studied an example potential in Sec.~\ref{su4disc}.
However, it remains an open question how the
four-fermion couplings originating from the EHC theory
can be arranged to give the desired result.
We note that we did not make any {\it ad hoc} assumptions about
the EHC sector.  It turns out that, in all cases considered here, the
most general form that the four-fermion lagrangian may take is quite
complicated, leading to many possibilities for the low-energy effective
theory (both the induced Higgs potential and the Yukawa couplings).
New ideas will be needed to simplify the situation, but those would
necessarily address the specific form of the EHC sector, and are
beyond the scope of this paper.

Let us briefly touch on another basic difficulty, which is the inherent tension
between fermion masses and flavor constraints.
Traditionally, fermion masses are generated in technicolor models
via four-fermion couplings that are induced by an extended technicolor
(ETC) dynamics, of which our extended hypercolor (EHC) dynamics is
a close cousin.  The main difference is the following.  If we generically use
$\j$ to denote a Standard Model fermion field, and $\J$ for
a fermion of the new strong dynamics (be it technicolor or hypercolor),
then ETC requires four-fermion interactions of the generic form $\j\j\J\J$,
whereas the EHC interactions are assumed to have the form $\j\J\J\J$.
The ETC four-fermion interactions induce a fermion mass term,
$\j\j\svev{\J\J}$, once the operator $\J\J$ acquires an expectation value.
By contrast, the EHC four-fermion interaction $\j\J\J\J$ allows for
a linear coupling of a Standard Model fermion to a hyperbaryon,
thereby giving rise to a partially composite state.\footnote{%
  In principle, a given EHC theory may induce both
  $\j\j\J\J$ and $\j\J\J\J$ type four-fermion interactions,
  in which case both mechanisms for fermion mass generation
  will be operative (see, \eg, Ref.~\cite{ferretti}).
}

The basic problem is that the same ETC or EHC dynamics
that gives rise to the desired four-fermion interactions can,
generically, also give rise to four-fermion interactions $\sim \j\j\j\j$,
namely interactions that involve four Standard Model fermions.
These interactions will trigger flavor-changing processes that,
if too strong, will be in conflict with experiment.  According to naive
power counting, fermion masses in ETC are suppressed relative
to the technicolor scale $\L_{TC}$ by $z_{TC}^2$, where $z_{TC}=\L_{TC}/\L_{ETC}$,
with $\L_{ETC}$ being the ETC scale.  Because of the flavor constraints
$\L_{ETC}$ must be quite large, making the ratio $z_{TC}$ small.
The resulting fermion masses, of order $\L_{TC}\,z_{TC}^2$, are then too small
in many cases.

A partial solution may be provided by walking technicolor,
where the technicolor dynamics is assumed to be nearly conformal.
Taking quantum effects into account, the induced fermion mass
in walking technicolor is $\sim\L_{TC}\,z_{TC}^{2-\g_m}$,
where $\g_m$ is the (approximately constant)
mass anomalous dimension of the technifermion $\J$.
Ideally, a very large anomalous dimension $\g_m \leqx 2$
would wipe out entirely the suppression factor $z_{TC}^{2-\g_m}$.
But various theoretical considerations suggest that such large values
of $\g_m$ are unlikely \cite{RC,BCS,PW}.
Lattice calculations in various models find that $\g_m$ does not exceed 1
(see the review articles \cite{TDreview,Pica}).  If indeed $\g_m\leqx 1$
then the induced fermion mass can only be as large as $\L_{TC}\,z_{TC}$,
\ie, still suppressed by one power of $z_{TC}$.
Thus, while near-conformality together a large $\g_m$ help in generating
larger fermion masses, it remains very difficult to generate a mass
as large as that of the top quark.
As an illustration, according to Ref.~\cite{PW}, $\L_{ETC}$ cannot be smaller
than about $10^5$~TeV,\footnote{%
  In the notation of Ref.~\cite{PW}, $\L_{ETC}$ is $\L_{UV}$.
}
so that $z_{TC}$ cannot be larger than $\sim 10^{-4}$.
With $\g_m \sim 1$ this might have allowed for generating the $\sim 1$~GeV mass
of the charm quark, but certainly not the top-quark mass.

If, instead, the top quark receives its mass via the partial compositeness
mechanism, this mass will be naively of order $\L_{HC}\,z_{HC}^4$,
where $z_{HC}=\L_{HC}/\L_{EHC}$, because, when measured in units
of the hypercolor theory, each four-fermion coupling
is naively of order $z_{HC}^2$, and two four-fermion couplings are needed
to generate a mass for the top: the top must transform into a hyperbaryon,
and then back into a top.  At tree level, the case for partial compositeness
is thus worse than traditional ETC.
Of course, one has to take into account quantum effects.
If again the theory is nearly conformal, the induced top mass is of order
$\L_{HC}\,z_{HC}^{4-2\g'}$, where $\g'$ is the (again, approximately constant)
anomalous dimension of the relevant four-fermion operators.
Once again, the suppression factor $z_{HC}^{4-2\g'}$ would be wiped out
when $\g'\leqx 2$.  The popularity of partial compositeness stems
from the fact that there are no theoretical considerations against such
large values of $\g'$. Thus, at least in principle, one could end up
with a suppression by a very small power of $z_{HC}$
\cite{RC,BCS,PW}.\footnote{%
  For a calculation of $\g_m$ and $\g'$ in a gauged Nambu--Jona-Lasinio model,
  see Ref.~\cite{BGR}.
}

We stress that in order to achieve a large enhancement,
be it in the context of extended technicolor
or in the context of a partially composite top,
the anomalous dimension must be approximately constant, and large,
over many energy decades.  This requires the dynamics to be nearly conformal.
In contrast, if the gauge dynamics is QCD-like, then this mechanism
is unlikely to be effective.  The reason is that
as we increase the energy scale, the gauge coupling quickly
becomes perturbative.  Existing perturbative calculations of the
anomalous dimension of various four-fermion operators always find
small values \cite{anomdim,PS}.
It remains an open question whether a realistic top-quark mass
can be achieved by invoking a strong near-conformal dynamics.
Lattice calculations of $\g'$ in candidate hypercolor theories
could help shed light on this important issue.

An alternative approach would be to assume that, while the top quark
receives its mass through partial compositeness from an
extended hypercolor dynamics, yet some other dynamics
(or, more generally, some additional high scales), are involved
in mass generation of all other Standard Model fermions.\footnote{%
  See, for example, Refs.~\cite{OM,Anarchic,PP}.
}
This approach is, obviously, less economic, but eventually it might
be forced upon us by the tension between flavor-changing processes
and quark masses.
In a way, in this paper we are following this approach,
because we study the interaction between the third-generation quarks
and the hypercolor theory, while disregarding the rest of
the fermions of the Standard Model.  In particular, we are in effect
allowing for the extended hypercolor scale $\L_{EHC}$ to be close enough
to the hypercolor scale $\L_{HC}$, so that the four-fermion couplings
will be large enough to generate phenomenologically viable mass
for the top quark and effective potential for the pNGBs.

In this paper we studied two $SO(d)$ gauge theories with $d=5,\ 11,$ where
chiral symmetry breaking gives rise to pNGBs in the $SU(4)/Sp(4)$ coset
(Sec.~\ref{su4sp4}); and three models where the coset is $SU(5)/SO(5)$,
two are again based on an $SO(d)$ gauge theory with $d=7,\ 9,$
and have a similar
set of top partners (Sec.~\ref{su5so5}), while the third is an $SU(4)$
gauge theory with a rather different set of top partners (Sec.~\ref{fix}).
Each model contains fermions in two different \irreps, leading
to a non-anomalous abelian axial symmetry, $U(1)_A$,
with an associated pNGB, $\z$, which is inert under
all the Standard-Model gauge interactions.
For each theory we first listed all the dimension-9/2 hyperbaryons
that can serve as top partners,
and wrote down the most general four-fermion lagrangian that couples
them to $t_L$, $b_L$ and $t_R$.  We then worked out the resulting
effective potential for the multiplet of pNGBs containing the Higgs
field together with the $U(1)_A$ pNGB.

We started with the $SU(4)/Sp(4)$ coset.  Its structure is simpler in that,
besides Higgs doublet $H$, this coset contains only one additional pNGB, $\h$,
which is inert under the Standard Model gauge interactions,
like the $U(1)_A$ pNGB $\z$.
We worked out the $O(\l^2)$ potential in closed form.
We found that it consists of a linear superposition of nine functions of
the variables $H^\dagger H$, $\h$ and $\z$ (\seef Sec.~\ref{su4disc}).
Thus, in general, a potential is generated for all the pNGBs,
including the $U(1)_A$ pNGB.
Each coefficient $c_i$ consists of a sum of terms, where each term is
the product of a low-energy constant and two four-fermion couplings.
The effective potential generated by the electro-weak gauge bosons
also depends on one of these functions, $f_6$, and so it contributes
only to its coefficient $c_6$.
Finally, there is an additional contribution to the effective potential
if a mass term for the $\c$ fermions is turned on in the hypercolor theory.
By itself, experimental constraints on the effective potential can be studied
directly in terms of the $c_i$'s.  But, if one wants
to incorporate also the top Yukawa coupling
into this analysis, then it has to be done in terms of the
four-fermion couplings, and requires knowledge of the low-energy constants.
The latter can, in principle, be calculated on the lattice.

Studying the minima of the full effective potential
as a function of all the relevant parameters is challenging.
Generically, minimizing the potential might give rise
to the condensation of not just the Higgs field,
but also the ``inert'' fields $\zeta$ and $\eta$.
Since these fields are pseudoscalars, the expectation values
$\svev{\h}$ or $\svev{\z}$ break $CP$ spontaneously,
and will thus be constrained by experiment.
We have discussed the conditions that these expectation values
are physical, and cannot be rotated away (Sec.~\ref{SCPB}).
Our discussion of the effective potential was limited
to a simple example, in which most of the
four-fermion couplings are turned off by hand (Sec.~\ref{su4disc}).
For this potential $\svev{\h}$ and $\svev{\z}$ are both physical.
We wrote down the conditions needed to have
$\svev{\h}=\svev{\z}=0$, at which point the potential reduces to
the familiar form of Eq.~(\ref{Vh}).\footnote{%
  For further discussion of the role of $\h$, see, for example,
  Refs.~\cite{diboson,JS,BBR}.
}

The low-energy constants of the $SO(d)$ models depend on
two-point functions of the hyperbaryons.  While,
as we have explained above, one can sometimes by-pass the calculation
of the low-energy constants by studying directly the $c_i$ coefficients
in Eq.~(\ref{Veff9}), the correct form of the effective potential cannot
be determined without the knowledge of the dimension-9/2 hyperbaryons.
In other words, if one starts directly from the non-linear sigma model
it is just not possible to determine the correct effective potential.
One can, of course, determine the structure of the effective potential
for a given set of spurion fields.  But the spurions must
match the top-partner hyperbaryons.
An {\em ad-hoc} list of spurions could amount to
arbitrarily setting some of the four-fermion couplings to zero,
in a manner that cannot be reproduced by any extended hypercolor theory.

Next let us discuss the models that yield the $SU(5)/SO(5)$ coset.
In addition to the pNGBs that are present in the $SU(4)/Sp(4)$ case,
there are nine additional pNGBs that fill a (3,3)-plet
of $SU(2)_L\times SU(2)_R$.
Because of this more complicated structure we were not able to obtain
the full potential in closed form.  Instead, we studied the potential
in various simplified cases.  First, we obtained the potential
to second order in all the pNGBs.  Some useful constraints can already
be obtained from this result because, ideally, we would like the curvature
at the origin to be negative in the direction of the Higgs field,
and to be positive in the direction of the triplet fields, to prevent any
triplet from condensing.\footnote{%
  In order to obtain the complete second-order potential one should add
  the contribution of the electro-weak gauge bosons, calculated
  in Ref.~\cite{topsect}.
}

We also considered third order terms.  These terms arise because
one can construct invariants of both $SU(2)_L$ and the $U(1)$
generated by $T_R^3$ from
a pair of Higgs fields and one triplet field.  For a concrete example,
see Eq.~(\ref{SMaL3rd}).  As we explained in the introduction,
and in more detail in Sec.~\ref{su5Veff}, these terms
are especially dangerous for phenomenology.  If the potential contains
cubic terms, then, once the Higgs field acquires an expectation value,
this induces a term linear in the triplet field.  This, in turn,
will necessarily drive the expectation value of the triplet field
away from zero.
The resulting triplet expectation value is different
from the one that preserves the custodial symmetry \cite{GM},
and so it will drive the $\r$-parameter
away from unity.  The magnitude of this triplet expectation value
is thus tightly constrained by experiment.

Studying this issue further, we have worked out
the full effective potential in the case that only
$h=\sqrt{2}\Re H_0$ and $\varphi=\sqrt{2}\, \Im \f_+^-$ are turned on.
We checked which ``templates'' for the effective potential
can give rise to odd-order terms, and, in particular, to the cubic term
$h^2\varphi$, finding that such contributions are possible
in all the $SU(5)/SO(5)$ models.\footnote{%
  No odd-order terms arise from the gauge bosons contribution
  in this case, see Eq.~(\ref{Vggphi}).
}
We then raised the question how likely it is that all cubic terms
(or, more generally, all odd-order terms) will be absent
from the effective potential thanks to cancellations.

As we explained in Sec.~\ref{su5Veff},
if all the four-fermion couplings are non-zero,
the vanishing of the coefficient of a particular (cubic) term in
the effective potential requires a ``conspiracy'' between the
four-fermion couplings and the low-energy constants.
What might be more natural is that the cubic terms will vanish thanks to
the vanishing of suitable (linear combinations of) four-fermion couplings.
The intrinsic parity transformation introduced in Sec.~\ref{su5Veff}
is a convenient device to determine which linear combinations of
the four-fermion couplings should vanish.  Unfortunately, we were unable
to conceive of any obvious symmetry at the level of the EHC theory
that would induce the intrinsic parity symmetry at the level of the
low-energy effective theory.  Still, one should remember that the
four-fermion couplings must be induced by integrating out
the heavy degrees of freedom of an EHC theory, and
a good candidate EHC theory will
conceivably induce only a small number of four-fermion couplings.

The $SU(4)$ model of Sec.~\ref{fix} was already studied in detail previously
\cite{ferretti,topsect}.  We found that if we allow for the most general
four-fermion lagrangian, an effective potential is induced
already at $O(\l^2)$.  While this potential contains no cubic terms,
it does have another serious phenomenological drawback.  If we set to zero
all the pNGB fields except for $h$, then the contribution from the $O(\l^2)$
potential is proportional to
$\cos(\a)$. Because the gauge bosons also contribute
to the same term, we would end up with the situation that $A\ne 0$ but $B=0$
in Eq.~(\ref{Vh}).  This appears to be incompatible with the requirement
of having small $h/f$.  A possible way out that we have discussed above
is that, when turning on also the inert pNGBs $\h$ and $\z$, this would
reveal new minima of the potential.

An alternative is that only a smaller subset of the four-fermion couplings
is actually induced by the EHC, and, as a result, the $O(\l^2)$ potential
vanishes.  We rederived the potential in the case that only
the two four-fermion couplings we considered in Ref.~\cite{topsect}
are non-zero, finding two more terms that we overlooked in Ref.~\cite{topsect}.
Like the other $SU(5)/SO(5)$ models, this $O(\l^4)$ potential
will generically have the undesired cubic terms $\propto h^2\varphi$,
so that, as explained above, further constraints must be satisfied
in order to achieve a phenomenologically viable minimum.

In this paper we discussed the non-linear field $\S$ associated with
the $SU(4)/Sp(4)$ or $SU(5)/SO(5)$ coset, and the field $\F$
that describes the pNGB of the non-anomalous $U(1)_A$ symmetry.
We did not discuss the other non-linear field containing the colored pNGBs,
which is associated with the $SU(6)/SO(6)$ coset in the case of
the $SO(d)$ theories of Sec.~\ref{su4sp4} and Sec.~\ref{su5so5},
or with $SU(3)\times SU(3)'/SU(3)_c$ in the case of the $SU(4)$ model
of Sec.~\ref{fix}.  While our results and conclusions are valid by themselves,
a more complete analysis that includes the potential for the remaining
non-linear effective field would allow for a more detailed study of the
phenomenological consequences.  The obvious additional constraint on
the complete potential is that the colored pNGBs are not allowed to condense.

\vspace{3ex}
\noindent {\bf Acknowledgments}
\vspace{2ex}

\noindent
We thank Gabriele Ferretti and Ben Svetitsky for many useful discussions.
We also thank one of the referees for comments on the first version
that led us to clarify some points and remedy some omissions.
This material is based upon work supported by the U.S. Department of
Energy, Office of Science, Office of High Energy Physics, under Award
Number DE-FG03-92ER40711 (MG).
YS is supported by the Israel Science Foundation
under grant no.~449/13.

\begin{table}[t]
\vspace*{2ex}
\begin{center}
\begin{tabular}{c|c|c|l|c} \hline\hline
2 & $C_d^T=-C_d$ & $C_d\G_{d+1}=-\G_{d+1}^T C_d$ & $C_{d+1}=C_d$ &
  $(C_{d+1}\G_I)^T=+C_{d+1}\G_I$ \\
4 & $C_d^T=-C_d$ & $C_d\G_{d+1}=+\G_{d+1}^T C_d$ & $C_{d+1}=C_d\G_{d+1}$ &
  $(C_{d+1}\G_I)^T=-C_{d+1}\G_I$ \\
6 & $C_d^T=+C_d$ & $C_d\G_{d+1}=-\G_{d+1}^T C_d$ & $C_{d+1}=C_d$  &
  $(C_{d+1}\G_I)^T=-C_{d+1}\G_I$ \\
0 & $C_d^T=+C_d$ & $C_d\G_{d+1}=+\G_{d+1}^T C_d$ & $C_{d+1}=C_d\G_{d+1}$ &
  $(C_{d+1}\G_I)^T=+C_{d+1}\G_I$
\\ \hline\hline
\end{tabular}
\end{center}
\begin{quotation}
\floatcaption{tabC}{Some properties of the charge conjugation
matrix.  These properties are periodic in the dimensionality $d$ modulo 8,
shown in the left column.  See App.~\ref{dconj}
for an explanation on the other columns.
}
\end{quotation}
\vspace*{-4ex}
\end{table}

\appendix

\begin{boldmath}
\section{\label{dconj} Charge conjugation matrix in $d$ dimensions}
\end{boldmath}
Here we review some properties of the charge conjugation matrix $C_d$
in euclidean space.  These properties are periodic in the
dimensionality $d$ modulo 8, as shown in Table~\ref{tabC}.
The basic relation satisfied by the charge conjugation matrix $C_d$
in $d=2n$ dimensions is
\begin{equation}
  C_d \G_I = - \G_I^T C_d \ , \qquad  I=1,2,\ldots,d \ .
\label{Cgamma}
\end{equation}
In addition, $C_d$ always satisfies $C_d^T=C_d^\dagger=C_d^{-1}$.
For the symmetry properties of $C_d$, see the second column of Table~\ref{tabC}.
The chirality matrix is defined by $\G_{2n+1}=e^{i\h_n}\G_1 \cdots \G_{2n}$,
where the phase $e^{i\h_n}$ is chosen such that $\G_{2n+1}^2=1$.
One has $C_{2n} \G_{2n+1} = \pm \G_{2n+1}^T C_{2n}$, where the sign
is given in the third column of Table~\ref{tabC}.  When this sign is negative,
Eq.~(\ref{Cgamma}) generalizes to include $\G_{2n+1}$, and we define
$C_{2n+1}=C_{2n}$.  When this sign is positive we define, instead,
$C_{2n+1}=C_{2n}\G_{2n+1}$, which implies
\begin{equation}
  C_{2n+1} \G_I = + \G_I^T C_{2n+1} \ , \qquad  I=1,2,\ldots,{2n+1}.
\label{Cgammaalt}
\end{equation}
In all cases, $C_{2n+1}$ has the same symmetry properties as $C_{2n}$.
It follows that for all $d$, the generators of $SO(d)$ rotations
on spinors, $\S_{IJ} = \frac{1}{4}[\G_I,\G_J]$, satisfy
\begin{equation}
  C_d \S_{IJ} = -\S_{IJ}^T C_d \ , \qquad  I,J=1,2,\ldots,d \ .
\label{CSig}
\end{equation}
For $d$ odd, the spinor \irrep\ is irreducible, and from the symmetry
properties of $C_d$ it follows that the spinor \irrep\ is real
for $d=1,7$ mod 8, and pseudoreal for $d=3,5$ mod 8.
For brevity, we will use the notation $C$ for the 4-dimensional
charge conjugation matrix, and $\cc$ for the charge conjugation matrix
in a given odd dimension.

The construction of the 4-component spinor $\c$ in Eq.~(\ref{majlike})
assumes the chiral representation of the Dirac matrices,
\begin{equation}
  \g_\m =
  \left(
  \begin{array}{cc}
    0      & \s_\m^\dagger  \\
    \s_\m & 0
  \end{array}
  \right) \ ,
\label{4dDirac}
\end{equation}
where $\s_4=1$, and $\s_\m$ is equal to $-i\s_k$ for $\m=k=1,2,3,$
where $\s_k$ are the Pauli matrices.
Also, $\g_5=diag(1,1,-1,-1)$, and,
as usual, $P_R=(1+\g_5)/2$ and $P_L=(1-\g_5)/2$.
The charge conjugation matrix is then
\begin{equation}
\label{Crep}
  C = \g_4\g_2 =
  \left(
  \begin{array}{cc}
   -\e & 0  \\
    0 & \e
  \end{array}
  \right) \ ,
\end{equation}
where $\e=i\s_2$.

\vspace{2ex}

\begin{boldmath}
\section{\label{appCP} Discrete symmetries}
\end{boldmath}
Here we discuss the discrete symmetries $C$, $P$ and $CP$
in $SO(d)$ gauge theories.
We first recall the familiar case of an $SU(N)$ gauge theory
with Dirac fermions in the fundamental \irrep.  Charge-conjugation symmetry
acts as
\begin{subequations}
\label{Csym}
\begin{eqnarray}
  \j &\to& C \bj^T \ , \qquad \bj \ \to \ \j^T C \ ,
\label{Csymj}\\
  A_\m &\to& -A_\m^T \ .
\label{CsymA}
\end{eqnarray}
\end{subequations}
Writing $A_\m=A_{\m a} T_a$ we infer the transformation rule
of the individual components, which is $A_{\m a}\to \mp A_{\m a}$
if $T_a^T=\pm T_a$.  Because all $SU(N)$ \irreps\ may be constructed
from tensor products of the fundamental \irrep,
these transformation rules remain valid for Dirac fermions in any \irrep.

We take parity to act as
\begin{subequations}
\label{Psym}
\begin{eqnarray}
  \j(x) &\to& i\g_4 \j(\tx) \ , \qquad
  \bj(x) \ \to -i \bj(\tx) \g_4 \ ,
\label{Psymj}\\
  A_\m(x) &\to& \tilde{A}_\m(\tx) \ .
\label{PsymA}
\end{eqnarray}
\end{subequations}
Here $\tx_\m=x_\m$ if $\m=4$, while $\tx_\m=-x_\m$ if $\m=1,2,3$.
A similar definition applies to $\tilde{A}_\m$.
The $C$ and $P$ fermion transformation rules, Eqs.~(\ref{Csymj}) and~(\ref{Psymj}),
both involve a choice of phase.  The reason for the particular choices
we have made is that we want the transformation rules to take the same
form for Majorana fermions.  If we replace the Dirac fermion $\j$
by a Majorana fermion $\c$, and $\bj$ by $\bc$, then $\bc$ is not
an independent field, but rather, it is related to $\c$ via Eq.~(\ref{maj}).
With the phases we have chosen in Eqs.~(\ref{Csymj}) and~(\ref{Psymj}),
these transformation rules are consistent with Eq.~(\ref{maj}).

Moving on to $SO(d)$ gauge theories, charge conjugation is still
given by Eq.~(\ref{Csymj}) for Dirac fermions in the fundamental, vector \irrep.
Because the generators in the vector \irrep\ are all antisymmetric,
the rule~(\ref{CsymA}) implies that the $SO(d)$ gauge field is
charge-conjugation invariant.
The rule for a Dirac fermion in a spinor \irrep\ is
\begin{equation}
  \h \ \to \ C \cc^T \bh^T \ , \qquad
  \bh \ \to \ \h^T C \cc \ ,
\label{Cspinor}
\end{equation}
where the presence of $\cc$ in Eq.~(\ref{Cspinor}) compensates for the fact
that the $SO(d)$ gauge field is invariant (note Eq.~(\ref{CSig})).
In the case of a real \irrep, the same rules (Eqs.~(\ref{CsymA}) or~(\ref{Cspinor}))
may be applied to Majorana fermions, because again our choice of phases
is consistent with Eq.~(\ref{maj}).  In fact, using Eq.~(\ref{maj}) it immediately
follows that Majorana fermions of $SO(d)$ gauge theories are
charge-conjugation invariant.  (For the case of a Majorana fermion
in the vector \irrep, the matrix $\cc$ in Eq.~(\ref{maj}) is replaced
by the identity matrix.)

We define $CP$ by first applying $P$ and then $C$.
The resulting transformation rules are given in Sec.~\ref{secCP}.
The rules for the gauge field, and for the Dirac and Majorana fermions
that we will encounter, follow from the transformation rules
we have already discussed above.

In the case of the $SU(4)/Sp(4)$ coset we have 4 Weyl fermions
in the pseudoreal spinor \irrep.  The discrete symmetries can be approached
in two ways.  First, we may assemble the 4 Weyl fermions into 2 Dirac fermions.
In this case, $P$ acts in the usual way, while $C$ acts as described above.
However, the Dirac formulation has the disadvantage that it obscures
the $SU(4)$ flavor symmetry of the pseudoreal Weyl fermions.\footnote{%
  A similar situation is discussed in Ref.~\cite{SU4sextet}.
}
The alternative we choose in this paper is to work in terms
of the 4-component fields $\c_i$ and $\bc_i$ introduced in Eqs.~(\ref{majlike})
and~(\ref{maj}), also for the pseudoreal case.
The advantage is that the flavor symmetry is manifest.
The separate $P$ and $C$ transformations will look more complicated
in terms of $\c_i$ and $\bc_i$, but, because of the properties of the
four-fermion lagrangian (Sec.~\ref{secCP}), we only need the explicit form of
the combined $CP$ transformation, which we can derive as follows.
We start from the observation that the Weyl action
\begin{equation}
\label{actionWeyl}
  S = \int d^4x\,\bU \s_\m D_\m \U \ ,
\end{equation}
is invariant under $CP$ symmetry where the $SO(d)$ gauge field
transforms as described above, and
\begin{equation}
\label{WeylCP}
  \U_i(x)\to i\cc\,\e\bU_i^T(\tx)\ ,\qquad
  \bU_i(x)\to i\U_i^T(\tx)\e\,\cc^T\ .
\end{equation}
In terms of the four-component fields $\c_i$ and $\bc_i$,
the transformation~(\ref{WeylCP}) takes the form
of Eq.~(\ref{CPd}) when the fermions belong to a pseudoreal \irrep.
For a real \irrep, we recover Eq.~(\ref{CPc}).

To avoid confusion, we recall that in the case of a real \irrep,
the action~(\ref{actionWeyl}) may be rewritten as
\begin{equation}
\label{Smaj}
  S = \half \int d^4x\, \bc_i \Sl{D} \c_i \ ,
\end{equation}
where the Majorana fermions are defined by Eqs.~(\ref{majlike})
and~(\ref{maj}).  But if we keep using the same 4-component fields
for a pseudoreal \irrep, then the right-hand side of Eq.~(\ref{Smaj})
will vanish identically.  Of course, for both real and pseudoreal \irreps\
we may recover Eq.~(\ref{actionWeyl}) from Eq.~(\ref{Smaj})
by inserting $2P_L$ between $\Sl{D}$ and $\c_i$.

\vspace{2ex}

\begin{boldmath}
\section{\label{su4sp4coset} The $SU(4)/Sp(4)$ coset}
\end{boldmath}
The $Sp(4)$ subgroup of $SU(4)$ is defined as the set of elements satisfying
\begin{equation}
\label{sp4def}
  g^T\e_0 g = \e_0\ ,
\end{equation}
where
\begin{equation}
\label{eps0}
\e_0=\t_3\times i\t_2 = \left( \begin{array}{cccc}
  0 & 1 & 0 & 0  \\
 -1 & 0 & 0 & 0  \\
  0 & 0 & 0 & -1 \\
  0 & 0 & 1 & 0  \\
  \end{array} \right) \ .
\end{equation}
(If $g\in Sp(4)$, then so is $g^T$.)
The 15 generators of $SU(4)$ split into 10 generators of $Sp(4)$,
\begin{equation}
\label{sp4basis}
  1\times\t_i\ ,\qquad
  \t_1\times 1\ ,\qquad
  \t_2\times\t_i\ ,
  \qquad\t_3\times\t_i\ ,
\end{equation}
and 5 generators for the coset $SU(4)/Sp(4)$,
\begin{equation}
\label{thecoset}
  \t_1\times\t_i\ ,\qquad\t_2\times 1\ ,\qquad\t_3\times 1\ ,
\end{equation}
where $\t_i$ are the Pauli matrices, and 1 stands for the $2\times 2$
identity matrix.  These generators satisfy
\begin{equation}
\label{sp4gendef}
  \e_0 T_a = \left\{ \begin{array}{ll}
    -T_a^T \e_0\ , & Sp(4)\ {\rm generators}\ , \\
    +T_a^T \e_0\ , & SU(4)/Sp(4)\ {\rm generators}\ .
  \end{array}\right.
\end{equation}
The tensor product of two fundamental $SU(4)$ \irreps\ contains
the six-dimensional anti-symmetric, and the ten-dimensional symmetric \irreps.
Under the reduction $SU(4)\to Sp(4)$, the {\bf 10} remains irreducible,
whereas the {\bf 6} reduces to a {\bf 5} and a singlet.
If $A_{ij}=-A_{ji}$ transforms in the {\bf 6} of $SU(4)$,
the singlet is $\tr(\e_0A)$, and the {\bf 5} is formed by
$A+\frac{1}{4}\e_0\,\tr(\e_0A)$.
The effective NGB field $\P$ introduced in Eq.~(\ref{vacsu4}) transforms
in the {\bf 5} of $Sp(4)$.

Following Ref.~\cite{ferretti16}, the Standard Model's $SU(2)_L$ and $SU(2)_R$
symmetries are identified with the subgroups of $Sp(4)$ with generators
\begin{eqnarray}
\label{su2s}
  T_L^i &=& \half(1+\t_3)\times\half\,\t_i\ ,
\\
  T_R^i &=& \half(1-\t_3)\times\half\,\t_i\ .
\nonumber
\end{eqnarray}
Correspondingly, the NGB field is parametrized as
\begin{eqnarray}
\label{pions}
  2\P &=& -(\Im{H_+})\,\t_1\times\t_1-(\Re{H_+})\,\t_1\times\t_2
\\
  && +(\Im{H_0})\,\t_1\times\t_3-(\Re{H_0})\,\t_2\times 1+\frac{1}{\sqrt{2}}\,
     \eta\,\t_3 \times 1
\nonumber\\
  &=&\left(
  \begin{array}{cccc}
    \eta/\sqrt{2} & 0 & iH_0^* & iH_+ \\
    0 & \eta/\sqrt{2} & -iH_+^*& iH_0 \\
    -iH_0 & iH_+ & -\eta/\sqrt{2} & 0 \\
    -iH_+^* & -iH_0^* & 0 & -\eta/\sqrt{2}
  \end{array} \right)\ .
\nonumber
\end{eqnarray}
The coset generators~(\ref{thecoset}) satisfy the 5-dimensional Dirac algebra.
(This property is closely related to the existence
of the isomorphisms $SU(4)/Z_2\simeq SO(6)$ and $Sp(4)/Z_2\simeq SO(5)$.)
Using
\begin{equation}
  \P^2 = \Big( \h^2 + 2H^\dagger H\Big)/8 \ ,
\label{Pisq}
\end{equation}
one can express $\S$ in closed form,
\begin{equation}
  \S = \left( \cos(\a) + \frac{2i}{\a f} \sin(\a)\P \right) \e_0\ ,
\label{Scsa}
\end{equation}
where
\begin{equation}
  \a^2 = ((1/2)\h^2 + H^\dagger H)/f^2 \ .
\label{Scsb}
\end{equation}

\vspace{2ex}

\begin{boldmath}
\section{\label{su5so5coset} The $SU(5)/SO(5)$ coset}
\end{boldmath}
The unbroken $SO(5)$ subgroup is generated by the 10 antisymmetric,
purely imaginary, generators of $SU(5)$.  We embed the generators of
$SU(2)_L\times SU(2)_R$, which is isomorphic to $SO(4)$,
in the upper-left $4\times4$ block.  They are given explicitly by
the following tensor products of the Pauli matrices \cite{ferretti16}
\begin{eqnarray}
  2\, T_L^1 &=&  \t_2 \times \t_1 \ ,
\label{tensorprod}\\
  2\, T_L^2 &=& -\t_2 \times \t_3 \ ,
\nonumber\\
  2\, T_L^3 &=& 1 \times \t_2 \ ,
\nonumber\\
  2\, T_R^1 &=& \t_1 \times \t_2 \ ,
\nonumber\\
  2\, T_R^2 &=& \t_2 \times 1 \ ,
\nonumber\\
  2\, T_R^3 &=& \t_3 \times \t_2 \ .
\nonumber
\end{eqnarray}
The non-linear field $\S\in SU(5)/SO(5)$ is expanded as
\begin{equation}
\label{Sigma}
  \S = \exp(i\P/f)\,\S_0\,\exp(i\P/f)^T
  = \exp\left(2i\Pi/f\right)\ ,
\end{equation}
where in the last equality we have set $\S_0=1$.  The pion field $\P$
is expanded in terms of the 14 real symmetric generators of $SU(5)$.
Its $SU(2)_L\times SU(2)_R$ content is
\begin{equation}
\label{Pi}
  \Pi = \Theta+\Theta^\dagger+\tPhi_0+\tPhi_++\tPhi_+^\dagger+\tileta\ ,
\end{equation}
where
\begin{equation}
\label{H}
  \Theta = \frac{1}{\sqrt{2}}\left(\begin{array}{ccccc}
    0&0&0&0&-iH_+\\
    0&0&0&0&H_+  \\
    0&0&0&0&iH_0 \\
    0&0&0&0&H_0  \\
    -iH_+&H_+&iH_0&H_0&0
  \end{array}\right)\ ,
\end{equation}
\begin{equation}
  \tPhi_0 = \rule{0ex}{11ex} \left(\begin{array}{ccccc}
       \f_0^0/\sqrt{2} & 0 & a & b & 0 \\
       0 & \f_0^0/\sqrt{2} & b & -a & 0 \\
       a & b & -\f_0^0/\sqrt{2} & 0 & 0 \\
       b & -a & 0 & -\f_0^0/\sqrt{2} & 0 \\
       0 & 0 & 0 & 0 & 0
  \end{array}\right)\ ,
\label{Phi0}
\end{equation}
\rule{0ex}{3ex}
with $a=(i/2)(\f_0^--\f_0^+)$ and $b=(1/2)(\f_0^-+\f_0^+)$,
\begin{equation}
  \tPhi_+ = \rule{0ex}{10ex} \left(\begin{array}{ccccc}
  \f_+^+/\sqrt{2}  & i\f_+^+/\sqrt{2} & i\f_+^0/2 & \f_+^0/2 & 0 \\
  i\f_+^+/\sqrt{2} & -\f_+^+/\sqrt{2} & -\f_+^0/2 & i\f_+^0/2 & 0 \\
  i\f_+^0/2 & -\f_+^0/2 & \f_+^-/\sqrt{2} & -i\f_+^-/\sqrt{2} & 0 \\
  \f_+^0/2  & i\f_+^0/2 & -i\f_+^-/\sqrt{2} & -\f_+^-/\sqrt{2} & 0 \\
       0 & 0 & 0 & 0 & 0
  \end{array}\right)\ ,
\label{Phip}
\end{equation}
and $\tileta=\eta\, {\rm diag}(1,1,1,1,-4)/\sqrt{20}$.
These conventions are the same as in Ref.~\cite{ferretti16}, except for
a slightly different normalization of the $\eta$ field.

The $\eta$ and $H$ fields, which we have already encountered in the
$SU(4)/Sp(4)$ case, constitute the $(1,1)$, respectively $(2,2)$,
representations of $SU(2)_L\times SU(2)_R$.  The $SU(5)/SO(5)$ coset
contains nine additional NGBs, the $\f$'s, that belong to $(3,3)$.
Their superscript and subscript label their $SU(2)_L$,
respectively $SU(2)_R$, quantum numbers.
The electric charge is $T_L^3+T_R^3$ for the coset fields,
hence the electric charge of each $\f$ field is the sum
of its superscript and subscript.
Complex conjugation works on the $\f$'s by interchanging
$+\leftrightarrow-$ for both the superscript and the subscript.
The $SU(2)_L$ triplets are $\F_0=\{\f_0^+,\f_0^0,\f_0^-\}$
(where $\f_0^0$ is real and $\f_0^- = (\f_0^+)^*$), with $T_R^3=0$,
$\F_+=\{\f_+^+,\f_+^0,\f_+^-\}$, with $T_R^3=+1$, and
$\F_-=(\F_+)^*=\{\f_-^-,\f_-^0,\f_-^+\}$, with $T_R^3=-1$.
The invariant bilinears are $\F_0^2 \equiv (\f_0^0)^2+2\f_0^+\f_0^-$ and
$\F_+\F_- \equiv \f_+^+ \f_-^- + \f_+^0 \f_-^0 + \f_+^- \f_-^+$.
We also introduce $2\times2$ matrix formats,
$\hF_0 = \f_0^0\, \t_3 + 2^{1/2} (\f_0^+ \t_+ + \f_0^- \t_-)$ and
$\hF_\pm = \f_\pm^0 \t_3 + 2^{1/2} ( -i\f_\pm^+ \t_+ + i\f_\pm^- \t_-)$,
where $\t_\pm = (\t_1 \pm i \t_2)/2$, which satisfy
$\tr(\hF_0^2) = 2\F_0^2$ and $\tr(\hF_+ \hF_-) = 2\F_+\F_-$.

\vspace{2ex}

\begin{boldmath}
\section{\label{Veff2nd} $\Veff$ at second order for $SU(5)/SO(5)$}
\end{boldmath}
In this appendix we list all the contributions to $\Veff$,
truncated to second order in the pNGB fields.
We use the expansion of the coset field $\S$ given in App.~\ref{su5so5coset},
and the expansion of the singlet NGB field $\F$ given in Eq.~(\ref{zeta}).
The $\svev{\cdot}$ notation is explained in Sec.~\ref{su4Veff},
and the list of templates for $\Veff$ may be found in Eq.~(\ref{V2}).

For template $\ct_1$ the spurions must be right-handed.  We get
\begin{subequations}
\label{SMn}
\begin{eqnarray}
  \svev{\F^{1-2q} \tr(\bS_R^1\S N_R) +\hc}
  &=& 8 -\frac{4}{f^2}
      \Big( (1/5)\h^2 + 2H^\dagger H + \F_0^2 + 2\F_+\F_-\Big) \hspace{5ex}
\label{SMnR4}\\
  && -\frac{8(1-2q)}{\sqrt{5}}\frac{\z\h}{f}
     -4(1-2q)^2\z^2 \ ,
\nonumber\\
  \svev{\F^{1-2q} \tr(\bS_R^2\S N_R) +\hc}
  &=& 2 -\frac{8}{f^2} \Big( (2/5)\h^2 + H^\dagger H\Big)
\label{SMnR5}\\
  && +\frac{8(1-2q)}{\sqrt{5}}\frac{\z\h}{f} -(1-2q)^2\z^2\ .
\nonumber
\end{eqnarray}
\end{subequations}
Since $t_R$ is embedded into the spurions $S_R$ and $S_R^c$
in the same way, the result for template $\ct_2$ is obtained from
the corresponding result for $\ct_1$ by replacing $1-2q$ with $1+2q$.
The results for template $\ct_3$ are
\begin{subequations}
\label{T3su5}
\begin{eqnarray}
\label{T3su5a}
\svev{\F^{1-2q}\tr(\bA_R\S D^{1T}_R)+\mbox{h.c.}}
&=& -2+\frac{1}{f^2}\left((1/5)\h^2+2H^\dagger H+2\F_0^2+4\F_+\F_-\right)
\hspace{7ex}\\
&& +\frac{2(1-2q)}{\sqrt{5}}\frac{\h\z}{f}+(1-2q)^2\z^2\ ,
\nonumber\\
\svev{\F^{1-2q}\tr(\bA_R\S D^{2T}_R)+\mbox{h.c.}}
&=&0\ .
\label{T3su5b}\\
\svev{\F^{1+2q}\tr(\bA_L\S D^{1T}_L)+\mbox{h.c.}}
&=& -8+\frac{32}{f^2}\left((2/5)\h^2+H^\dagger H\right)
\label{T3su5c}\\
&& -\frac{32(1+2q)}{\sqrt{5}}\frac{\h\z}{f}+4(1+2q)^2\z^2\ ,
\nonumber\\
\svev{\F^{1+2q}\tr(\bA_L\S D^{2T}_L)+\mbox{h.c.}}
&=& 8-\frac{4}{f^2}\left((1/5)\h^2+2H^\dagger H+2\F_0^2+4\F_+\F_-\right)
\label{T3su5d}\\
&& -\frac{8(1+2q)}{\sqrt{5}}\frac{\h\z}{f}-4(1+2q)^2\z^2\ ,
\nonumber
\end{eqnarray}
\end{subequations}
for template $\ct_4$,
\begin{subequations}
\label{T4su5}
\begin{eqnarray}
\svev{\F^{1-2q}\tr(\bS^1_R\S D^{1T}_R)+\mbox{h.c.}}
&=& 0 \ ,
\label{T4su5a}\\
\svev{\F^{1-2q}\tr(\bS^1_R\S D^{2T}_R)+\mbox{h.c.}}
&=& 8 -\frac{4}{f^2}
      \Big( (1/5)\h^2 + 2H^\dagger H + 2\F_0^2 + 4\F_+\F_-\Big) \hspace{5ex}
\label{T4su5b}\\
  && -\frac{8(1-2q)}{\sqrt{5}}\frac{\z\h}{f}
     -4(1-2q)^2\z^2 \ ,
\nonumber\\
\svev{\F^{1-2q}\tr(\bS^2_R\S D^{1T}_R)+\mbox{h.c.}}
&=& 0 \ ,
\label{T4su5c}\\
\svev{\F^{1-2q}\tr(\bS^2_R\S D^{2T}_R)+\mbox{h.c.}}
&=& -8 +\frac{32}{f^2} \Big( (2/5)\h^2 + H^\dagger H\Big)
\label{T4su5d}\\
  && -\frac{32(1-2q)}{\sqrt{5}}\frac{\z\h}{f} +4(1-2q)^2\z^2 \ .
\nonumber\\
\svev{\F^{1+2q}\tr(\bS_L\S D^{1T}_L)+\mbox{h.c.}}
&=& 8-\frac{32}{f^2}\left((2/5)\h^2+H^\dagger H\right)
\label{T4su5e}\\
&& +\,\frac{32(1+2q)}{\sqrt{5}}\frac{\h\z}{f}-4(1+2q)^2\z^2\ .
\nonumber\\
\svev{\F^{1+2q}\tr(\bS_L\S D^{2T}_L)+\mbox{h.c.}}
&=& 8-\frac{4}{f^2}\left((1/5)\h^2+2H^\dagger H+2\F_0^2+4\F_+\F_-\right)
\label{T4su5f}\\
&& -\frac{8(1+2q)}{\sqrt{5}}\frac{\h\z}{f}-4(1+2q)^2\z^2\ ,
\nonumber
\end{eqnarray}
\end{subequations}
for template $\ct_5$,
\begin{subequations}
\label{T5su5}
\begin{eqnarray}
\svev{\F^{-1-2q}\tr(\bA_R^c\S^* D^1_R)+\mbox{h.c.}}
&=& 2-\frac{1}{f^2}\left((1/5)\h^2+2H^\dagger H+2\F_0^2+4\F_+\F_-\right)
\hspace{7ex}
\label{T5su5a}\\
&& -\frac{2(1+2q)}{\sqrt{5}}\frac{\h\z}{f}-(1+2q)^2\z^2\ ,
\nonumber\\
\svev{\F^{-1-2q}\tr(\bA_R^c\S^* D^2_R)+\mbox{h.c.}}
&=&0\ .
\label{T5su5b}\\
\svev{\F^{-1+2q}\tr(\bA_L^c\S^* D^1_L)+\mbox{h.c.}}
&=& 8-\frac{4}{f^2}\left((1/5)\h^2+2H^\dagger H+2\F_0^2+4\F_+\F_-\right)
\label{T5su5c}\\
&& -\frac{8(1-2q)}{\sqrt{5}}\frac{\h\z}{f}-4(1-2q)^2\z^2\ ,
\nonumber\\
\svev{\F^{-1+2q}\tr(\bA_L^c\S^* D^2_L)+\mbox{h.c.}}
&=&-8+\frac{32}{f^2}\left((2/5)\h^2+H^\dagger H\right)
\label{T5su5d}\\
&& -\frac{32(1-2q)}{\sqrt{5}}\frac{\h\z}{f}+4(1-2q)^2\z^2\ ,
\nonumber
\end{eqnarray}
\end{subequations}
and for template $\ct_6$,
\begin{subequations}
\label{T6su5}
\begin{align}
\svev{\F^{-1-2q}\tr(\bS^{1c}_R\S^* D^1_R)+\mbox{h.c.}}
&= 0 \ ,
\allowdisplaybreaks
\label{T6su5a}\\
\svev{\F^{-1-2q}\tr(\bS^{1c}_R\S^* D^2_R)+\mbox{h.c.}}
&= 8 -\frac{4}{f^2}
      \Big( (1/5)\h^2 + 2H^\dagger H + 2\F_0^2 + 4\F_+\F_-\Big)
\hspace{7ex}
\label{T6su5b}\\
& \quad -\frac{8(1+2q)}{\sqrt{5}}\frac{\z\h}{f}
     -4(1+2q)^2\z^2 \ ,
\allowdisplaybreaks
\nonumber\\
\svev{\F^{-1-2q}\tr(\bS^{2c}_R\S^* D^1_R)+\mbox{h.c.}}
&= 0 \ ,
\label{T6su5c}\\
\svev{\F^{-1-2q}\tr(\bS^{2c}_R\S^* D^2_R)+\mbox{h.c.}}
&= -8 +\frac{32}{f^2} \Big( (2/5)\h^2 + H^\dagger H\Big)
\label{T6su5d}\\
& \quad -\frac{32(1+2q)}{\sqrt{5}}\frac{\z\h}{f} +4(1+2q)^2\z^2 \ .
\allowdisplaybreaks
\nonumber\\
\svev{\F^{-1+2q}\tr(\bS^c_L\S^* D^1_L)+\mbox{h.c.}}
&= 8-\frac{4}{f^2}\left((1/5)\h^2+2H^\dagger H+2\F_0^2+4\F_+\F_-\right)
\label{T6su5e}\\
& \quad -\frac{8(1-2q)}{\sqrt{5}}\frac{\h\z}{f}-4(1-2q)^2\z^2\ ,
\allowdisplaybreaks
\nonumber\\
\svev{\F^{-1+2q}\tr(\bS^c_L\S^* D^2_L)+\mbox{h.c.}}
&= 8-\frac{32}{f^2}\left((2/5)\h^2+H^\dagger H\right)
\label{T6su5f}\\
& \quad +\,\frac{32(1-2q)}{\sqrt{5}}\frac{\h\z}{f}-4(1-2q)^2\z^2\ .
\allowdisplaybreaks
\nonumber
\end{align}
\end{subequations}
For $\ct_{7}$ we obtain
\begin{subequations}
\label{SMa}
\begin{eqnarray}
  \svev{\tr(\bS_L\S)\tr(S_L\S^*)}
  &=& \frac{32}{f^2}\, H^\dagger H \ ,
\label{SMaL}\\
  \svev{\tr(\bS_R^1\S)\tr(S_R^1\S^*)}
  &=& 16-\frac{32}{f^2}\, \left( H^\dagger H + \F_0^2 + 2\F_+\F_- \right) \ ,
\label{SMaR44}\\
  \svev{\tr(\bS_R^2\S)\tr(S_R^2\S^*)}
  &=& 1-\frac{8}{f^2}\, H^\dagger H \ , \hspace{8ex}
\label{SMaR55}\\
  \svev{\tr(\bS_R^1\S)\tr(S_R^2\S^*)+\hc}
  &=& 8-\frac{4}{f^2} \left(5\h^2 +10H^\dagger H +2\F_0^2 +4\F_+\F_- \right) \ .
  \hspace{5ex}
\label{SMaR45}
\end{eqnarray}
\end{subequations}
Again the second-order results for template $\ct_{8}$ can be obtained
by replacing $S$ with $S^c$ and $\S$ with $\S^*$ on the left-hand sides,
while keeping the right-hand sides unchanged.
The results for template $\ct_{9}$ are
\begin{subequations}
\label{SMc}
\begin{eqnarray}
  \svev{\F^2 \tr(S_L^c\S) \tr(\bS_L\S)+\hc}
  &=& -\,\frac{64}{f^2}\, H^\dagger H \ ,
\label{SMcL}\\
  \svev{\F^2 \tr(S_R^{1c}\S) \tr(\bS_R^1\S)+\hc}
  &=& -\frac{64}{f^2}\,\Big( (1/5)\eta^2 + H^\dagger H
  + \Phi_0^2 +2\Phi_+\Phi_- \Big)
\label{SMcR44}\\
  && -\frac{128}{\sqrt{5}}\frac{\z\h}{f} -64\z^2 \ ,
\nonumber\\
  \svev{\F^2 \tr(S_R^{2c}\S) \tr(\bS_R^2\S)+\hc}
  &=& -\frac{16}{f^2}\,\Big( (4/5)\eta^2 + H^\dagger H \Big)
  +\frac{32}{\sqrt{5}}\frac{\z\h}{f} -4\z^2 \ ,
\label{SMcR55}\\
  \svev{\F^2 \tr(S_R^{1c}\S) \tr(\bS_R^2\S)+\hc}
  &=& -\frac{8}{f^2}\,\Big( (9/10)\eta^2 + 5H^\dagger H
  + \Phi_0^2 +2\Phi_+\Phi_- \Big)
\label{SMcR45}\\
  && +\frac{48}{\sqrt{5}}\frac{\z\h}{f} -16\z^2 \ ,
\nonumber\\
  \svev{\F^2 \tr(S_R^{2c}\S) \tr(\bS_R^1\S)+\hc}
  &=& -\frac{8}{f^2}\,\Big( (9/10)\eta^2 + 5H^\dagger H
  + \Phi_0^2 +2\Phi_+\Phi_- \Big)
  \hspace{8ex}
\label{SMcR54}\\
  && +\frac{48}{\sqrt{5}}\frac{\z\h}{f} -16\z^2 \ ,
\nonumber
\end{eqnarray}
\end{subequations}
for template $\ct_{10}$,
\begin{subequations}
\label{SMd}
\begin{eqnarray}
  \svev{\F^2\tr(S_L^c\S \bS_L\S)+\hc}
  &=& -\frac{16}{f^2}\, \Big( (9/10)\h^2 + 7 H^\dagger H
  +\Phi_0^2 +2\Phi_-\Phi_+ \Big)
\label{SMdL}\\
  && +\frac{96}{\sqrt{5}}\frac{\z\h}{f} -32\z^2 \ ,
\nonumber\\
  \svev{\F^2\tr(S_R^{1c}\S \bS_R^1\S)+\hc}
  &=&  -\frac{16}{f^2}\, \Big( (1/5)\h^2 + H^\dagger H
  +2\Phi_0^2 +4\Phi_-\Phi_+ \Big)
\label{SMdR44}\\
  && -\frac{32}{\sqrt{5}}\frac{\z\h}{f} -16\z^2 \ ,
\nonumber\\
  \svev{\F^2\tr(S_R^{2c}\S \bS_R^2\S)+\hc}
  &=& -\frac{1}{f^2}\, \Big( (64/5)\h^2 + 16H^\dagger H \Big)
      +\frac{32}{\sqrt{5}}\frac{\z\h}{f} -4\z^2 \ , \hspace{5ex}
\label{SMdR55}\\
  \svev{\F^2\tr(S_R^{1c}\S \bS_R^2\S)+\hc}
  &=& -\frac{16}{f^2}\, H^\dagger H \ ,
\label{SMdR45}\\
  \svev{\F^2\tr(S_R^{2c}\S \bS_R^1\S)+\hc}
  &=& -\frac{16}{f^2}\, H^\dagger H \ ,
\label{SMdR54}
\end{eqnarray}
\end{subequations}
for template $\ct_{11}$,
\begin{subequations}
\label{AM}
\begin{eqnarray}
  \svev{\F^2\tr(A_L^c\S \bA_L\S) +\hc}
  &=& -\frac{16}{f^2} \left((9/10)\eta^2+3H^\dagger H
  +\Phi_0^2+2\Phi_+\Phi_-\right)
\label{AML}\\
  && +\frac{96}{\sqrt{5}}\frac{\z\h}{f} -32\z^2 \ ,
\nonumber\\
  \svev{\F^2\tr(A_R^c\S \bA_R\S) +\hc}
  &=& -\frac{4}{f^2} \left((1/5)\eta^2+H^\dagger H+2\Phi_0^2\right)
      -\frac{8}{\sqrt{5}}\frac{\z\h}{f} -4\z^2 \ ,
  \hspace{7ex} \ .
\label{AMR}
\end{eqnarray}
\end{subequations}
and for template $\ct_{12}$,
\begin{subequations}
\label{DM}
\begin{eqnarray}
  \svev{\tr(D_L^1\S \bD_L^{1T}\S^*)}
  &=& \frac{8}{f^2}\,H^\dagger H\ ,
\label{DMvv}\\
  \svev{\tr(D_L^2\S \bD_L^{2T}\S^*)}
  &=& \frac{8}{f^2}\,H^\dagger H\ ,
\label{DMhh}\\
  \svev{\tr(D_L^1\S \bD_L^{2T}\S^*)+\hc}
  &=& -\frac{8}{f^2}\left((5/2)\eta^2+5H^\dagger H
  +\Phi_0^2+2\Phi_+\Phi_-\right)\ ,
  \hspace{8ex}
\label{DMvh}\\
  \svev{\tr(D_R^1\S \bD_R^{1T}\S^*)}
  &=&-\frac{2}{f^2}\left(H^\dagger H+4\Phi_+\Phi_-\right)\ ,
\label{DMaa}\\
  \svev{\tr(D_R^2\S \bD_R^{2T}\S^*)}
  &=&-\frac{200}{f^2}\,H^\dagger H\ ,
\label{DMdd}\\
  \svev{\tr(D_R^1\S \bD_R^{2T}\S^*)+\hc} &=& 0 \ .
\label{DMad}
\end{eqnarray}
\end{subequations}
Notice that $\svev{\tr(D_L^2\S \bD_L^{1T}\S^*)+\hc}
=\svev{\tr(D_L^1\S \bD_L^{2T}\S^*)+\hc}$.

\vspace{2ex}

\begin{boldmath}
\section{\label{Veffphi} $\Veff$ for the $h$ and $\varphi$ fields}
\end{boldmath}
In this appendix we obtain the exact form of the effective potential,
assuming that the $SU(5)/SO(5)$ pion field is given by $\P=V$, where
\begin{equation}
\label{Vdef}
V=\left(\begin{array}{ccccc}
    0&0&0&0&0\\
    0&0&0&0&0  \\
    0&0&0&\varphi&0 \\
    0&0&\varphi&0&h  \\
    0&0&0&h&0
  \end{array}\right)\ .
\end{equation}
A comparison to Eq.~(\ref{Pi}) shows that this corresponds to keeping
$h=\sqrt{2} \Re H_0$ and $\varphi=\sqrt{2}\, \Im \f_+^-$ arbitrary,
while turning off the other 12 pNGBs.  (Like $H_0$,
also $\f_+^-$ is electrically neutral.)  The coset field is then given by
\begin{eqnarray}
\label{Sigphi}
  \S &=& 1 + \frac{2iV}{f}\,\frac{\sin(\a)}{\a}
      -\frac{4V^2}{f^2}\,\frac{1-\cos(\a)}{\a^2}
\\
     &=& 1 + \frac{2iV}{f}\,\hs -\frac{4V^2}{f^2}\, \hatc \ ,
\nonumber
\end{eqnarray}
where
\begin{equation}
\label{Vsq}
V^2=\left(\begin{array}{ccccc}
    0&0&0&0&0\\
    0&0&0&0&0  \\
    0&0&\varphi^2&0&\varphi h \\
    0&0&0&\varphi^2+h^2&0  \\
    0&0&\varphi h&0&h^2
  \end{array}\right)\ ,
\end{equation}
$\a^2=4(h^2+\varphi^2)/f^2$, and we have introduced the shorthands
$\hs=\sin(\a)/\a$ and $\hatc=(1-\cos(\a))/\a^2$.

The contributions to the effective potential from template $\ct_1$ are
\begin{subequations}
\label{phiSMn}
\begin{eqnarray}
  \svev{\F^{1-2q} \tr(\bS_R^1\S N_R) +\hc}
  &=& 8 \cos((1-2q)\z) \left(1-\hatc\,\frac{2\varphi^2+h^2}{f^2} \right) \ ,
\label{phiSMnR4}\\
  \svev{\F^{1-2q} \tr(\bS_R^2\S N_R) +\hc}
  &=& \cos((1-2q)\z) \left(2-8\hatc\,\frac{h^2}{f^2} \right) \ .
\label{phiSMnR5}
\end{eqnarray}
\end{subequations}
The contributions of template $\ct_2$ are again obtained from those of $\ct_1$
by replacing $1-2q$ with $1+2q$.
For $\ct_3$ we have
\begin{subequations}
\label{T3su5V}
\begin{eqnarray}
\svev{\F^{1-2q}\tr(\bA_R\S D^{1T}_R)+\mbox{h.c.}}
&=& -2 \cos((1-2q)\z) \left(1-\hatc\,\frac{2\varphi^2+h^2}{f^2} \right) \ ,
\label{T3su5Va}\\
\svev{\F^{1-2q}\tr(\bA_R\S D^{2T}_R)+\mbox{h.c.}}
&=& 0 \ ,
\label{T3su5Vb}\\
\svev{\F^{1+2q}\tr(\bA_L\S D^{1T}_L)+\mbox{h.c.}}
&=& -8\cos((1+2q)\z) \left(1-4\hatc\,\frac{h^2}{f^2} \right) \ ,
\label{T3su5Vc}\\
\svev{\F^{1+2q}\tr(\bA_L\S D^{2T}_L)+\mbox{h.c.}}
&=& 8 \cos((1+2q)\z) \left(1-\hatc\,\frac{2\varphi^2+h^2}{f^2} \right) \ ,
\label{T3su5Vd}
\end{eqnarray}
\end{subequations}
for $\ct_4$,
\begin{subequations}
\label{T4su5V}
\begin{eqnarray}
\svev{\F^{1-2q}\tr(\bS^1_R\S D^{1T}_R)+\mbox{h.c.}}
&=& 0 \ ,
\label{T4su5Va}\\
\svev{\F^{1-2q}\tr(\bS^1_R\S D^{2T}_R)+\mbox{h.c.}}
&=& 8 \cos((1-2q)\z) \left(1-\hatc\,\frac{2\varphi^2+h^2}{f^2} \right) \ ,
\label{T4su5Vb}\\
\svev{\F^{1-2q}\tr(\bS^2_R\S D^{1T}_R)+\mbox{h.c.}}
&=& 0 \ ,
\label{T4su5Vc}\\
\svev{\F^{1-2q}\tr(\bS^2_R\S D^{2T}_R)+\mbox{h.c.}}
&=& -8\cos((1-2q)\z) \left(1-4\hatc\,\frac{h^2}{f^2} \right) \ ,
\label{T4su5Vd}\\
\svev{\F^{1+2q}\tr(\bS_L\S D^{1T}_L)+\mbox{h.c.}}
&=& 8\cos((1-2q)\z) \left(1-4\hatc\,\frac{h^2}{f^2} \right) \ ,
\label{T4su5Ve}\\
\svev{\F^{1+2q}\tr(\bS_L\S D^{2T}_L)+\mbox{h.c.}}
&=& 8 \cos((1-2q)\z) \left(1-\hatc\,\frac{2\varphi^2+h^2}{f^2} \right) \ ,
\label{T4su5Vf}
\end{eqnarray}
\end{subequations}
for $\ct_5$
\begin{subequations}
\label{T5su5V}
\begin{eqnarray}
\svev{\F^{-1-2q}\tr(\bA^c_R\S^* D^1_R)+\mbox{h.c.}}
&=& 2 \cos((1+2q)\z) \left(1-\hatc\,\frac{2\varphi^2+h^2}{f^2} \right) \ ,
\label{T5su5Va}\\
\svev{\F^{-1-2q}\tr(\bA^c_R\S^* D^2_R)+\mbox{h.c.}}
&=& 0 \ ,
\label{T5su5Vb}\\
\svev{\F^{-1+2q}\tr(\bA^c_L\S^* D^1_L)+\mbox{h.c.}}
&=& 8 \cos((1-2q)\z) \left(1-\hatc\,\frac{2\varphi^2+h^2}{f^2} \right) \ ,
\label{T5su5Vc}\\
\svev{\F^{-1+2q}\tr(\bA^c_L\S^* D^2_L)+\mbox{h.c.}}
&=& -8\cos((1-2q)\z) \left(1-4\hatc\,\frac{h^2}{f^2} \right) \ ,
\label{T5su5Vd}
\end{eqnarray}
\end{subequations}
and for $\ct_6$,
\begin{subequations}
\label{T6su5V}
\begin{eqnarray}
\svev{\F^{-1-2q}\tr(\bS^{1c}_R\S^* D^1_R)+\mbox{h.c.}}
&=& 0 \ ,
\label{T6su5Va}\\
\svev{\F^{-1-2q}\tr(\bS^{1c}_R\S^* D^2_R)+\mbox{h.c.}}
&=& 8 \cos((1+2q)\z) \left(1-\hatc\,\frac{2\varphi^2+h^2}{f^2} \right) \ ,
\label{T6su5Vb}\\
\svev{\F^{-1-2q}\tr(\bS^{2c}_R\S^* D^1_R)+\mbox{h.c.}}
&=& 0 \ ,
\label{T6su5Vc}\\
\svev{\F^{-1-2q}\tr(\bS^{2c}_R\S^* D^2_R)+\mbox{h.c.}}
&=& -8\cos((1+2q)\z) \left(1-4\hatc\,\frac{h^2}{f^2} \right) \ ,
\label{T6su5Vd}\\
\svev{\F^{-1+2q}\tr(\bS^c_L\S^* D^1_L)+\mbox{h.c.}}
&=& 8 \cos((1-2q)\z) \left(1-\hatc\,\frac{2\varphi^2+h^2}{f^2} \right) \ ,
\label{T6su5Ve}\\
\svev{\F^{-1+2q}\tr(\bS^c_L\S^* D^2_L)+\mbox{h.c.}}
&=& 8\cos((1-2q)\z) \left(1-4\hatc\,\frac{h^2}{f^2} \right) \ .
\label{T6su5Vf}
\end{eqnarray}
\end{subequations}
The results for template $\ct_{7}$ are
\begin{subequations}
\label{phiSMa}
\begin{align}
  \svev{\tr(\bS_L\S)\tr(S_L\S^*)}
  &= 16\, \frac{h^2}{f^2} \left(\hs-2\hatc\,\frac{\varphi}{f}\right)^2 \ ,
\allowdisplaybreaks
\label{phiSMaL}\\
  \svev{\tr(\bS_R^1\S)\tr(S_R^1\S^*)}
  &= 16 \left(-1+\hatc\,\frac{2\varphi^2+h^2}{f^2}\right)^2 \ ,
\allowdisplaybreaks
\label{phiSMaR44}\\
  \svev{\tr(\bS_R^2\S)\tr(S_R^2\S^*)}
  &= \left(-1+4\hatc\,\frac{h^2}{f^2}\right)^2 \ ,
\allowdisplaybreaks
\label{phiSMaR55}\\
  \svev{\tr(\bS_R^1\S)\tr(S_R^2\S^*)}
  &= 4\left(-1+4\hatc\,\frac{h^2}{f^2}\right)
       \left(-1+\hatc\,\frac{2\varphi^2+h^2}{f^2}\right) \ .
\label{phiSMaR45}
\end{align}
\end{subequations}
The expansion of Eq.~(\ref{phiSMaL}) contains the cubic term
$h^2\varphi$ (compare Eq.~(\ref{SMaL3rd})).

In the case of template $\ct_{8}$,
the results for the right-handed spurions may again be obtained from
those for $\ct_{7}$ as in App.~\ref{Veff2nd}.
For the contribution of $q_L$ we now find
\begin{equation}
  \svev{\tr(S^c_L\S)\tr(\bS^c_L\S^*)}
  = 16\, \frac{h^2}{f^2} \left(\hs+2\hatc\,\frac{\varphi}{f}\right)^2 \ .
\label{phiSMbL}
\end{equation}
This differs from Eq.~(\ref{phiSMaL}) by the relative sign inside the square,
and again contains a cubic term.   We observe that a cancellation of
the cubic terms between the contributions of  Eqs.~(\ref{phiSMaL})
and~(\ref{phiSMbL}), while technically possible, is unlikely.
In $\Veff$, each contribution gets multiplied by two coupling constants
from $\cl_{EHC}$, and by a low-energy constant.
In order for this cancellation to happen,
the ratio of the relevant low-energy constants, which is a feature of the
stand-alone hypercolor theory, would have to be equal
to the ratio of the coupling constants squared,
which are features of the EHC theory.

The results for template $\ct_{9}$ are
\begin{subequations}
\label{phiSMc}
\begin{eqnarray}
  \svev{\F^2 \tr(S_L^c\S) \tr(\bS_L\S)+\hc}
  &=& 32\cos(2\z)\, \frac{h^2}{f^2}
  \left(4\hatc^2\,\frac{\varphi^2}{f^2} - \hs^2\right) \ ,
\label{phiSMcL}\\
  \svev{\F^2 \tr(S_R^{1c}\S) \tr(\bS_R^1\S)+\hc}
  &=& 32\cos(2\z) \left(-1+\hatc\,\frac{2\varphi^2+h^2}{f^2}\right)^2 \ ,
\label{phiSMcR44}\\
  \svev{\F^2 \tr(S_R^{2c}\S) \tr(\bS_R^2\S)+\hc}
  &=& 2\cos(2\z) \left(-1+4\hatc\,\frac{h^2}{f^2}\right)^2 \ ,
\label{phiSMcR55}\\
  \svev{\F^2 \tr(S_R^{1c}\S) \tr(\bS_R^2\S)+\hc}
  &=& 8\cos(2\z) \left(-1+4\hatc\,\frac{h^2}{f^2}\right)
       \left(-1+\hatc\,\frac{2\varphi^2+h^2}{f^2}\right) ,
  \hspace{8ex}
\label{phiSMcR45}
\end{eqnarray}
\end{subequations}
for template $\ct_{10}$,
\begin{subequations}
\label{phiSMd}
\begin{eqnarray}
  \svev{\F^2\tr(S_L^c\S \bS_L\S)+\hc}
  &=& 16\cos(2\z)\,\left(1 - \frac{(\hs^2+5\hatc)h^2 + 2\hatc\varphi^2}{f^2}
  \right.
\label{phiSMdL}\\
  && \left. +4\hatc^2\,\frac{h^2(h^2+3\varphi^2)}{f^4} \right) \ ,
\nonumber\\
  \svev{\F^2\tr(S_R^{1c}\S \bS_R^1\S)+\hc}
  &=& 8\cos(2\z)\,\left(1 - \frac{2\hatc h^2+(4\hatc+2\hs^2)\varphi^2}{f^2}
  \right.
\label{phiSMR44}\\
  && \left. +4\hatc^2\,\frac{2\varphi^4+2\varphi^2h^2+h^4}{f^4} \right) \ ,
\nonumber\\
  \svev{\F^2\tr(S_R^{2c}\S \bS_R^2\S)+\hc}
  &=& 2\cos(2\z)\,\left(-1+4\hatc\,\frac{h^2}{f^2}\right)^2 \ ,
\label{phiSMdR55}\\
  \svev{\F^2\tr(S_R^{1c}\S \bS_R^2\S)+\hc}
  &=& 8\cos(2\z)\,\frac{h^2}{f^2}
  \left(-\hs^2+4\hatc^2\,\frac{\varphi^2}{f^2} \right) \ ,
\label{phiSMdR45}
\end{eqnarray}
\end{subequations}
for template $\ct_{11}$,
\begin{subequations}
\label{phiAM}
\begin{eqnarray}
  \svev{\F^2\tr(A_L^c\S \bA_L\S) +\hc}
  &=& 16\cos(2\z)\,\left(1 + \frac{(\hs^2-5\hatc)h^2 -2\hatc\varphi^2}{f^2}
  \right.
\label{phiAML}\\
  && \left.
  + 4\hatc^2\frac{h^2(\varphi^2+h^2)}{f^4} \right) \ ,
\nonumber\\
  \svev{\F^2\tr(A_R^c\S \bA_R\S) +\hc}
  &=& 2\cos(2\z)\,\left(1 + 2\frac{(\hs^2-2\hatc)\varphi^2-\hatc h^2}{f^2}
  \right.
\label{phiAMR}\\
  && \left.
  + 8\hatc^2\frac{\varphi^2(\varphi^2+h^2)}{f^4} \right) \ ,
\nonumber
\end{eqnarray}
\end{subequations}
and finally, for template $\ct_{12}$,
\begin{subequations}
\label{phiDM}
\begin{eqnarray}
  \svev{\tr(D_L^1\S \bD_L^{1T}\S^*)}
  &=& 4\,\frac{h^2}{f^2} \left(\hs - 2\hatc\,\frac{\varphi}{f} \right)^2 \ ,
\label{phiDMvv}\\
  \svev{\tr(D_L^2\S \bD_L^{2T}\S^*)}
  &=& 4\,\frac{h^2}{f^2} \left(\hs + 2\hatc\,\frac{\varphi}{f} \right)^2 \ ,
\label{phiDMhh}\\
  \svev{\tr(D_L^2\S \bD_L^{1T}\S^*)}
  &=& 4\left(-1+4\hatc\,\frac{h^2}{f^2}\right)
  \left(-1+\hatc\,\frac{2\varphi^2+h^2}{f^2}\right) \ ,
\label{phiDMhv}\\
  \svev{\tr(D_L^1\S \bD_L^{2T}\S^*)}
  &=& 4\left(-1+4\hatc\,\frac{h^2}{f^2}\right)
  \left(-1+\hatc\,\frac{2\varphi^2+h^2}{f^2}\right) \ ,
\label{phiDMvh}\\
  \svev{\tr(D_R^1\S \bD_R^{1T}\S^*)}
  &=& -1 +(2\hs^2+4\hatc)\frac{\varphi^2}{f^2} +2\hatc\,\frac{h^2}{f^2}
  -8\hatc^2\,\frac{\varphi^2(\varphi^2+h^2)}{f^4} \ , \hspace{7ex}
\label{phiDMaa}\\
  \svev{\tr(D_R^2\S \bD_R^{2T}\S^*)}
  &=& 20 +8(\hs^2-2\hatc)\frac{\varphi^2}{f^2} -8(4\hs^2+17\hatc)\frac{h^2}{f^2}
\label{phiDMdd}\\
  && + 16\hatc^2\,\frac{2\varphi^4-6\varphi^2h^2+17h^4}{f^4} \ ,
\nonumber\\
  \svev{\tr(D_R^1\S \bD_R^{2T}\S^*)}
  &=& -40\hatc\hs\,\frac{h^2\varphi}{f^3} \ ,
\label{phiDMad}\\
  \svev{\tr(D_R^2\S \bD_R^{1T}\S^*)}
  &=& -40\hatc\hs\,\frac{h^2\varphi}{f^3} \ .
\label{phiDMda}
\end{eqnarray}
\end{subequations}

An interesting property is that, for the parametrization~(\ref{Sigphi}),
only one odd function of the pNGB fields occurs
in the potential, namely, $\hatc\hs h^2\varphi$ (see Sec.~\ref{su5Veff}
for further discussion).  By contrast, the effective potential
depends on a large number of even functions of the pNGBs.
For completeness, the gauge sector's contribution for the
parametrization~(\ref{Sigphi}) is\footnote{%
  For a lattice calculation of $C_{LR}$ in a lattice theory which is
  closely related to the $SU(4)$ model of Sec.~\ref{fix}, see Ref.~\cite{CLR}.
}
\begin{eqnarray}
  V_{\rm eff}^{\rm gauge}
  &=& -g^2 C_{LR} \left( 3 -\frac{(12\hatc+2\hs^2)\varphi^2+6\hatc h^2}{f^2}
  +\frac{8\hatc^2\varphi^2(\varphi^2+h^2)}{f^4}\right)
\label{Vggphi}\\
  && -{g'}^2 C_{LR} \left( 1 -\frac{(4\hatc+2\hs^2)\varphi^2+2\hatc h^2}{f^2}
  +\frac{8\hatc^2\varphi^2(\varphi^2+h^2)}{f^4}\right) \ .
\nonumber
\end{eqnarray}
This contribution is the same for all the $SU(5)/SO(5)$ model,
including in particular the models we consider in Sec.~\ref{su5so5}
and in Sec.~\ref{fix}.

\vspace{3ex}

\end{document}